\documentclass{article}
\usepackage{arxiv}
\usepackage[utf8]{inputenc} % allow utf-8 input
\usepackage[T1]{fontenc}    % use 8-bit T1 fonts
\usepackage{hyperref}       % hyperlinks
\usepackage{url}            % simple URL typesetting
\usepackage{booktabs}       % professional-quality tables
\usepackage{amsfonts}       % blackboard math symbols
\usepackage{nicefrac}       % compact symbols for 1/2, etc.
\usepackage{microtype}      % microtypography
\usepackage{lipsum}
\usepackage{graphicx}

\usepackage{authblk}
\usepackage[numbers]{natbib}
\usepackage{amsmath}
\usepackage{multirow}
\usepackage{diagbox}
\usepackage{amssymb}
\usepackage{lscape}
\usepackage{threeparttable}
\usepackage{lineno,bm}
\usepackage{makecell}
\usepackage{color}

\graphicspath{{./depth_images/}, {./ssimmaps/}}

\usepackage{soul}  %underlining, overstriking and highlighting...
\setul{}{0.2ex}  % \setul{underline depth}{underline thickness}

\title{Quality Assessment of DIBR-synthesized views: An Overview}

\author[a,b]{Shishun Tian}
\author[c,d]{Lu Zhang}
\author[a,b]{Wenbin Zou}
\author[a,b]{Xia Li}
\author[e]{Ting, Su}
\author[c,d]{Luce, Morin}
\author[c,d]{Olivier, D\'eforges}
\affil[a]{College of Electronics and Information Engineering, Shenzhen University, Shenzhen 518060, China.}
\affil[b]{Guangdong Key Laboratory of Intelligent Information Processing, Shenzhen University, Shenzhen 518060, China.}
\affil[c]{National Institute of Applied Sciences of Rennes (INSA Rennes), Rennes, France.}
\affil[d]{IETR (Institut d'Electronique et des Technologies du num\'Rique),\\
 UMR CNRS 6164, Rennes, France.}
\affil[e]{Research Center for Medical Articial Intelligence, Shenzhen Institutes of Advanced Technology, \\
Chinese Academy of Sciences, Shenzhen, Guangdong 518055, China.}

\begin{document}

\maketitle

\begin{abstract}
The Depth-Image-Based-Rendering (DIBR) is one of the main fundamental technique to generate new views in 3D video applications, such as Multi-View Videos (MVV), Free-Viewpoint Videos (FVV) and Virtual Reality (VR). However, the quality assessment of DIBR-synthesized views is quite different from the traditional 2D images/videos. In recent years, several efforts have been made towards this topic, but there {is a lack of} detailed survey in {the} literature. In this paper, we provide a comprehensive survey on various current approaches for DIBR-synthesized views. The current accessible datasets of DIBR-synthesized views are firstly reviewed{, followed} by a summary analysis of the representative state-of-the-art objective metrics. Then, the performances of different objective metrics are evaluated and discussed on all available datasets. Finally, we discuss the potential challenges and suggest possible directions for future research.
\end{abstract}



\section{Introduction}

{Providing} more immersive experiences with depth perception to the observers, the 3D applications, such as the Multi-View Video (MVV) and Free-Viewpoint Video (FVV), have drawn great public attention in recent years. These 3D applications allow the users to view the same scene at various angles which may result in a huge information redundancy and cost tremendous bandwidth or storage space. To reduce these limitations, researchers attempt to transmit and store only a subset of these views and synthesize the others at {the} receiver by using the Multiview-Video-Plus-Depth (MVD) data format and Depth-Image-Based-Rendering (DIBR) techniques \cite{fehn2004depth, sun2010overview}. Only limited viewpoints (both texture images and depth maps) are included in the MVD data format, the other view images are synthesized through DIBR. 

\begin{figure}%[!htbp]
  \centering
  \includegraphics[width=\linewidth]{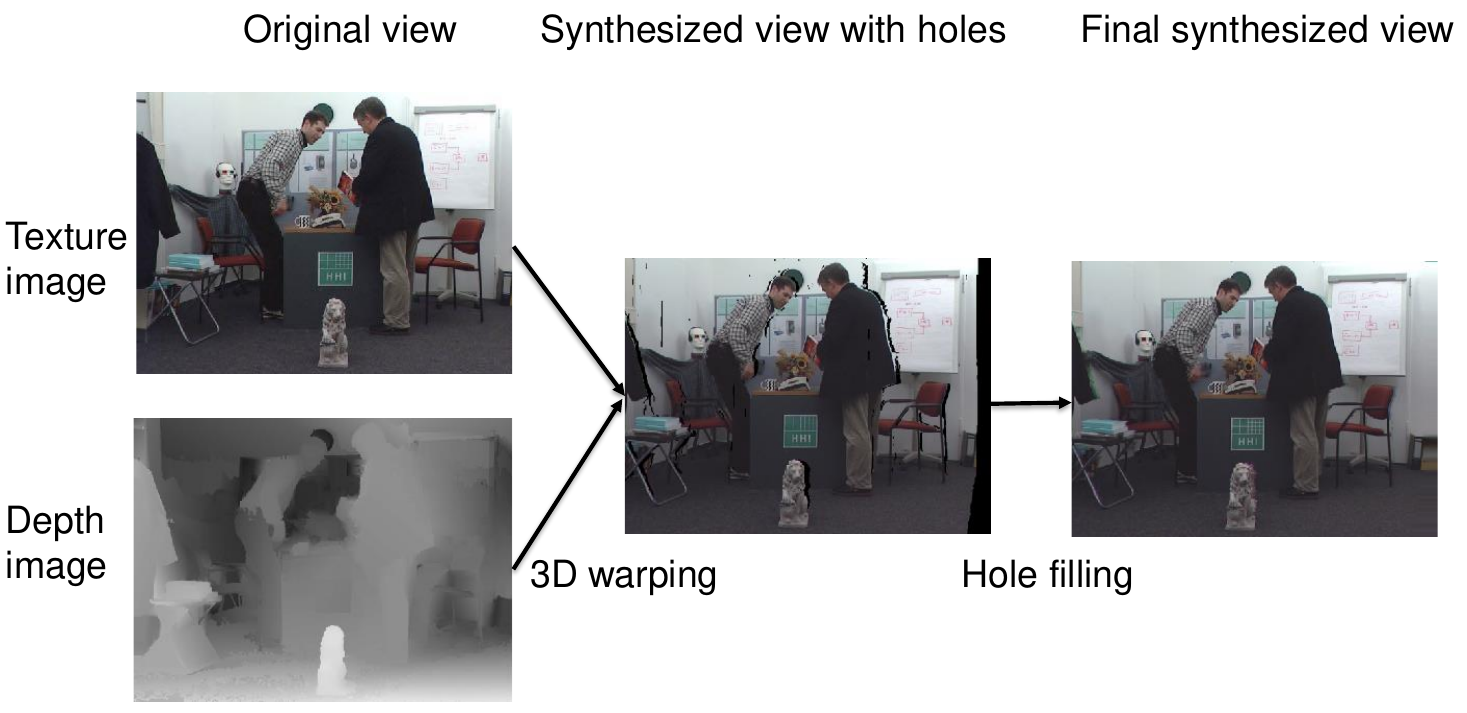}
%  \vspace{2.0cm}
  \caption{Procedure of DIBR.}
  \label{Fig: DIBR}
\end{figure}

This MVD plus DIBR scenario greatly reduces the burden on the storage and transmission of 3D video content{s}. 
%Rather than store or transmit the images from all the viewpoints, only a subset of them are needed to be stored or transmitted. 
%DIBR can not only be used in 
However, the DIBR view synthesis technique also raises new challenges in the quality assessment of virtual synthesized views. During the DIBR process, the pixels in the texture image at the original viewpoint are back-projected to the real 3D space, and then re-projected to the target virtual viewpoint using the depth map, which is called 3D image warping in the literature. As shown in Fig.~\ref{Fig: DIBR}, DIBR view synthesis can be divided into two parts: 3D image warping and hole filling. During the 3D image warping procedure, the pixels in the original view are warped to the corresponding position in the target view. {Because of the change in the} viewpoint, some objects which are invisible in the original view may become visible in the target one, which is called dis-occlusion and causes black holes in the synthesized view. Then, the second step is to fill the black holes. The holes can be filled by typical image in-painting algorithms \cite{jiao2019multi}. Most of the image in-painting algorithms use the pixels around the ``black holes" to search the similar regions in the same image, and then use this similar region to fill the ``black holes". Due to the imprecise depth map and imperfect image in-painting method, various distortions, which are quite different from the traditional ones in 2D images/videos, may be caused. Most of the 2D objective quality metrics \cite{Wangssim, wang2003multiscale, li2016no, jiang2019no, li2019blind} which focus on the traditional distortions will fail to evaluate the quality of DIBR-synthesized views. Subjective test is the most accurate and reliable way to assess the quality of media content since the human observers are the ultimate users in most applications. The subjective tests offer the datasets along with subjective quality scores. The objective metrics are designed to mathematically model and predict the subjective quality scores. In other words, an ideal objective model is expected to be consistent with the subjective results. Since the subjective test is time consuming and practically not suitable for real-time applications, effective objective metrics are highly desired.

Although several efforts have been made targeting at the objective quality assessment of DIBR-synthesized views in recent years, to the best of our knowledge, there is not a detailed survey on these works in {the} current literature. In this paper, we provide a comprehensive survey on the quality assessment approaches for DIBR-synthesized views ranging from the subjective to objective methods. The main contributions can be summarized as follows: (1) the state-of-the-art metrics are introduced and classified based on their approaches; (2) the metrics {are analyzed deeply in terms of the contributions, advantages and disadvantages}; (3) the performances of these metrics are evaluated on different datasets, and their performance{s} on different type of distortions are {reasoned}; (4) furthermore, the limitations of current works are discussed and the possible directions for future research are given.

The rest of this paper is organized as follows. Firstly, Section {2} introduces the DIBR view synthesis technique and analyses the view synthesis distortions. Secondly, the subjective methods are surveyed in Section {3}. Section {4} introduces the state-of-the-art objective quality metrics in detail. The experimental results are presented and discussed in Section {5}. Finally, the conclusions are given in Section {6}.

\section{Depth-Image-Based-Rendering (DIBR) and distortion analysis}
{As introduced in the previous section, the DIBR view synthesis procedure consists of two parts: 3D warping and hole filling \textsl{cf.} Fig.~\ref{Fig: DIBR}. Due to the lack of original texture information, various distortions may be induced in the DIBR-synthesized views which significantly degrade the image quality. In this section, we give a review of the algorithms that are designed to improve the visual quality of DIBR-synthesized views, and then analyze the distortions that may occur in the DIBR-synthesized views.} 

\subsection{Review of state-of-the-art DIBR algorithms}
\begin{figure}
\begin{minipage}[b]{0.48\linewidth}
  \centering
  \centerline{\includegraphics[width=\textwidth]{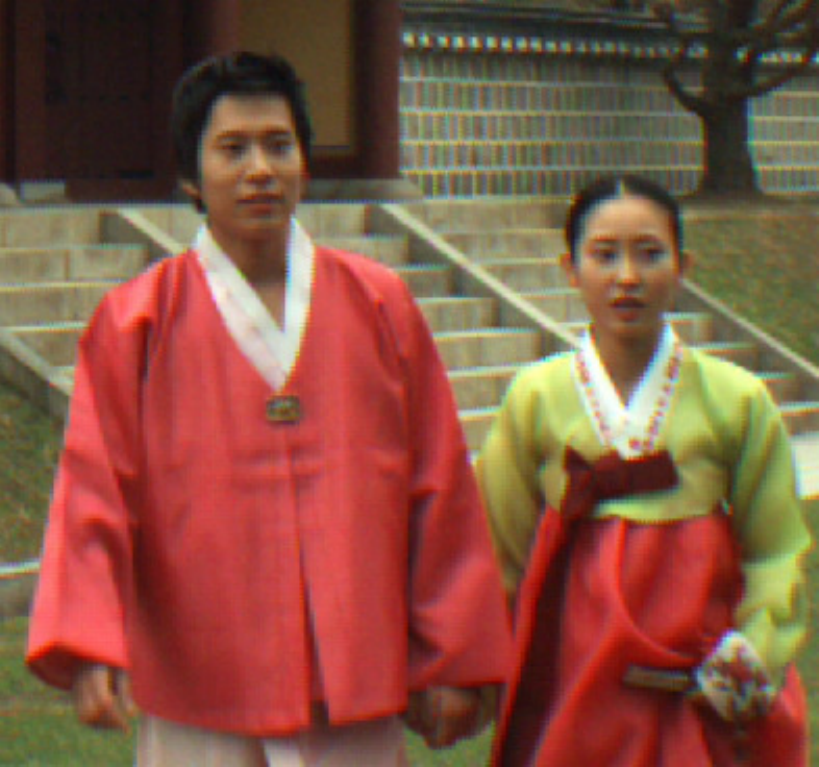}}
%  \vspace{2.0cm}
  \centerline{(a) Reference view}\medskip
  \end{minipage}
  \begin{minipage}[b]{0.48\linewidth}
  \centering
  \centerline{\includegraphics[width=\textwidth]{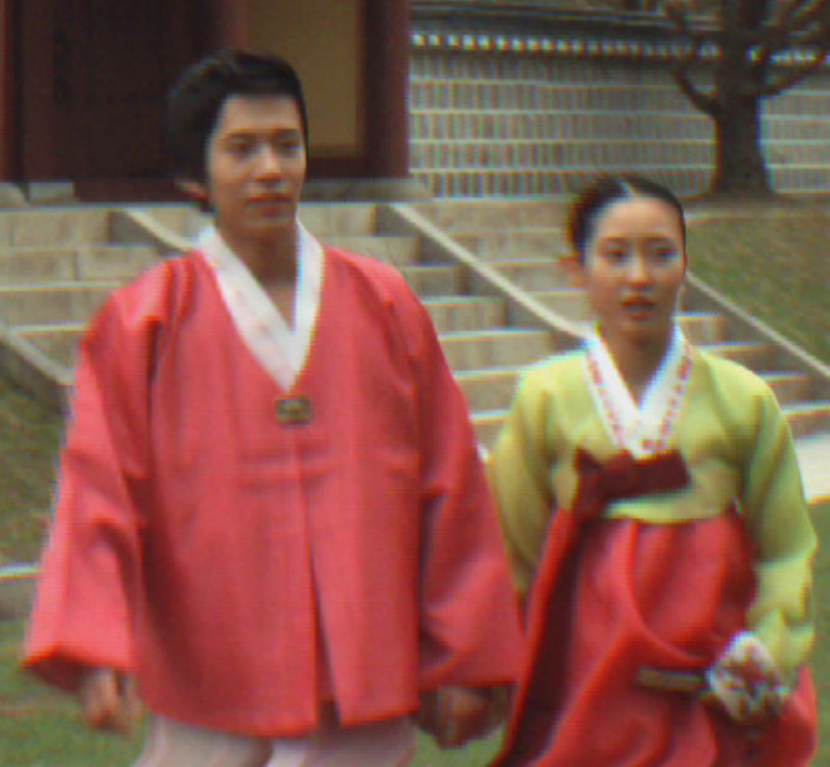}}
%  \vspace{2.0cm}
  \centerline{(b) Synthesized view}\medskip
  \end{minipage}
  \caption{Object shifting caused by depth low-passing filter, the right borders of the character's faces are slightly modified. These images are from IVC DIBR image dataset \cite{bosc2011towards}.}
  \label{Fig_Shifting}
\end{figure}

{During the 3D warping process, a large number of small cracks may be induced by the numerical rounding operations of pixel positions since the corresponding pixel position in the target viewpoint may be not an integer. These distortions mainly happen in the regions where the depth values are significantly different from their neighbours. Normally, these small cracks are handled by filtering the warped depth map with a low-pass filter \cite{ahn2013novel, jantet2011object, tanimoto2008reference}. However, this may also cause slightly object shift in the synthesized views \textsl{cf.} Fig.~\ref{Fig_Shifting}.}

%Besides, various distortions also results from the inaccurate depth map and imperfect hole filling methods. They mainly happen in the {dis-occluded} regions which are non-visible in the previous view but become visible in the target view. 

{Dis-occlusion hole filling also plays an important role in generating a high quality synthesized view. 
Many image in-painting algorithms have been used to fill the dis-occlusion holes, such as the Criminisi's Examplar based algorithm \cite{criminisi2004region} and the Telea's algorithm \cite{telea2004image}. However, these in-painting algorithms do not consider the view synthesis characteristics. For example, the dis-occlusion regions are non-visible background objects in the original viewpoint but become visible in the target viewpoint. In other words, the dis-occlusion regions should be filled with background content. 
Face this issue, many studies \cite{oliveira2015selective, muddala2016virtual, ahn2013novel} tried to extend the main idea of these image in-painting methods to DIBR view synthesis.
Oliveira \cite{oliveira2015selective} extends the Criminisi's image in-painting method by changing the hole filling order with depth information. The texture propagation is enforced from the background to the foreground. Muddala \cite{muddala2016virtual} constrains the confidence and data terms to the background areas and local information.
Ahn \cite{ahn2013novel} improves the Criminisi's image in-painting method by optimizing the filling priority and the patch-matching measure. The optimized matched patch is selected only through the data term on the background areas which are extracted using warped depth map. It greatly reduces the ghost effect in the DIBR-synthesized views.}

{Instead of optimizing the priorities and searching regions of in-painting method, \cite{jantet2011object, luo2016hole} try to reconstruct the background content and then use the reconstructed background to eliminate the dis-occlusion holes in the virtual viewpoint.
Jantet \textsl{et al.} proposed an object-based Layered Depth Image (LDI) representation to improve the quality of virtual synthesized views \cite{jantet2011object}. They firstly segment the foreground and background based on a region growing algorithm, which allows organising the LDI pixels into two object-based layers. Once the extracted foreground is obtained, an in-painting method is used to reconstruct the complete background image on both depth and texture images. 
Luo \textsl{et al.} proposed a hole filling approach for DIBR systems based on background reconstruction \cite{luo2016hole}. The foreground is firstly removed by using morphological operations and random walker segmentation. Then, the background is reconstructed based on motion compensation and a modified Gaussian Mixture model. }

{All the DIBR view synthesis algorithms introduced above are single view based synthesis method. They use only one neighbouring view to extrapolate the synthesized views. Differently, the interview algorithms use two neighbouring views to synthesize the virtual viewpoint images. The most popular interview synthesis method would be the View Synthesis Reference Software (VSRS) \cite{tanimoto2008reference} which has been adopted by the MPEG 3D video Group. The depth discontinuity artefacts are firstly solved by performing a post-filter on the projected depth map. Then, the in-painting method proposed in \cite{telea2004image} is used to fill the holes in the dis-occluded regions. Note that this approach is primarily used in the inter-view synthesis applications which only have small holes to be filled, but it can also be used in single view based rendering cases.}
{Instead of in-painting the warped images directly, \cite{zhu2016depth} focuses on the use of the occluded information to identify the relevant background pixels around the holes. Firstly, the occluded background information is registered in both texture and depth during 3D warping. Then, the un-occluded background information around the holes is extracted based on the depth map. After that, a virtual image is generated by integrating the occluded background and un-occluded background information. The dis-occluded holes are filled based on this generated image with the help of a depth-enhanced Criminisi's in-painting method and a simplified block-averaged filling method.}

\begin{figure}
\centering
  \begin{minipage}[b]{0.48\linewidth}
  \centering
  \centerline{\includegraphics[width=\linewidth]{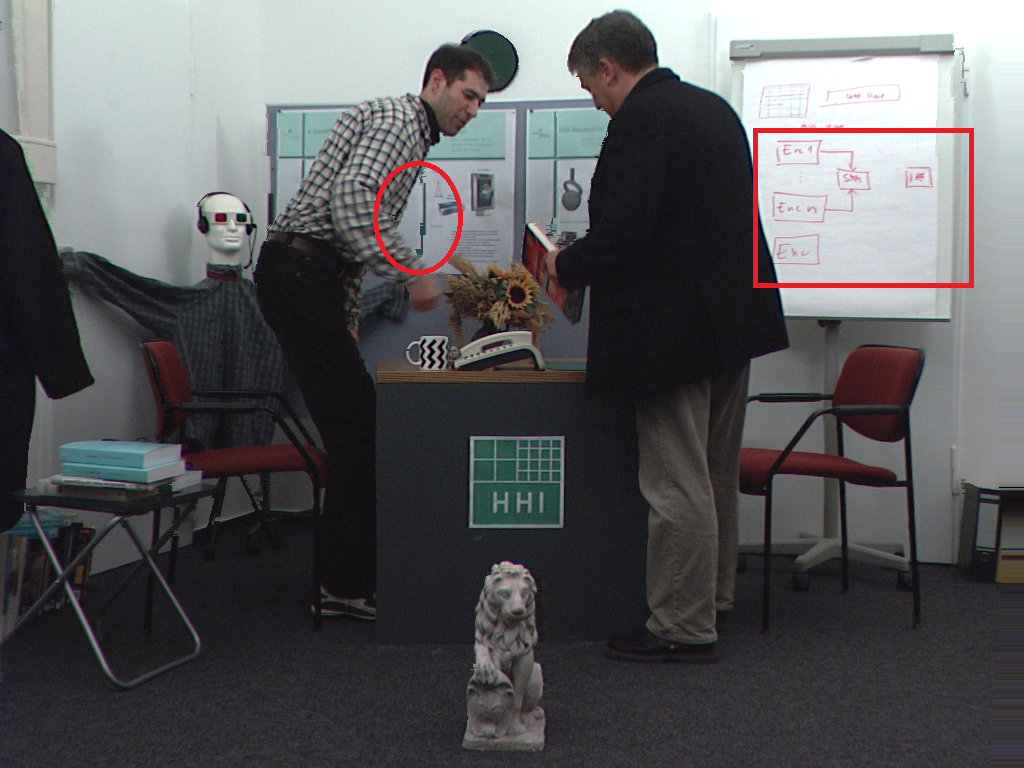}}
  \centerline{(a) single view based mode}\medskip
  \end{minipage}
  \begin{minipage}[b]{0.48\linewidth}
  \centering
  \centerline{\includegraphics[width=1\linewidth]{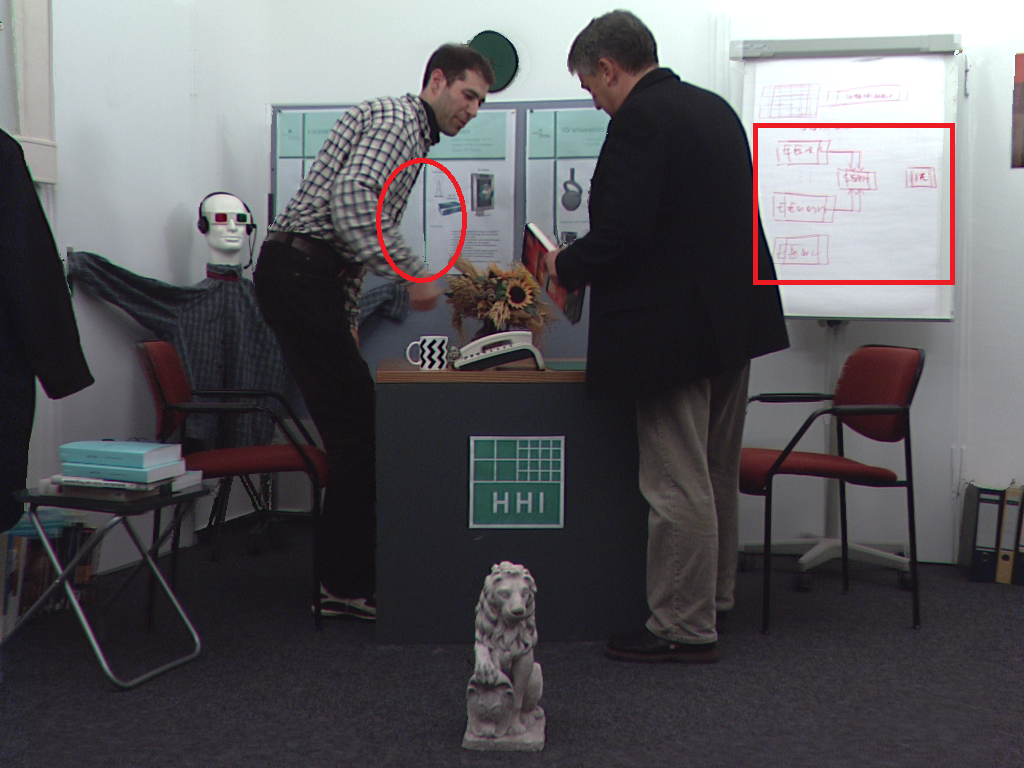}}
  \centerline{(b) interview mode}\medskip
  \end{minipage}
\caption{Examples of images synthesized by VSRS using single view based synthesis mode and interview synthesis mode. These images are from IETR DIBR image dataset \cite{tian2019benchmark}.}
\label{fig:examples}
\end{figure}

{With more information, the interview synthesis cases only have smaller dis-occlusion regions to be filled, they thus outperform the single view based view synthesis methods in most circumstances. However, due to the inaccuracy of depth map, the same object in the two base views could be rendered to different positions which results in a ``ghost'' effect in the synthesized view. This phenomenon does not happen in the single view base synthesis method. As shown in Fig.~\ref{fig:examples}, there exists a ``ghost'' effect of the ``chat flow'' on the board marked by red block in (b); but according to the synthesized content marked by red circle, the interview synthesis method (b) works better than the single view based one (a) in generating the object texture. }

\subsection{Distortion analysis}

\begin{figure}
  \centering
  \begin{minipage}[b]{0.24\linewidth}
  \centering
  \centerline{\includegraphics[width=\linewidth]{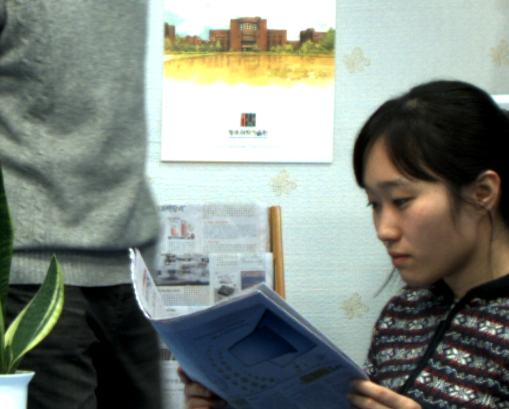}}
%  \vspace{1.5cm}
  \centerline{(a) Reference view}\medskip
  \end{minipage}
  \begin{minipage}[b]{0.24\linewidth}
  \centering
  \centerline{\includegraphics[width=\linewidth]{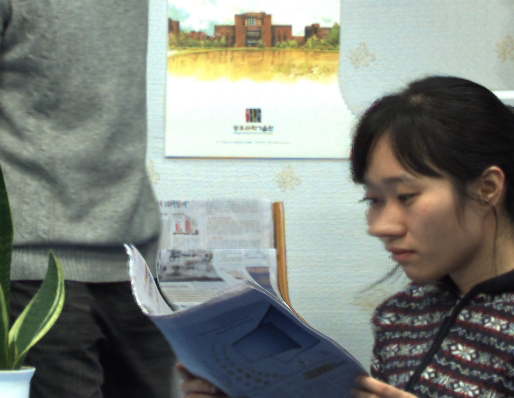}}
%  \vspace{1.5cm}
  \centerline{(b) Synthesized view (object warping)}\medskip
  \end{minipage} 
  
  \begin{minipage}[b]{0.24\linewidth}
  \centering
  \centerline{\includegraphics[width=\textwidth]{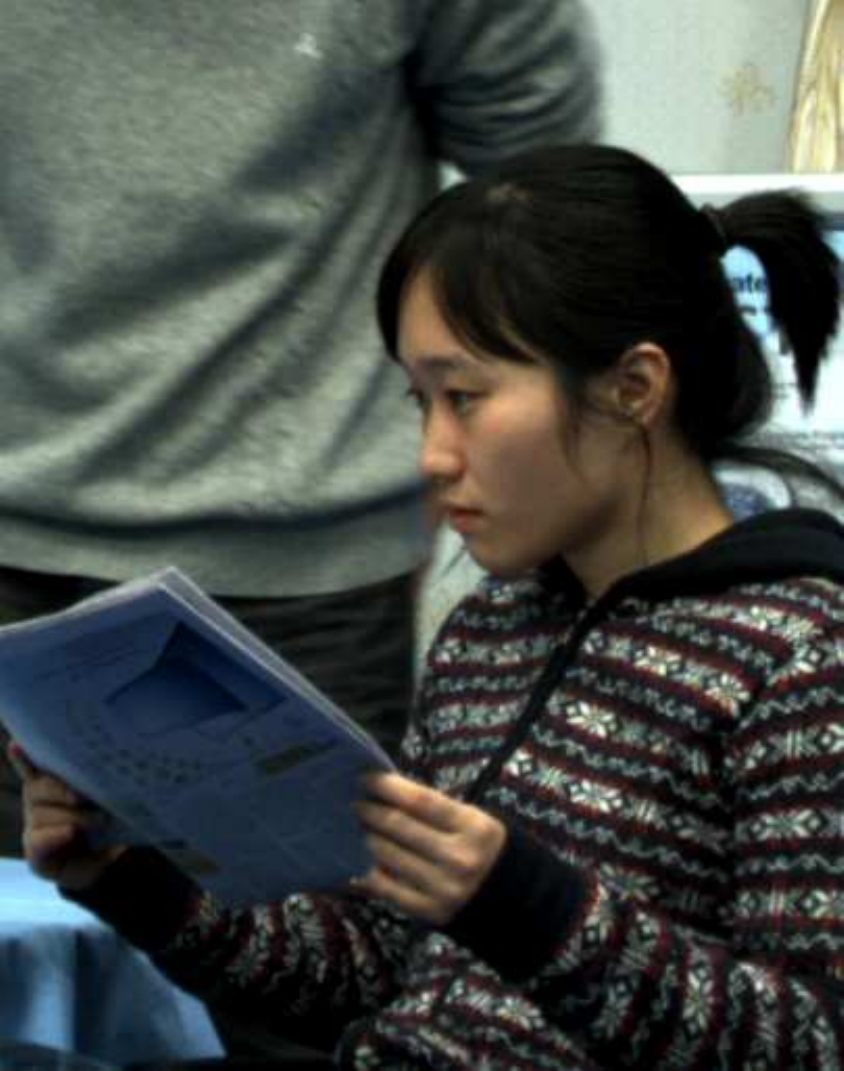}}
%  \vspace{2.0cm}
  \centerline{(c) Reference view}\medskip
  \end{minipage}
  \begin{minipage}[b]{0.24\linewidth}
  \centering
  \centerline{\includegraphics[width=\textwidth]{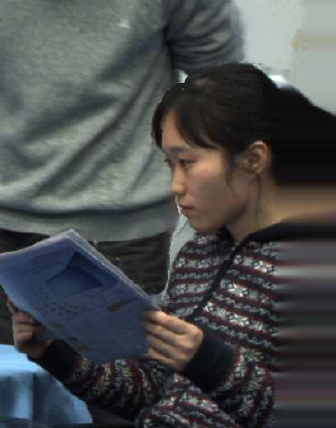}}
%  \vspace{2.0cm}
  \centerline{(d) Synthesized view (stretching)}\medskip
  \end{minipage}
  \begin{minipage}[b]{0.24\linewidth}
  \centering
  \centerline{\includegraphics[width=\textwidth]{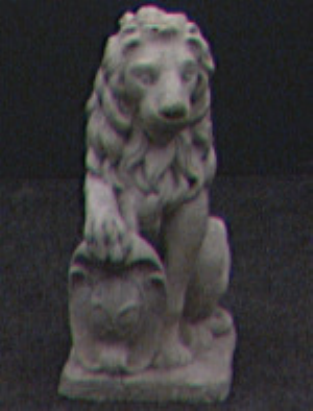}}
%  \vspace{2.0cm}
  \centerline{(e) Reference view}\medskip
  \end{minipage}
  \begin{minipage}[b]{0.24\linewidth}
  \centering
  \centerline{\includegraphics[width=\textwidth]{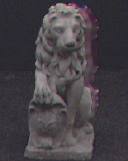}}
%  \vspace{2.0cm}
  \centerline{(f) Synthesized view (blurry region)}\medskip
  \end{minipage}
\caption{Example of distortions caused by imperfect image in-painting method. These images are from IVC DIBR-image dataset \cite{bosc2011towards}.}
  \label{Fig: dis_inpainting}
\end{figure}

{Imperfect hole filling methods may induce various distortions in the DIBR-synthesized views, such as object warping, stretching and blurry regions, \textsl{cf.} Fig.~\ref{Fig: dis_inpainting}. Fig.~\ref{Fig: dis_inpainting} (a), (b) give an example of object warping distortion caused by Telea's image in-painting algorithm \cite{telea2004image}. It could be observed that the ``newspaper'' and the ``girl's nose'' are extreme warped. The stretching distortion (the ``girl's  hair and clothes'') mainly happen in the out-of-field areas \textsl{cf.} Fig.~\ref{Fig: dis_inpainting} (c), (d). The blurry regions can be noticed around the sculpture in Fig.~\ref{Fig: dis_inpainting} (e), (f).}

\begin{figure*}
  \centering
  \begin{minipage}[b]{0.16\linewidth}
  \centering
  \centerline{\includegraphics[width=\linewidth]{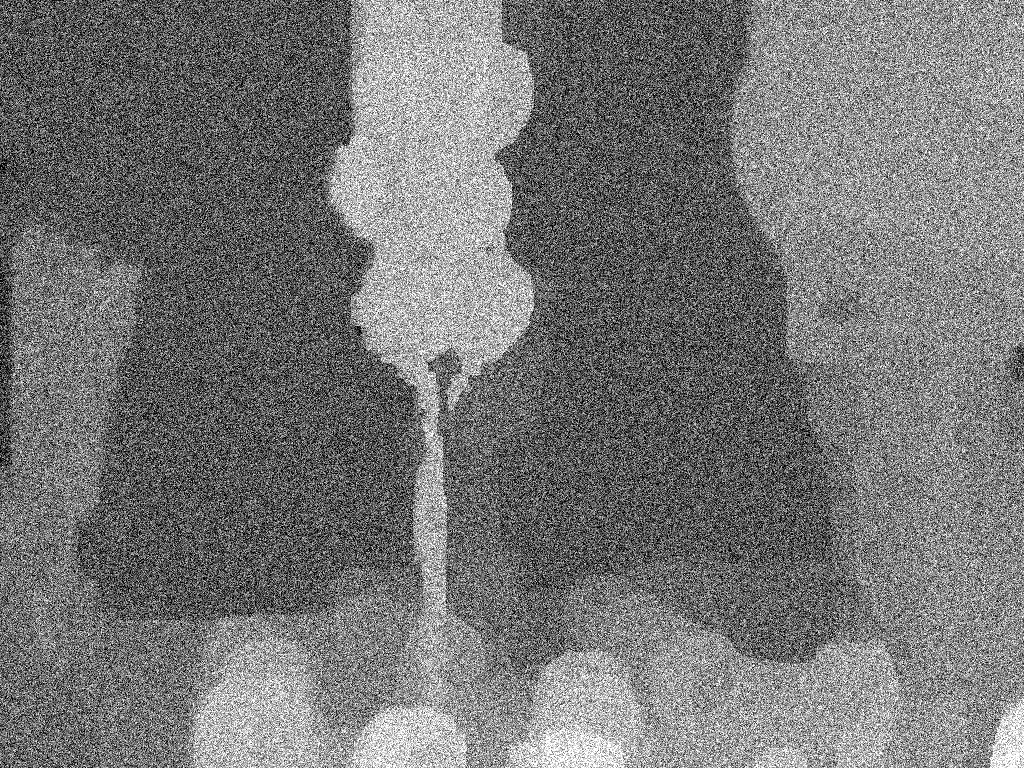}}
%  \vspace{1.5cm}
%  \centerline{(a) AWN}\medskip
  \end{minipage}
  \begin{minipage}[b]{0.16\linewidth}
  \centering
  \centerline{\includegraphics[width=\linewidth]{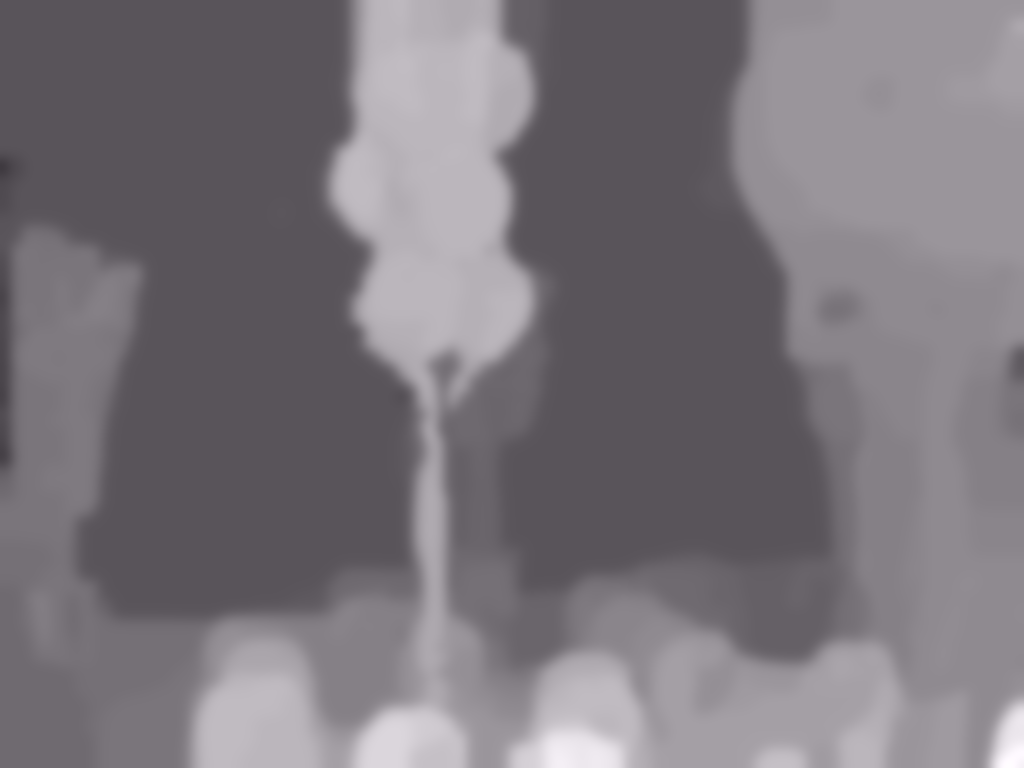}}
%  \vspace{1.5cm}
%  \centerline{(b) Gaussian blur (object warping)}\medskip
  \end{minipage}
  \begin{minipage}[b]{0.16\linewidth}
  \centering
  \centerline{\includegraphics[width=\linewidth]{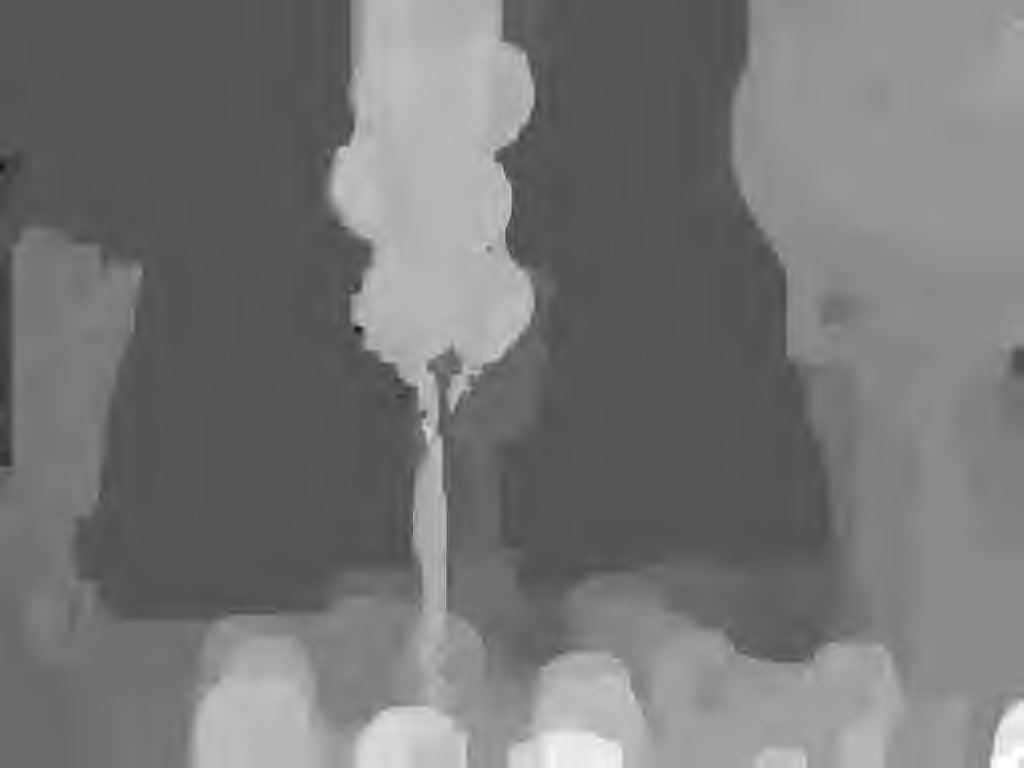}}
%  \vspace{1.5cm}
%  \centerline{(b) Gaussian blur (object warping)}\medskip
  \end{minipage}
  \begin{minipage}[b]{0.16\linewidth}
  \centering
  \centerline{\includegraphics[width=\linewidth]{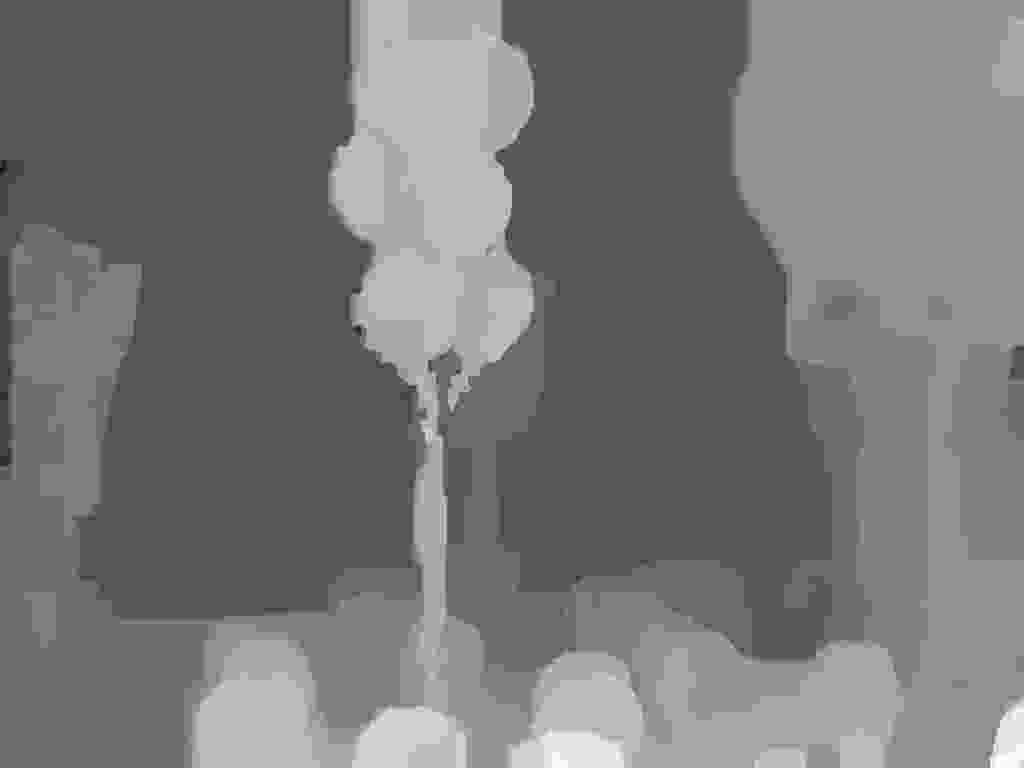}}
%  \vspace{1.5cm}
%  \centerline{(b) Gaussian blur (object warping)}\medskip
  \end{minipage}
  \begin{minipage}[b]{0.16\linewidth}
  \centering
  \centerline{\includegraphics[width=\linewidth]{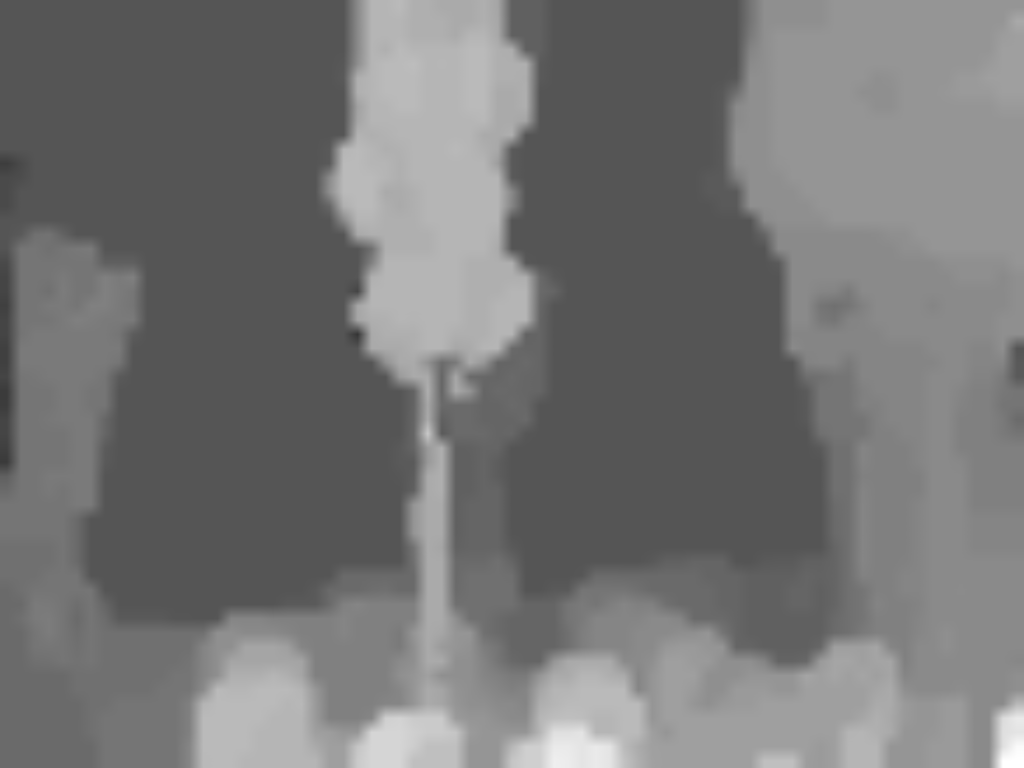}}
%  \vspace{1.5cm}
%  \centerline{(b) Gaussian blur (object warping)}\medskip
  \end{minipage}
  \begin{minipage}[b]{0.16\linewidth}
  \centering
  \centerline{\includegraphics[width=\linewidth]{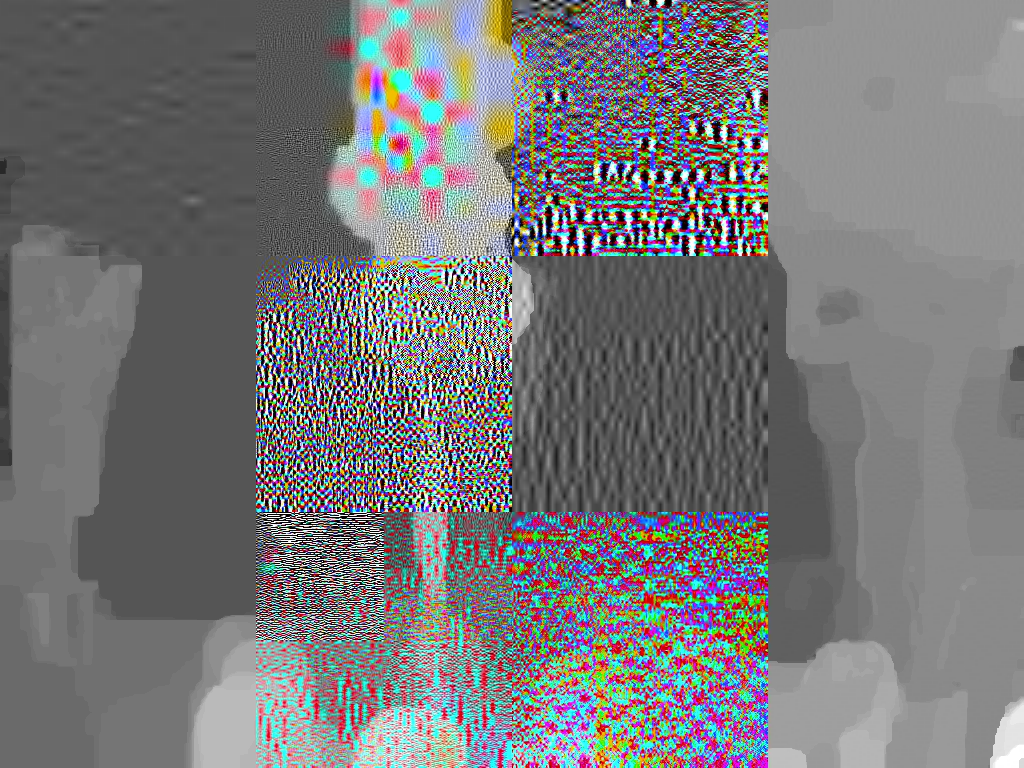}}
%  \vspace{1.5cm}
%  \centerline{(b) Gaussian blur (object warping)}\medskip
  \end{minipage}
  
  \centering
  \begin{minipage}[b]{0.16\linewidth}
  \centering
  \centerline{\includegraphics[width=\linewidth]{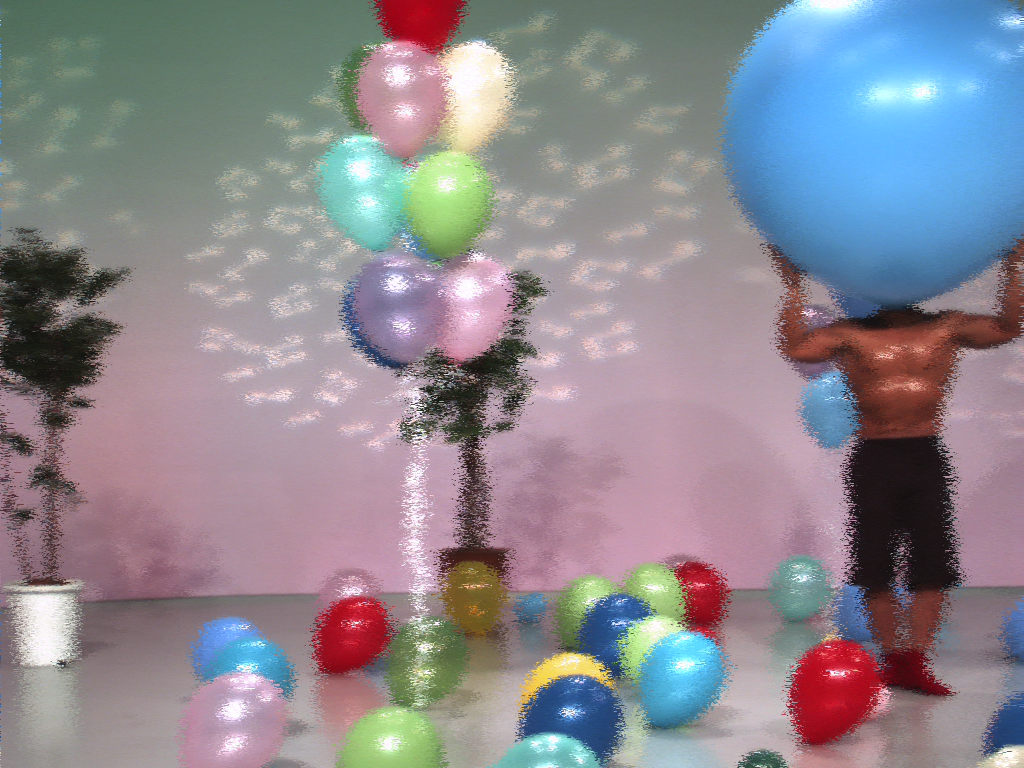}}
%  \vspace{1.5cm}
%  \centerline{(a) AWN}\medskip
  \end{minipage}
  \begin{minipage}[b]{0.16\linewidth}
  \centering
  \centerline{\includegraphics[width=\linewidth]{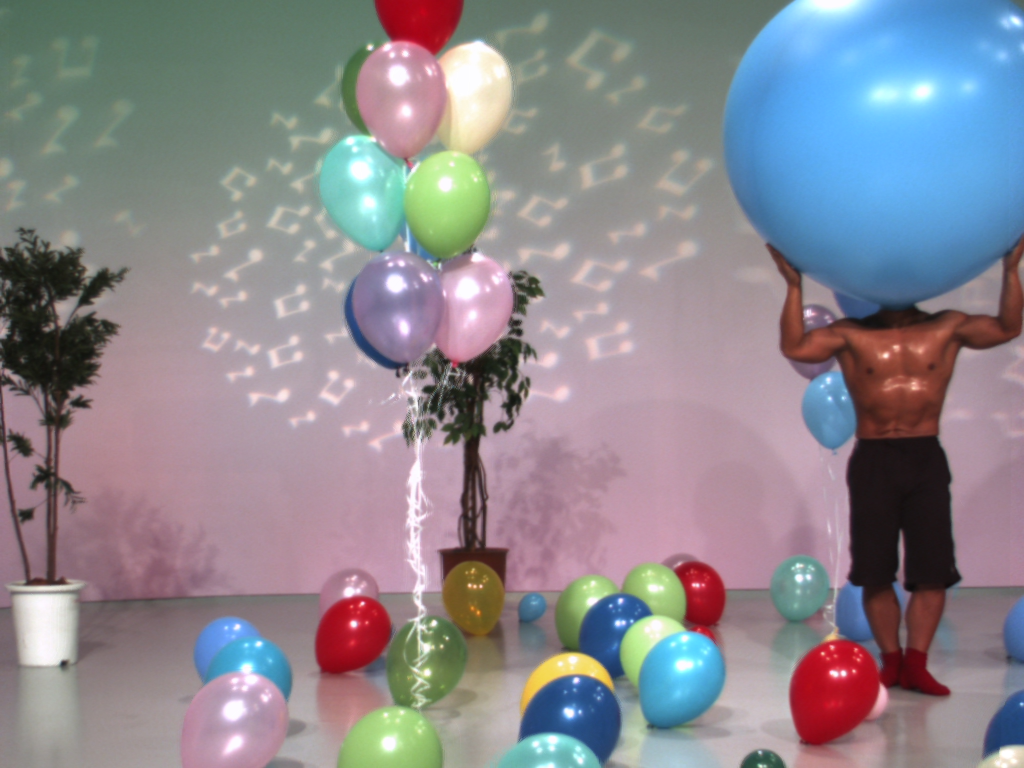}}
%  \vspace{1.5cm}
%  \centerline{(b) Gaussian blur (object warping)}\medskip
  \end{minipage}
  \begin{minipage}[b]{0.16\linewidth}
  \centering
  \centerline{\includegraphics[width=\linewidth]{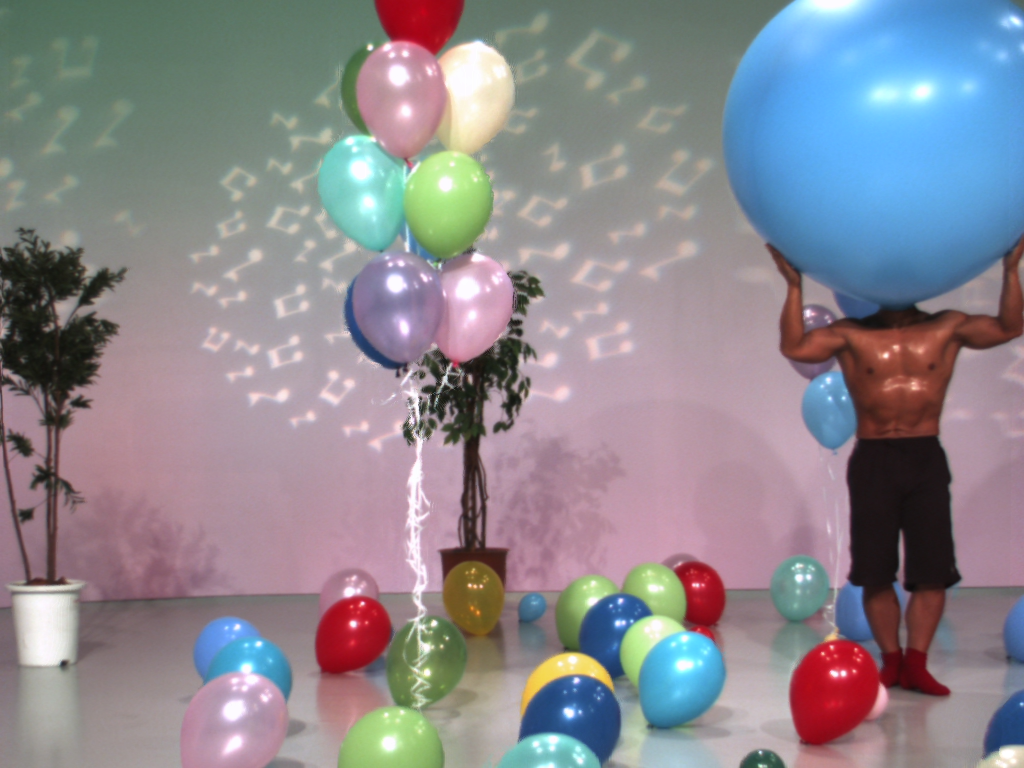}}
%  \vspace{1.5cm}
%  \centerline{(b) Gaussian blur (object warping)}\medskip
  \end{minipage}
  \begin{minipage}[b]{0.16\linewidth}
  \centering
  \centerline{\includegraphics[width=\linewidth]{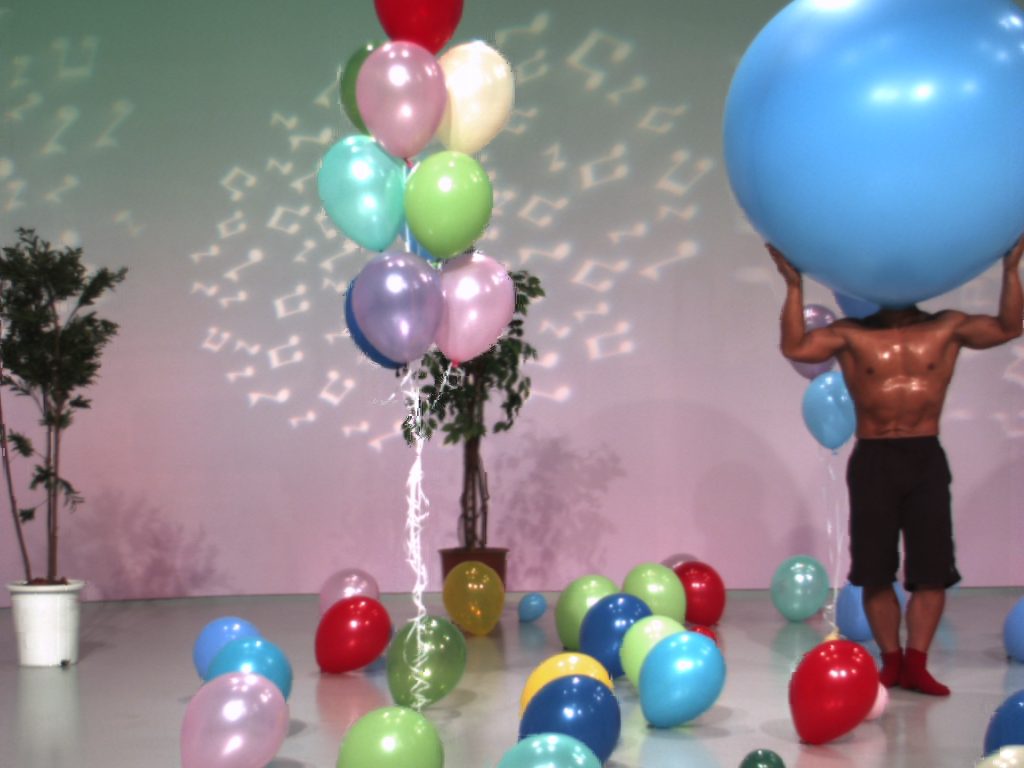}}
%  \vspace{1.5cm}
%  \centerline{(b) Gaussian blur (object warping)}\medskip
  \end{minipage}
  \begin{minipage}[b]{0.16\linewidth}
  \centering
  \centerline{\includegraphics[width=\linewidth]{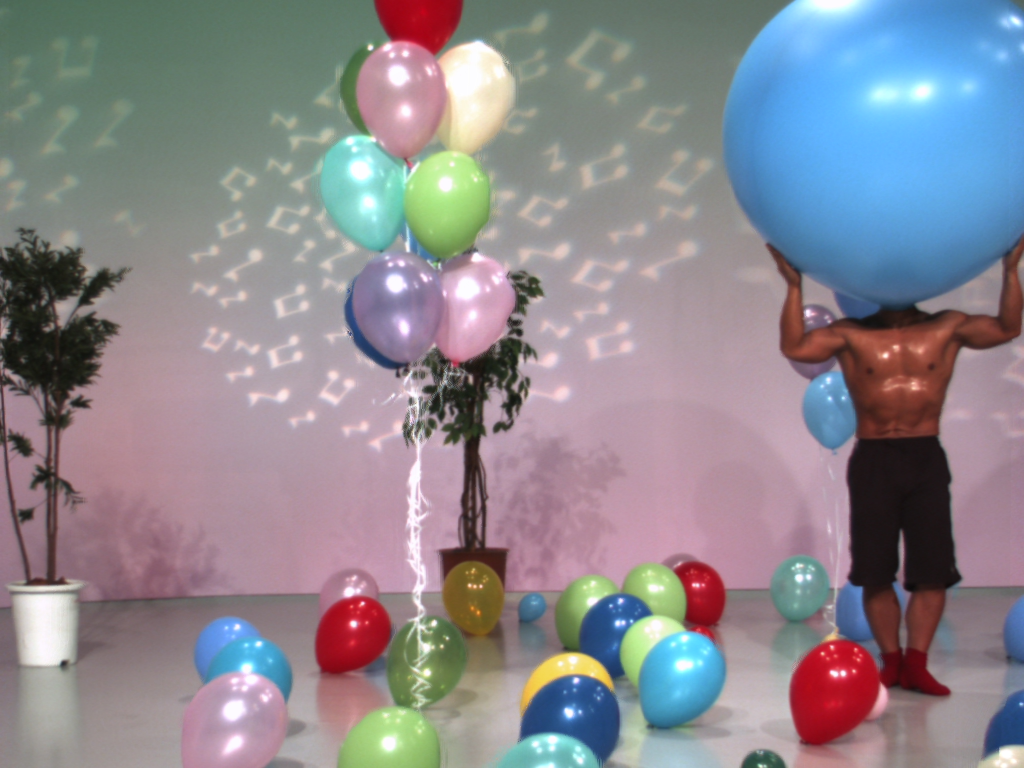}}
%  \vspace{1.5cm}
%  \centerline{(b) Gaussian blur (object warping)}\medskip
  \end{minipage}
  \begin{minipage}[b]{0.16\linewidth}
  \centering
  \centerline{\includegraphics[width=\linewidth]{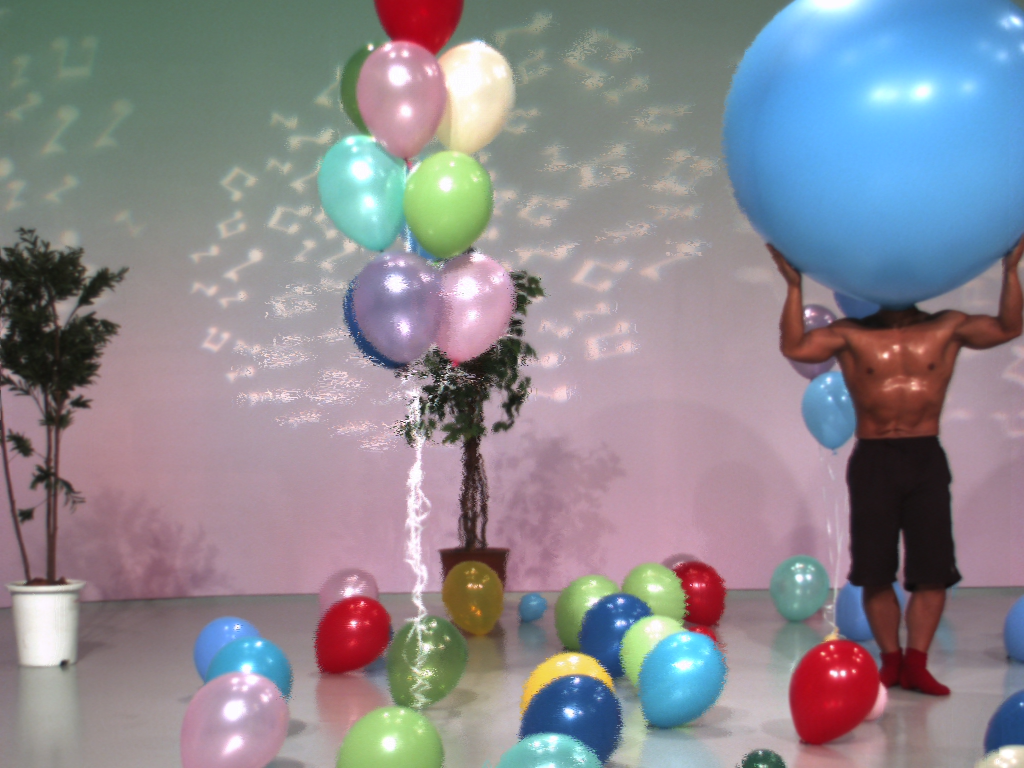}}
%  \vspace{1.5cm}
%  \centerline{(b) Gaussian blur (object warping)}\medskip
  \end{minipage}
  
  \centering
  \begin{minipage}[b]{0.16\linewidth}
  \centering
  \centerline{\includegraphics[width=\linewidth]{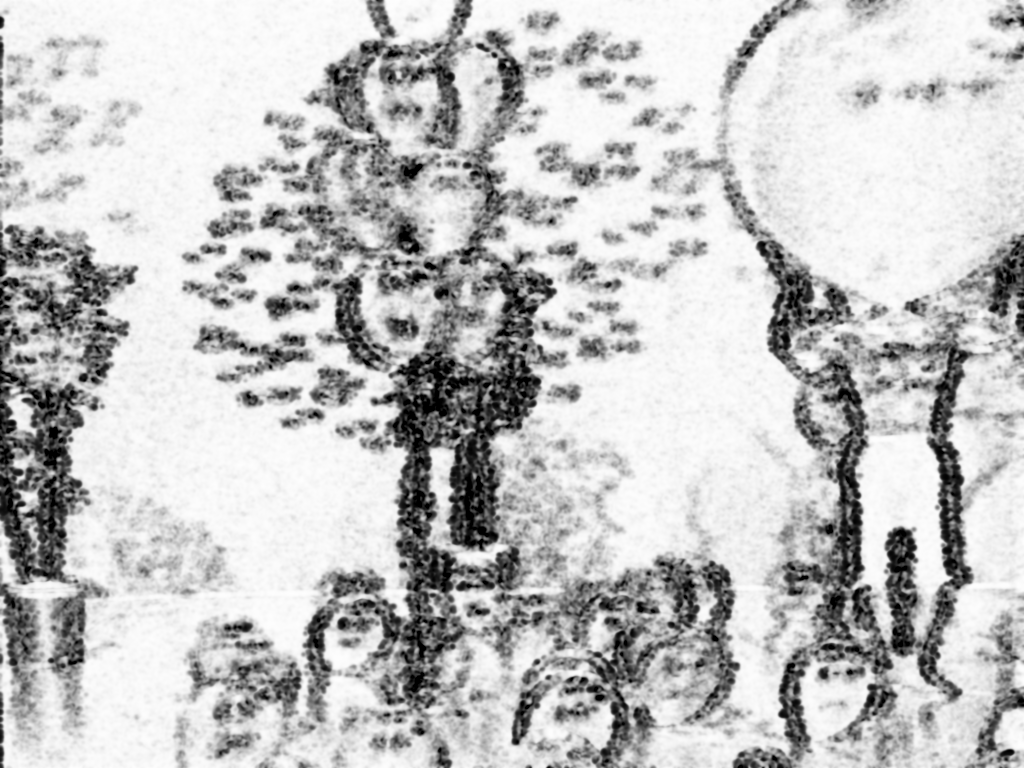}}
%  \vspace{1.5cm}
  \centerline{(a) AWN}\medskip
  \end{minipage}
  \begin{minipage}[b]{0.16\linewidth}
  \centering
  \centerline{\includegraphics[width=\linewidth]{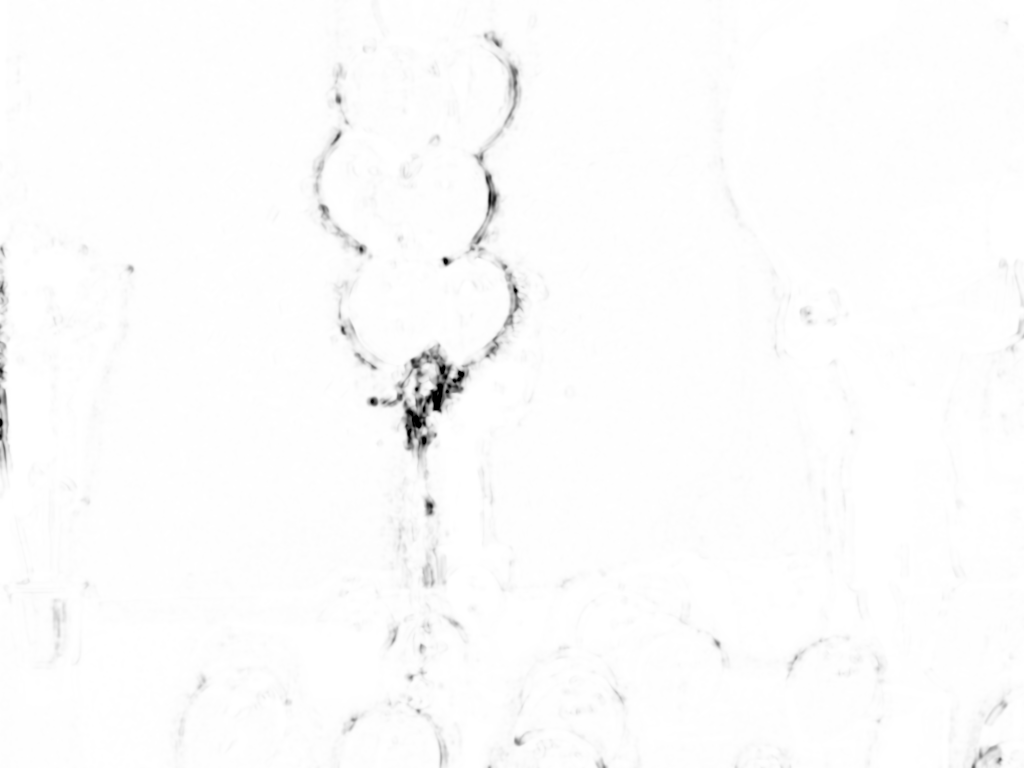}}
%  \vspace{1.5cm}
  \centerline{(b) Gaussian blur }\medskip
  \end{minipage}
  \begin{minipage}[b]{0.16\linewidth}
  \centering
  \centerline{\includegraphics[width=\linewidth]{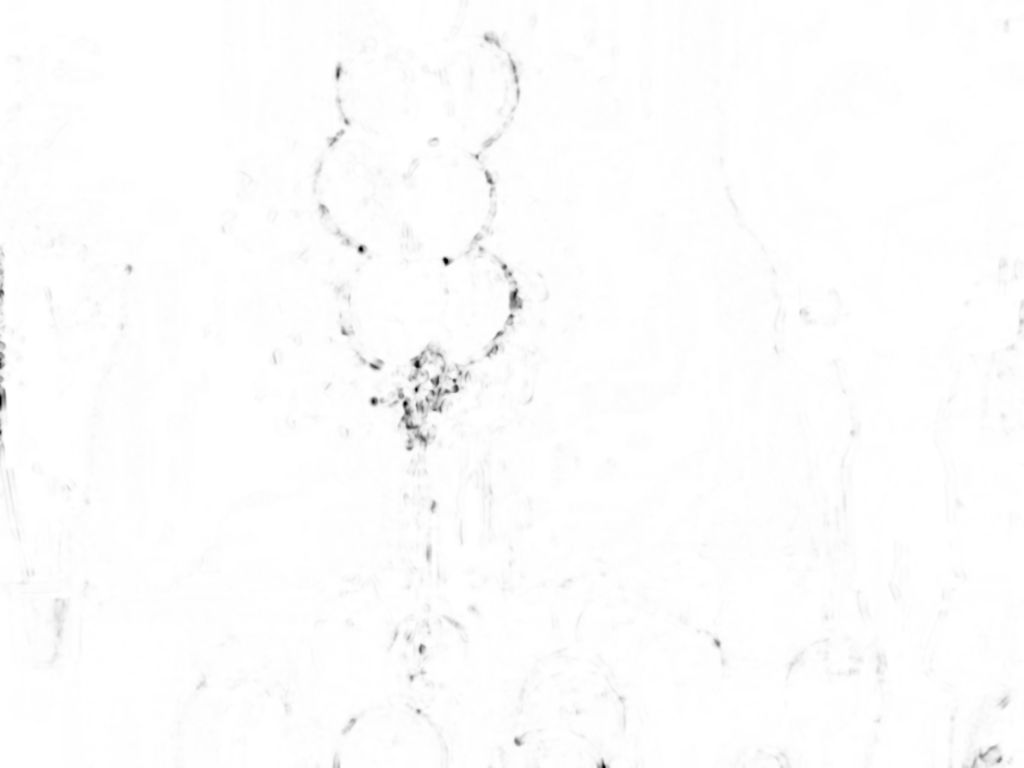}}
%  \vspace{1.5cm}
  \centerline{(c) JP2K}\medskip
  \end{minipage}
  \begin{minipage}[b]{0.16\linewidth}
  \centering
  \centerline{\includegraphics[width=\linewidth]{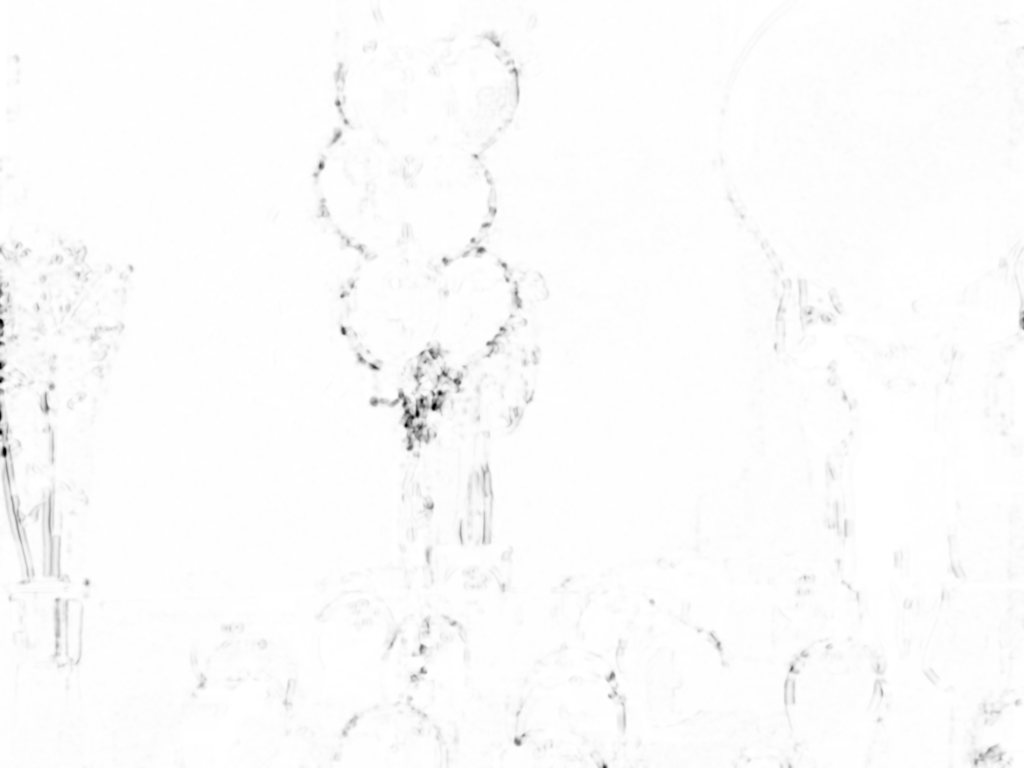}}
%  \vspace{1.5cm}
  \centerline{(d) JPEG}\medskip
  \end{minipage}
  \begin{minipage}[b]{0.16\linewidth}
  \centering
  \centerline{\includegraphics[width=\linewidth]{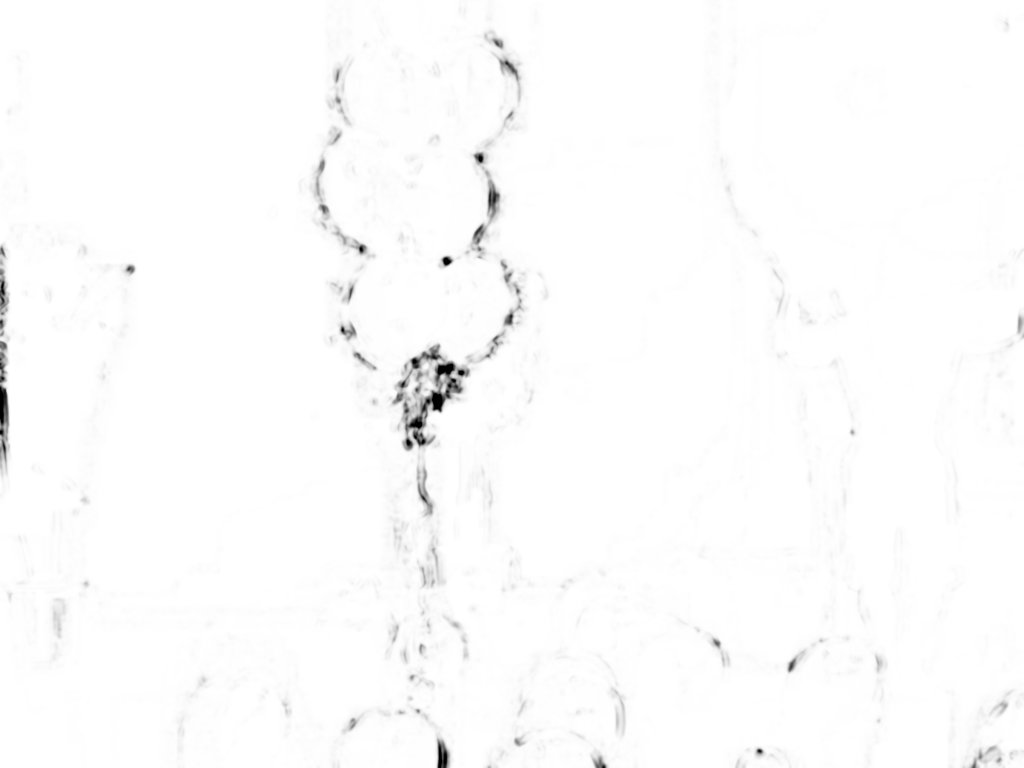}}
%  \vspace{1.5cm}
  \centerline{(e) Sampling blur}\medskip
  \end{minipage}
  \begin{minipage}[b]{0.16\linewidth}
  \centering
  \centerline{\includegraphics[width=\linewidth]{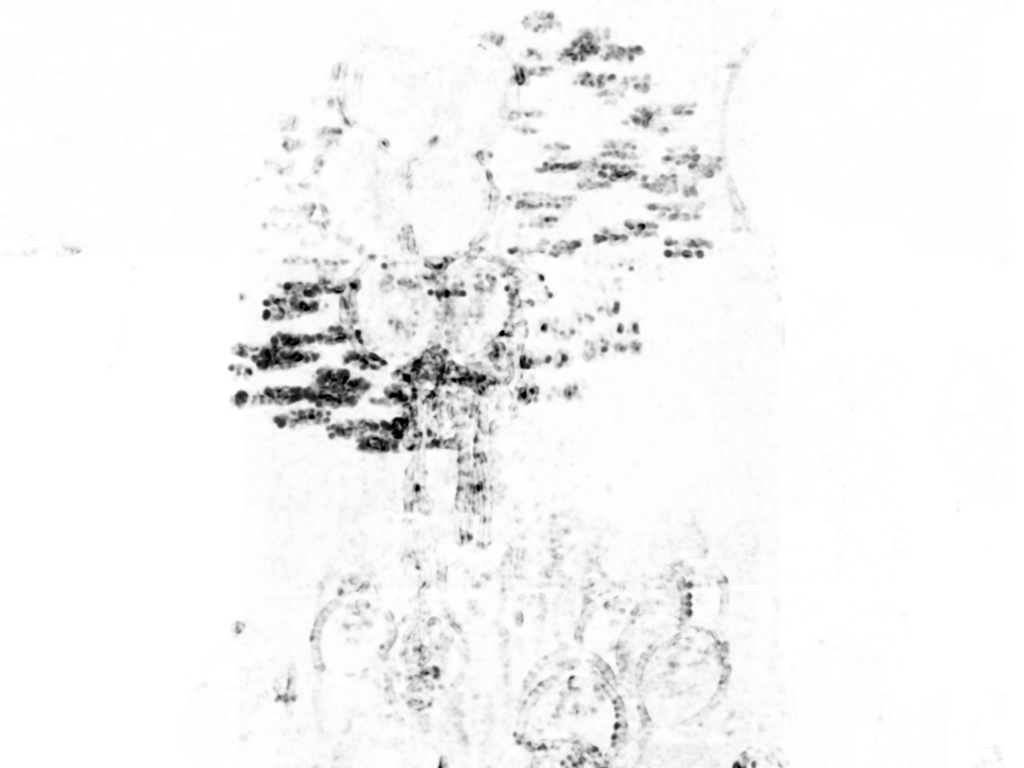}}
%  \vspace{1.5cm}
  \centerline{(f) Transmission loss}\medskip
  \end{minipage}
\caption{Example of synthesized images using depth map with different distortions. The first row shows the distorted depth maps while the second row gives the DIBR-synthesized images using the corresponding distorted depth maps. The third row presents the SSIM maps between the synthesized and reference images. Note that, the reference images are the images synthesized with undistorted depth maps. These images are from MCL-3D image dataset \cite{song2014mcl}.}
  \label{Fig: dis_depth}
\end{figure*}

{Depth map represents the distance of objects to the camera. It is composed of a series of flat homogeneous regions and sharp edges. The flat areas indicate the objects at a certain distance while the edges relate to the transition of foreground and background objects. This is quite different from the natural scene images. In DIBR view synthesis, depth maps are used to guide the 3D warping. The distortions in the depth map will certainly induce degradations in the DIBR-synthesized views. In order to analyze the effect of depth distortions on the quality of DIBR-synthesized views, we compare the images that are synthesized with undistorted depth map and depth maps with various distortions.
As shown in Fig.~\ref{Fig: dis_depth}, we can easily observe that most of the distortions distribute around the edge regions of the depth map. It is logical that the edge of depth map represents the transition of foreground and background objects, the noise in these edge regions will certainly cause aliasing of foreground and background texture. Besides, we also notice that the synthesized view quality is more sensitive to high-frequency distortions (\textsl{e.g.} additive white noise (AWN), transmission loss) in the depth map compared to the low-frequency distortions (\textsl{e.g.} Gaussian blur). The main reason would be that high-frequency distortions in depth map will cause great local shift in the synthesized view, which is much more annoying to human vision system and can be easily penalized by pixel-based IQA metrics. }

\section{Subjective image/video quality assessment of DIBR-synthesized views}
Subjective test is the most direct method for image/video quality assessment. During the test, a group of human observers are asked to rate the quality of each tested image or video. 
%Normally, the rating scale is graded to five equal intervals: excellent, good, fair, poor, bad, which are associated to their annoyance degree to the observers. 
The subjective test results obtained from the subjective ratings are recognized as the quality of the tested images/videos. In different subjective test methodologies, the acquisition of subjective scores {is} also different.
%The subjective scores are obtained differently in different datasets since different subjective test methodologies are used \textsl{cf.} Table~\ref{Tab:dibrdatabase}. 

The Absolute Category Rating (ACR) method used in IVC image / video datasets \cite{bosc2011towards, bosc2013visual} randomly present the test sequences to the observers and ask them to rate on five-scale\st{s} quality judgement (excellent, good, fair, poor, bad). The subjective quality scores are calculated by simply averaging the ratings.
The Single Stimulus Continuous Quality Evaluation (SSCQE) in SIAT \cite{liu2015subjective} dataset allows the observer to rate on a continuous scale instead of a discrete five-scale\st{s} evaluation. 
The IVY \cite{jung2016critical} image dataset uses the Double Stimulus Continuous Quality Scale (DSCQS). The test image along with its associated reference image are presented in succession. It is usually used when the test and reference images are similar.
Pairwise Comparison (PC) method directly performs a one-to-one comparison of every image pair in the dataset. It is the most accurate and reliable way to get the subjective quality scores, but it {takes} too much time since all the image pairs need to be tested. 
The Subjective Assessment Methodology for VIdeo Quality (SAMVIQ) method used in IETR dataset can achieve much higher accuracy than ACR method for the same number of observers{. It takes} less time than PC since it allows the observer to freely view several image multiple times and {adopts} a continuous rating scale.
Besides, the IVY \cite{jung2016critical}, IETR \cite{tian2019benchmark} and SIAT datasets normalize the obtained scores to \textsl{z-score} to make the results more intuitive. The IVC and MCL-3D \cite{song2014mcl} datasets directly use the average scores. Except for the subjective test methodology, as shown in Table~\ref{Tab:dibrdatabase}, they use different sequences, DIBR algorithms{,} etc. In the following part, we will introduce them respectively in detail.
%latex
%As shown in Table~\ref{Tab:dibrdatabase}, towards the quality assessment of DIBR-synthesized views, several publicly accessible datasets were proposed with different targets. 

\subsection{IVC DIBR datasets}

The IVC DIBR-image dataset \cite{bosc2011towards} was proposed by Bosc \textsl{et al.} in 2011. It contains 84 DIBR-synthesized view images synthesized by 7 DIBR algorithms \cite{fehn2004depth, telea2004image, mori2009view, mueller2009view, ndjiki2010depth, ndjiki2011depth, koppel2010temporally}. 3 Multi-view plus Depth (MVD) sequences, \textsl{BookArrival}, \textsl{Lovebird1} and \textsl{Newspaper}, are extracted as the source contents. For each sequence, 4 virtual views are synthesized from the adjacent viewpoint by using the above algorithms. Note that in this dataset, virtual views were only generated by single-view-based synthesis, which means that the virtual view is synthesized with only one image and its associated depth map. The IVC DIBR-video dataset \cite{bosc2013visual} uses almost the same contents and methodologies except that it adds the H.264 compression (with 3 quantization levels) distortion for each test sequence. In other words, there are 93 distorted videos in this dataset, among which 84 ones only contain the DIBR view synthesis distortions. As one of the first DIBR related image datasets, the IVC datasets play an important role in the first research phase of this topic. However, because of the fast development of DIBR view synthesis algorithms, some of the distortions in these datasets do not exist {any more} in the state-of-the-art view synthesis algorithms. 

%Another limitation of these datasets could be the dataset size, it only contains 84 synthesized images/videos in each dataset. 

%Make one sentence for input images " These MVD sequences contain several (how many?) view-points acquired with real cameras in rectified configuration capturing the same scene, together with a depth image for each acquired view".
%one sentence for the reference views
%"some input images are used as full reference views"
%and then another for virtual viwes.
%"From the other input views, virtual views located on same view-points as reference views are synthesized by using the above algorithms"
%Then explain what is shown to the observers (still images? virtual sequences on fixed view-point? virtual sequences wth changing viewpoint?...) -give this information for the other datasets-

\subsection{IETR DIBR image dataset}

Similar to the IVC datasets, the IETR dataset \cite{tian2019benchmark} is dedicated to investigate the DIBR view synthesis distortions as well. Compared to the IVC datasets, it uses more and newer DIBR view synthesis algorithms \cite{criminisi2004region, jantet2011object, ahn2013novel, luo2016hole, solh2012hierarchical, mori2009view, zhu2016depth}, {includes} both interview synthesis and single view based synthesis, {and excludes} some ``old fashioned'' distortions, \textsl{e.g.} ``black holes''. 
%The interview DIBR algorithms use two neighbouring views to synthesize the virtual viewpoint. 
In addition, the IETR dataset also uses more MVD sequences, of which 7 sequences (\textsl{Balloons}, \textsl{BookArrival}, \textsl{Kendo}, \textsl{Lovebird1}, \textsl{Newspaper}, \textsl{Poznan {Street}} and \textsl{Poznan{Hall}}) are natural images and 3 sequences (\textsl{Undo {Dancer}}, \textsl{Shark} and \textsl{Gt\_Fly}) are computer animation images. It contains 140 synthesized view images and their associated 10 reference images which are also captured by real cameras at the virtual viewpoints. 

%\begin{landscape}
\begin{table*}[htbp]
\footnotesize
\centering
\caption{Summary of existing DIBR related datasets.}
%\begin{sideways}
\label{Tab:dibrdatabase}
\begin{threeparttable}
\begin{tabular}{|c|c|c|c|c|c|c|c|c|c|}\toprule
\hline
\multirow{2}{*}{Name} & \multirow{2}{*}{Sequence} & \multirow{2}{*}{Resolution} &\multirow{2}{*}{Method.} & \multicolumn{2}{c|}{DIBR algos}& {Other} &{No.} & \multirow{2}{*}{Ref.} & \multirow{2}{*}{Disp.}\\
\cline{5-6}
& & & &Name &Year & distortions & PVS \tnote{1} & & \\
\hline
\multirow{7}{*}{IVC DIBR-image} & {BookArrival} & 1024 $\times$ 768 & \multirow{7}{*}{ACR\tnote{2 }} & Fehn's &2004& \multirow{7}{*}{None}&\multirow{7}{*}{84}& \multirow{7}{*}{Ori.} & \multirow{7}{*}{2D}\\
\cline{2-3}\cline{5-6}
& Lovebird1 & 1024 $\times$ 768 & & Telea's &2003& & & & \\
\cline{2-3}\cline{5-6}
& Newspaper & 1024 $\times$ 768 & & VSRS &2009& & & & \\
\cline{2-3}\cline{5-6}
& & & & Müller &2008& & & & \\
\cline{5-6}
& & &PC\tnote{3}& Ndjiki-Nya &2010 & & & & \\
\cline{5-6}
& & & & Köppel &2010& & & & \\
\cline{5-6}
& & & & Black hole &---& & & & \\
\hline
\multirow{2}{*}{IVC DIBR-video} & \multicolumn{2}{c|}{\multirow{2}{*}{idem}} & \multirow{2}{*}{ACR\tnote{2}} & \multicolumn{2}{c|}{\multirow{2}{*}{idem}} & \multirow{2}{*}{H.264} &\multirow{2}{*}{93}&\multirow{2}{*}{Ori.} & \multirow{2}{*}{2D} \\
%\cline{7-7}
& \multicolumn{2}{c|}{} & & \multicolumn{2}{c|}{} & & & & \\
\hline
\multirow{10}{*}{IETR-image} & BookArrival & 1024 $\times$ 768 & \multirow{10}{*}{SAMVIQ\tnote{4}} &Criminisi &2004 & \multirow{10}{*}{None}& \multirow{10}{*}{140}& \multirow{10}{*}{Ori.} & \multirow{10}{*}{2D}\\
\cline{2-3}\cline{5-6}
& Lovebird1 & 1024 $\times$ 768 & & VSRS &2009& & & & \\
\cline{2-3}\cline{5-6}
& Newspaper & 1024 $\times$ 768 & & LDI &2011& & & & \\
\cline{2-3}\cline{5-6}
& Balloons & 1024 $\times$ 768  & & HHF &2012& & & & \\
\cline{2-3}\cline{5-6}
& Kendo  & 1024 $\times$ 768 & & Ahn's &2013& & & & \\
\cline{2-3}\cline{5-6}
& Dancer  & 1920 $\times$ 1088 & & Luo's &2016 & & & & \\
\cline{2-3}\cline{5-6}
& Shark  & 1920 $\times$ 1088 & & Zhu's &2016& & & & \\
\cline{2-3}\cline{5-6}
& Poznan\_Street  & 1920 $\times$ 1088 & &  & & & & & \\
\cline{2-3}\cline{5-6}
& PoznanHall2  & 1920 $\times$ 1088 & &  & & & & & \\
\cline{2-3}\cline{5-6}
& GT\_fly & 1920 $\times$ 1088 &  & & & & & &\\
\hline
\multirow{7}{*}{IVY image} & Aloe  & 1280 $\times$ 1100 & \multirow{7}{*}{DSCQS\tnote{5}} &Criminisi &2004 & \multirow{7}{*}{None} & \multirow{7}{*}{84} & \multirow{7}{*}{Ori.} & \multirow{7}{*}{Stereo.} \\
\cline{2-3}\cline{5-6}
& Dolls & 1300 $\times$ 1100  & & Ahn's &2013& & & & \\
\cline{2-3}\cline{5-6}
& Reindeer & 1300 $\times$ 1100 & & VSRS &2009& & & & \\
\cline{2-3}\cline{5-6}
& Laundry & 1300 $\times$ 1100 & & Yoon &2014 & & & & \\
\cline{2-3}\cline{5-6}
& Lovebird1 & 1024 $\times$ 768 & & & & & & & \\
\cline{2-3}
& Newspaper & 1024 $\times$ 768 & & & & & & & \\
\cline{2-3}
& BookArrival & 1024 $\times$ 768 & & & & & & & \\
\hline
\multirow{9}{*}{MCL-3D image} & Kendo  & 1024 $\times$ 768 & \multirow{9}{*}{PC\tnote{3}} & Fehn's & 2004 & {Additive White Noise} &\multirow{9}{*}{684} & \multirow{9}{*}{Syn.} & \multirow{9}{*}{Stereo.} \\
\cline{2-3}\cline{5-7}
& Lovebird1 & 1024 $\times$ 768 & & Telea's &2003& Blur & & & \\
\cline{2-3}\cline{5-7}
& Balloons & 1024 $\times$ 768 & & HHF &2012& Down Sampling& & & \\
\cline{2-3}\cline{5-7}
& Dancer & 1920 $\times$ 1088 & & Black hole &---& JPEG& & & \\
\cline{2-3}\cline{5-7}
& Shark & 1920 $\times$ 1088 & & & & JPEG2k& & & \\
\cline{2-3}\cline{7-7}
& Poznan\_Street  & 1920 $\times$ 1088 & & & & Trans. Loss \tnote{9}& & & \\
\cline{2-3}\cline{7-7}
& PoznanHall2 & 1920 $\times$ 1088 & & & & & & & \\
\cline{2-3}
& GT\_fly & 1920 $\times$ 1088 & & & & & & & \\
\cline{2-3}
& Microworld & 1920 $\times$ 1088 & & & & & & & \\
\hline
\multirow{10}{*}{SIAT video} & BookArrival & 1024 $\times$ 768  & \multirow{10}{*}{SSCQE\tnote{6}} &\multirow{10}{*}{VSRS} &\multirow{10}{*}{2009} & \multirow{10}{*}{3DV-ATM} & \multirow{10}{*}{140} & \multirow{10}{*}{Ori.} & \multirow{10}{*}{2D} \\
\cline{2-3}
& Balloons & 1024 $\times$ 768 & & & & & & & \\
\cline{2-3}
& Kendo & 1024 $\times$ 768 & & & & & & & \\
\cline{2-3}
& Lovebird1 & 1024 $\times$ 768 & & & & & & & \\
\cline{2-3}
& Newspaper & 1024 $\times$ 768 & & & & & & & \\
\cline{2-3}
& Dancer &1920 $\times$ 1088 & & & & & & & \\
\cline{2-3}
& PoznanHall2 & 1920 $\times$ 1088 & & & & & & & \\
\cline{2-3}
& Poznan\_Street & 1920 $\times$ 1088 & & & & & & & \\
\cline{2-3}
& GT\_fly & 1920 $\times$ 1088 & & & & & & & \\
\cline{2-3}
& Shark & 1920 $\times$ 1088 & & & & & & & \\
\hline
\end{tabular}

\begin{tablenotes}
\footnotesize
\item[1] PVS: Processed Video Sequences.
\item[2] ACR: Absolute Categorical Rating.
\item[3] PC: Pairwise Comparison.
\item[4] SAMVIQ: Subjective Assessment Methodology for VIdeo Quality.
\item[5] DSCQS: Double Stimulus Continuous Quality Scale.
\item[6] SSCQ: Single Stimulus Continuous Quality Scale.
%\item[7] AWN: Additive White Noise.
%\item[8] DS: Down Sampling.
%\item[9] TL: Translation Loss.
\end{tablenotes}
\end{threeparttable}
%\tabnote{$^{\rm 1}$ PVS: Processed Video Sequences.}
%\tabnote{$^{\rm 2}$ ACR: Absolute Categorical Rating.}
%\tabnote{$^{\rm 3}$ PC: Pair Comparison.}
%\tabnote{$^{\rm 4}$ SAMVIQ: Subjective Assessment Methodology for VIdeo Quality.}
%\tabnote{$^{\rm 5}$ DSCQS: Double Stimulus Continuous Quality Scale.}
%\tabnote{$^{\rm 6}$ SSCQ: Single Stimulus Continuous Quality Scale.}
%\end{sideways}
\end{table*}
%\end{landscape}

\subsection{IVY stereoscopic image dataset}

Jung \textsl{et al.} proposed the IVY stereoscopic 3D image dataset for the quality assessment of DIBR-synthesized stereoscopic images \cite{jung2016critical}. Different from the above two datasets, in addition to the DIBR view synthesis distortion, the IVY dataset explores binocular perception \cite{wang2019jointly, yang2018stereoscopic} by showing the synthesized image pairs on a stereoscopic display. A total of 7 sequences and three MVD sequences are selected. 84 stereo images are synthesized by four DIBR algorithms \cite{criminisi2004region}, \cite{ahn2013novel}, \cite{tanimoto2008reference}, \cite{yoon2014inter} in this dataset. All the virtual view images in the IVY dataset are generated by single-view-based synthesis methods.

\subsection{MCL-3D image dataset}

Song \textsl{et al.} proposed the MCL-3D stereoscopic image dataset \cite{song2014mcl} to evaluate the quality of DIBR-synthesized stereoscopic images. Although 4 DIBR algorithms are included, the number of images synthesized by these algorithms is quite limited (36 pairs). The major part of this dataset focuses on the traditional distortions in the synthesized views. 6 types of traditional distortions are considered in this dataset: additive white noise, Gaussian blur, down sampling blur, JPEG, JPEG2000 and transmission loss. Nine MVD sequences are collected, among which \textsl{Kendo}, \textsl{Lovebird1}, \textsl{Balloons}, \textsl{PoznanStreet} and \textsl{PoznanHall2} are natural images; \textsl{Shark}, \textsl{Microworld}, \textsl{GT\_Fly} and \textsl{Undodancer} are Computer Graphics images. For each sequence, these traditional distortions are first applied on the base views. Then, the left and right view images are synthesized from these distorted base view images by using the view synthesis reference software (VSRS) \cite{mori2009view}. Different from the above IVC, IETR and IVY datasets, the reference images in the MCL-3D dataset are the images synthesized from undistorted base view images instead of the ones captured by real cameras. 

\subsection{SIAT synthesized video dataset}

The SIAT synthesized video dataset \cite{liu2015subjective} focuses on the distortions caused by compressed texture and depth images in the synthesized views. It uses the same 10 MVD sequences as the IETR image dataset. For each sequence, 4 different texture and depth image quantization levels and their combinations are applied on the base views. Then, the videos at the virtual viewpoints are synthesized using the VSRS-1D-Fast software \cite{sullivan2013standardized}. This dataset uses the real images (captured by real cameras at the virtual viewpoint) as references. Only interview synthesis is used in this dataset.

In the above datasets, the distortions in the DIBR-synthesized views come from not only the DIBR view synthesis algorithms, but also from the distorted texture and depth images.
The IVC \cite{ivcDIBRimagedatabase, bosc2011towards, bosc2013visual}, IVY \cite{jung2016critical} and IETR \cite{tian2019benchmark} datasets focus on the distortions caused by different DIBR view synthesis algorithms; while the MCL-3D \cite{song2014mcl} and SIAT \cite{liu2015subjective} datasets explore the influence of traditional 2D distortions of original texture and depth map on the DIBR-synthesized views. 
These datasets were usually used to evaluate and validate several quality metrics. In the next section, we will introduce the objective approaches for the quality assessment of DIBR-synthesized views.
%The datasets introduced in this section provide the ground truth of DIBR-synthesized view images/videos. 
%In the next section, we'll introduce the obejective approaches for the quality assessment of DIBR-synthesized views.

\section{Objective image/video quality assessment of DIBR-synthesized views}

\begin{figure*}%[!htbp]
\centering
  \includegraphics[width=\linewidth]{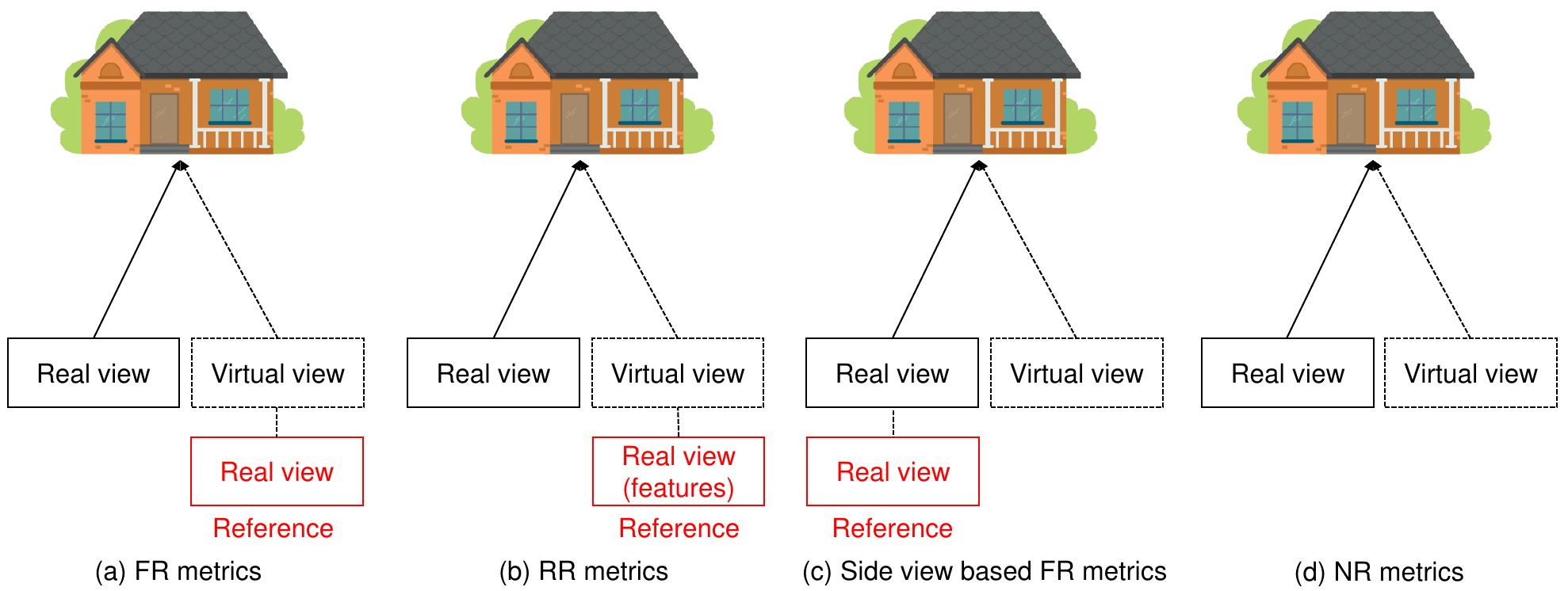}
\caption{Categories of quality assessment metrics for DIBR-synthesized views.}
\label{Fig: FRNR}
\end{figure*}

Several methods have been proposed to evaluate the quality of DIBR-synthesized views in the past decade. Based on the amount of reference information, these methods can be divided into 4 categories: Full-reference (FR), Reduced-reference (RR), Side View based Full-reference (SV-FR) and No-reference (NR), as shown in Fig.~\ref{Fig: FRNR}. The FR methods use the original undistorted image/video at the virtual viewpoint as reference to assess the quality of synthesized views, while the RR methods only use some features extracted from the original reference. Especially, the SV-FR methods use the undistorted image/video at the original viewpoint, from which the virtual view is synthesized, as the reference. The NR methods need no access to the original image/video.

%\begin{figure*}%[!htbp]
%\centering
%\begin{minipage}[b]{0.24\linewidth}
%  \centering
%  \includegraphics[width=0.7\linewidth]{FR.png}
%  \centerline{(a) {FR} metrics}
%\end{minipage}
%\begin{minipage}[b]{0.24\linewidth}
%  \centering
%  \includegraphics[width=0.75\linewidth]{RR.png}
%  \centerline{(b) {RR} metrics }
%  \end{minipage}
%\begin{minipage}[b]{0.24\linewidth}
%  \centering
%  \includegraphics[width=0.8\linewidth]{SVFR.png}
%  \centerline{(c) Side view based FR metrics}
%\end{minipage}
%\begin{minipage}[b]{0.24\linewidth}
%  \centering
%  \includegraphics[width=0.9\linewidth]{NR.png}
%  \centerline{(d) {NR} metrics }
%\end{minipage}
%\caption{Categories of quality assessment metrics for DIBR-synthesized views.}
%\label{Fig: FRNR}
%\end{figure*}

Table~\ref{tab:metrics} {classifies} the metrics based on their \st{used} approaches. Most of them (VSQA, MP-PSNR, MW-PSNR, EM-IQA and CT-IQA) evaluate the quality of synthesized views by considering the contour or gradient degradation between the synthesized and the reference images which is one of the most annoying characteristics of geometric distortions. Meanwhile some metrics (DSQM, 3DSwIM) calculate the quality score by comparing the extracted perceptual features between the synthesized and the reference images. Especially, the APT metric uses a local image description model to reconstruct the synthesize image, and evaluates the quality of the synthesized view based on the reconstruction error. These metrics are introduced as follows.

\begin{table*}%[!htbp]
\centering
\caption{Overview of the existing metrics. The features in the first column indicate hand-craft feature (HF), deep feature (DF), contour/gradient (C/G), JND, Multi-scale decomposition (MSD), local image description (LID), depth estimation (DE), dis-occlusion Region (DR), Sharpness Evaluation (SE), Shift compensation (SC), Image complexity (IC), ML (Machine Learning).}
\label{tab:metrics}
\begin{tabular}{|c|c|c|c|c|c|c|c|c|c|c|c|c|c|}\toprule
\hline
\multicolumn{2}{|c|}{\diagbox{Metric}{Approach}} & HF & DF & C/G & JND & MSD & LID & DE & DR & SE & SC & IC & ML\\
\hline
\multirow{19}{*}{FR}&
{Bosc \textsl{et al.} 2012 \cite{bosc2012edge}}& - & - & \checkmark & - &- &- &- &- &-&- &- &-\\
%\hline
\cline{2-14}
&{{VSQA} \cite{conze2012objective}}& - & - & \checkmark & - &- &- &- &- &- &- &- &-\\
%\hline
\cline{2-14}
&{3DSwIM \cite{battisti2015objective}}& \checkmark & - & - & - &- &- &- &- &- & \checkmark &- &-\\
%\hline
\cline{2-14}
&{MW-PSNR \cite{sandic2016free,sandic2015dibrw}}& \checkmark & - & - & - & \checkmark &- &- &-&- &- &- &-\\
\cline{2-14}
&{MP-PSNR \cite{sandic2016multi}}& \checkmark & - & - & - & \checkmark &- &- &- &- &- &-&-\\
\cline{2-14}
&{CT-IQA \cite{ling2017quality}} & \checkmark & - & - &- &- &- &- &- &- &- &-&-\\
\cline{2-14}
&{ST-SIAQ \cite{ling2017icme}}& \checkmark & - & \checkmark & - &- &- &- &- &- &\checkmark &- &-\\
\cline{2-14}
&{EM-IQA \cite{ling2017image}}& \checkmark & - & \checkmark & - &- &- &- &- &- &\checkmark &- &-\\
\cline{2-14}
&{PSPTNR \cite{zhao2010perceptual}}& - & - & - & \checkmark &- &- &- &- &- &- &-&-\\
\cline{2-14}
&{VQA-SIAT \cite{liu2015subjective}}& - & - & \checkmark & - &- &- &- &- &- &\checkmark &- &-\\
\cline{2-14}
&{SR-3DVQA \cite{Zhang2019Sparse}}& - & - & \checkmark & - &- &- &- &- &- &\checkmark  &- & \checkmark \\
\cline{2-14}
& SDRD \cite{zhou2016quality}& - & - & - & - &- &- &- &\checkmark &- &\checkmark &- &-\\
\cline{2-14}
&{SCDM \cite{tian2018dibrfr}}& - & - & - & - &- &- &- &\checkmark &- &\checkmark &- &-\\
\cline{2-14}
&{SC-IQA \cite{tian2018sc}}& - & - & - & - &- &- & - &- &- &\checkmark &- &-\\ 
\cline{2-14}
&{CBA \cite{jung2016critical}}& - & - & - & - &- &- & - &\checkmark &- &- &- &-\\ 
\cline{2-14}
&{Zhou \cite{zhou2019quality}}& \checkmark & - & \checkmark & - &\checkmark &- & - &- &- &- &- &\checkmark \\
\cline{2-14}
& {Ling \cite{ling2019perceptual}} & \checkmark & - & \checkmark & - &- & - &- & - & - &- &- & \checkmark \\
\cline{2-14}
& {Wang \cite{wang2019qualityaccess}} & \checkmark & - & \checkmark & - &- & - &- & \checkmark & \checkmark &- &- & - \\
\hline
\multirow{4}{*}{RR}&
{MP-PSNRr \cite{sandic2016dibr}} & \checkmark & - & - & - & \checkmark &- &- &- &- &- &- &-\\
\cline{2-14}
&{MW-PSNRr \cite{sandic2016dibr}}& \checkmark & - & - & - & \checkmark &- &- &- &- &- &- &-\\
\cline{2-14}
&{RRLP \cite{jakhetiya2017prediction}} & - & - & \checkmark & - &- & \checkmark &- & - & \checkmark &-  &- &-\\
\hline

\multirow{5}{*}{{\makecell*[c]{Depth IQA \\ FR/RR/NR}}}&
{(FR) Li \cite{li2019depth}}& - & - & \checkmark & - &\- &- & - &- &- &- &- &- \\
\cline{2-14}
&{(RR) RR-DQM \cite{le2017new}} & - & - & \checkmark & - &- & \checkmark &- & - & - &-  &- &-\\
\cline{2-14}
& {(NR) BDQM \cite{farid2015blind}} & - & - & \checkmark & - &- & - &- & - & - &- &- &-\\
\cline{2-14}
& {(FR) Xiang \cite{xiang2015no}} & - & - & \checkmark & - &- & - &- & - & - &- &- &-\\
\cline{2-14}
& {(NR) SEP \cite{li2020no}} & \checkmark & - & \checkmark & - &- & - &- & - & - &- &- &\checkmark\\
\hline

\multirow{5}{*}{SV-FR}&
{3VQM \cite{solh20113vqm}}& - & - & - & - &- &- & \checkmark &- &- &- &- &-\\
\cline{2-14}
&{LOGS \cite{li2018quality}} & - & - & - & - &- &- &- & \checkmark & \checkmark &\checkmark &- &- \\
\cline{2-14}
&{DSQM \cite{farid2017perceptual}}& \checkmark & - & - & - &- &- &- & - & - &\checkmark &- &-\\
\cline{2-14}
&{SIQE \cite{farid2015objective}}& \checkmark & - & - & - &- &- &- & - & - &\checkmark &- &-\\
\cline{2-14}
&{SIQM \cite{farid2018evaluating}}& \checkmark & - & - & - &- &- &- & - &- & \checkmark &- &-\\
\hline

\multirow{16}{*}{NR}&
{APT \cite{gu2017model}} & - & - & - & - &- & \checkmark &- & - & - &- &- &-\\
\cline{2-14}
&{OUT \cite{jakhetiya2018highly}} & - & - & - & - &- & \checkmark &- & - &- & - &- &-\\
\cline{2-14}
&{MNSS \cite{gu2019multiscale}} & \checkmark & - & - & - & \checkmark & - &- & - &- & - &- &\checkmark\\
\cline{2-14}
&{NR\_MWT \cite{sandic2019fast}} & \checkmark & - & \checkmark & - &\checkmark & - &- & - &\checkmark & - &- &-\\
\cline{2-14}
&{NIQSV \cite{shishun2017}} & - & - & \checkmark & - &- & \checkmark &- & - & - &- &-&- \\
\cline{2-14}
&{NIQSV+ \cite{tian2018niqsv+}} & - & - & \checkmark & - &- & \checkmark &- & - & - &- &- &-\\
\cline{2-14}
&{HEVSQP\cite{shao2017no}} & - & - & \checkmark & - &- & - &- & - & - &- &- &\checkmark\\
\cline{2-14}
& {CLGM \cite{yue2018combining}} & - & - & \checkmark & - &- & - &- & \checkmark & \checkmark &- &- &-\\
\cline{2-14}
& {GDIC \cite{wang2019blind}} & \checkmark & - & \checkmark & - &- & - &- & - & - & - &\checkmark &-\\
\cline{2-14}
& {Wang \cite{wang2019blindtip}} & \checkmark & - & \checkmark & - &- & - &- & - & \checkmark &- &\checkmark &-\\
\cline{2-14}
& {SET \cite{zhou2019no}} & \checkmark & - & \checkmark & - &\checkmark & - &- & - & - &- &- &\checkmark\\
\cline{2-14}
& {CTI \cite{kim2016measurement}} & - & - & - & - &- & - &- & - & - &\checkmark  &- &-\\
\cline{2-14}
& {FDI \cite{zhou2018no}} & - & - & \checkmark & - &- & - &- & - & - &\checkmark  &- &-\\
\cline{2-14}
& {CSC-NRM \cite{ling2018learn}} & - & - & - & - &- & - &- & - & - &- &- & \checkmark\\
\cline{2-14}
& {SIQA-CFP \cite{wang2019deepicip}} & - & \checkmark & - & - &- & - &- & - & - &- &- & \checkmark \\
\cline{2-14}
& {GANs-NRM \cite{ling2019gannrm}} & - & \checkmark & - & - &- & - &- & - & - &- &- & \checkmark \\
\hline
\end{tabular}
\end{table*}

\subsection{FR and RR metrics}
%The FR metrics evaluate the quality of synthesized views by comparing the difference between the synthesized view image/video and the reference image/video while the RR metrics only uses their associate features. 
In this subsection, we review 20 well-known FR metrics and 4 RR metrics. 

\subsubsection{{Edge/Contour based FR metrics}}
The distortions in DIBR-synthesized views are mostly geometrical and structural distortions, which may degrade the object shape in the synthesized image. It can be measured by the change of object edges. In addition, the sharp edges in the depth map may also induce large dis-occlusions in the synthesized views which may result in dramatic distortions. Thus, a few edge-based methods have been proposed to evaluate the quality of DIBR-synthesized views. 

The FR metric proposed by Bosc \textsl{et al.} in \cite{bosc2012edge} indicates the structural degradations by calculating the contour displacement between the synthesized and the reference images. Firstly, a Canny edge detector is used to extract the image contours; then, the contour displacements between the synthesized and reference images are estimated. Based on the contour displacement map, three parameters are computed: the mean ratio of inconsistent displacement vectors per contour pixel, the ratio of inconsistent vectors, the ratio of new contours. The final quality score is obtained as a weighted sum of these three parameters.

In \cite{ling2017icme}, Ling \textsl{et al.} proposed a contour-based FR metric ST-SIAQ for the quality assessment of DIBR-synthesized views. Instead of directly using the contour information in \cite{bosc2012edge}, ST-SIAQ uses mid-level contour descriptor called ``Sketch Token'' \cite{Lim2013sketch}. The ``Sketch Token'' stands as a codebook of image contour representation, of which each dimension can be recognized as the possibility which indicates how likely the current patch belongs to one certain category of contour from the codebook. To reduce the shifting effect in the feature comparison stage, the patches in the reference image are firstly matched to the synthesized image. The ``Sketch Token'' is clustered into 151 categories, which means the ``Sketch Token'' descriptor has 151 dimensions. A Random Forests decision model associated with a set of low-level features (including oriented gradient channels \cite{dollar2009integral}, color channels, and self-similarity channels \cite{shechtman2007matching}) {is} used to obtain the ``Sketch Token'' descriptor. The geometric distortion strength in the synthesized view is calculated as the Kullback Leibler divergence of ``Sketch Token'' descriptors between the synthesized and reference images. In \cite{ling2019prediction}, this metric is improved to evaluate the quality of DIBR-synthesized videos by considering the temporal dissimilarity.

Ling \textsl{et al.} also proposed another contour-based FR metric EM-IQA in \cite{ling2017image}. Different from ST-SIAQ metric, EM-IQA uses {an} interest points matching and {an} elastic metric \cite{mio2007shape}, instead of block matching and ``Sketch Token'' descriptor, to compensate the shifting and evaluate the contour degradation respectively. After {the} interest points matching, a Simple Linear Iterative Clustering (SLIC) is used to extract the contours in the image. SLIC is originally proposed for image segmentation{. In} the EM-IQA metric, the boundaries of the segmented objects are considered as contours. Then, the elastic metric proposed in \cite{mio2007shape,srivastava2010shape} is used to finally measure the degradation between the contours of synthesized and reference images, which provides the quality score of DIBR-synthesized view.

In \cite{ling2017quality}, Ling \textsl{et al.} proposed a variable-length context tree based image quality assessment metric CT-IQA, dedicated to quantify the overall structure dissimilarity and dissimilarities in various contour characteristics. Firstly, the contours of the reference and synthesized images are converted to differential chain code (DCC) \cite{freeman1978application} which represents the direction of object contours. Then, an optimal context tree \cite{zheng2016context} is learned from the DCC in the reference image. The overall structural dissimilarity is calculated by subtracting the encoding cost of DCC in the synthesized image and reference images. In addition, the overall dissimilarity in contour characteristics is also obtained by measuring the difference of total contour number, total contour start information and total number of symbols between the reference and synthesized image. The final quality score is calculated by combining the overall structure dissimilarity and contour characteristics dissimilarity.

Liu \textsl{et al.} proposed a gradient-based FR video quality assessment metric VQA-SIAT \cite{liu2015subjective} by considering the ``Activity'' and ``Flickering'' which is the most annoying temporal distortion in the DIBR-synthesized views. The main contribution of this metric is the two following proposed structures: Quality Assessment Group of Pictures (QA-GoP) and Spatio-Temporal (S-T) tube. The QA-GoP acts as a process unit on a whole video sequence, it contains a group of 2N+1 frames (N frames before and N frames after the central frame). Besides, a block matching method is used to search the corresponding blocks of the central frame blocks in the forward and backward frames. The 2N + 1 blocks along the motion trajectory construct a S-T tube. The distortion of ``Activity'' is calculated from the difference of the spatial gradient in the (S-T) tube and (QA-GoP) between the synthesized and reference videos. The ``Flickering'' distortion is measured from the difference of temporal gradient, which is defined below:
\begin{equation}
\vec {\bigtriangledown} I_{x,y,i}^{temporal} = I(x,y,i) - I(x', y',i-1),
\end{equation}
where ($x', y'$) is the coordinate in frame $i-1$ corresponding to ($x,y$) along the motion trajectory in {the} previous frame $i$. The final quality score of DIBR-synthesized view video is obtained by integrating both ``Activity'' and ``Flickering'' distortions.

Furthermore, in \cite{Zhang2019Sparse}, Zhang \textsl{et al.} proposed a FR metric SR-3DVQA combining the ``Activity'' measurement module in VQA-SIAT with a sparse representation-based flicker estimation method. In the SR-3DVQA metric, a DIBR-synthesized video is treated as a 3D volume data by stacking the frames sequentially. Then, the volume data is decomposed as a number of spatially neighboring temporal layers i.e. X-T or Y-T planes{,} where X, Y are the spatial coordinate and T is the temporal coordinate. In order to effectively evaluate the flicker distortion in the synthesized video, the gradient in the temporal layers and sharp edges in the associate depth map are extracted as key features for the dictionary learning and sparse representation. The rank-based method in \cite{li2018quality} is used to pool the flicker score from the temporal layers. The final quality score is calculated by combining the flicker score and ``Activity'' score in the previous VQA-SIAT \cite{liu2015subjective}.

Jakhetiya \textsl{et al.} proposed a free-energy-principle-based IQA metric RRLP for Screen Content and DIBR-synthesized view images based on prediction model and distortion categorization \cite{jakhetiya2017prediction}. The image quality is measured by calculating the disorder and sharpness similarity between the distorted and reference images. The disorder is obtained from a prediction model. As shown in Eq.~\ref{eq:obf}, an observation-model-based bilateral filter (OBF) \cite{Jakhetiya2014fast} is firstly used to divide the predicted and disorder parts.
\begin{equation}
\hat{X_{d_i}} = \frac{X_{d_i}\lambda + \sum_{k \in N_i}\omega_{k_i}I_{k_i}}{\lambda + \sum_{k \in N_i}\omega_{k_i}}
\label{eq:obf}
\end{equation}
where $\hat{X_{d_i}}$ represents the predicted part, $I_{k_i}$ and $\omega_{k_i}$ are respectively the pixels and their associated weights in the surrounding $3 \times 3$ window $N_i$ of the $i$th pixel, $\lambda$ is a parameter. The disorder part is computed as the difference between the predicted part and the original image:
\begin{equation}
R_{d_i} = |\hat{X_{d_i}} - {X_{d_i}}|
\end{equation}
Then, the sharpness (edge structures) is calculated by four filters in\cite{Wu2013reduce}. Finally, the disorder and sharpness similarity between the distorted and reference images are estimated by using the similarity function in SSIM \cite{Wangssim}.

\subsubsection{{Wavelet transform based FR metrics}}

In the previous part, we introduced the metrics that use the edge/contour in luminance domain to evaluate the geometric distortions in DIBR-synthesized views. According to previous research, the wavelet transform representation can not only capture the image edges, but also some other texture unnaturalness. In this part, the wavelet transform based FR metrics will be reviewed. 

Battisti \textsl{et al.} proposed an FR metric (3DSwIM) for DIBR-synthesized views based on the comparison of statistical features of wavelet sub-bands \cite{battisti2015objective, 3dswimsoft}. The same as EM-IQA \cite{ling2017image} and VQA-SIAT \cite{liu2015subjective}, 3DSwIM uses a block matching to ensure the ``shifting-resilience''. The distortions in each block of the synthesized view is measured by the Kolmogorov-Smirnov \cite{lilliefors1967kolmogorov} distance between {the histograms of the matched blocks in the synthesized and reference images.} In addition, since the Human Vision System (HVS) pays more attention on the human body, a skin detector is used to weight the skin regions in the matched blocks. 

Sandi\'c-Stankovi\'c \textsl{et al.} proposed another multi-scaled decomposition based FR metric MW-PSNR \cite{sandic2015dibrw, sandic2016free}. The MW-PSNR uses morphological wavelet filters for decomposition. Then a multi-scale wavelet mean square error (MW-MSE) is calculated as the average MSE of all sub-bands and finally the MW-PSNR is calculated from it. 

The wavelet transform based FR metrics can be recognized as a kind of edge/contour based metrics. For example, the higher sub-bands of the wavelet transformed image represent the edge information of the original image. Compared to the pixel level edge/contour used in the previous subsection, the metrics in this subsection use the features in wavelet transformed domain to represent both the image edges and other characteristics.

\subsubsection{{Morphological operation based FR metrics}}
\begin{figure}%[!htbp]
  \centering
  \begin{minipage}[b]{0.5\linewidth}
  \centering
  \centerline{\includegraphics[width=1\linewidth]{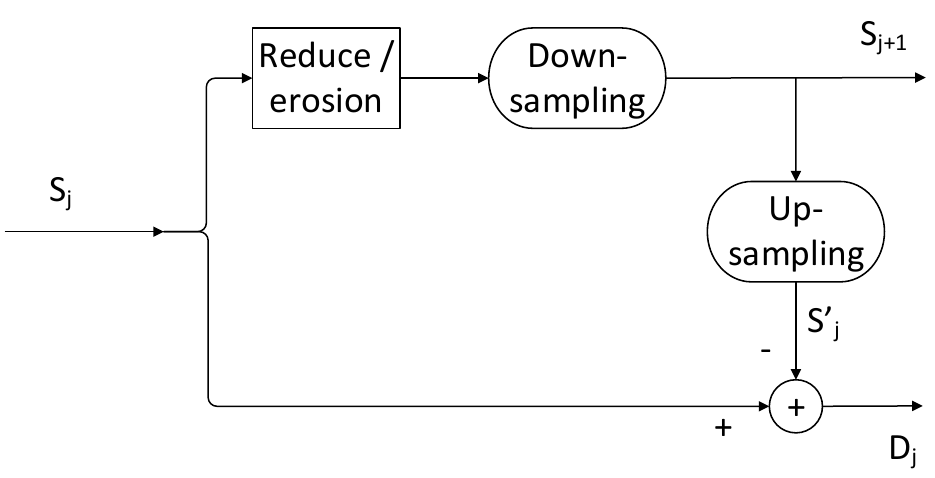}}
%  \vspace{1.5cm}
%  \centerline{(a) different / similar}\medskip
  \end{minipage}
  \caption{Decomposition scheme of MP-PSNR. $S_j$ represents the image at scale $j$ ($j \in [1,5]$), $D_j$ represent the detail image at scale $j$ \cite{sandic2016multi}.}
\label{fig:mppsnr}
\end{figure}

Morphological operations are widely used in image processing, especially a couple of erosion and dilation operations can be used to detect the image edges \cite{maragos1990morphological}. In \cite{sandic2016multi}, Sandi\'c-Stankovi\'c \textsl{et al.} proposed the MP-PSNR based on multi-scaled pyramid decomposition using morphological filters. The basic erosion and dilation operations used in MP-PSNR are calculated as maximum and minimum in the neighbourhood defined by the structure element, as shown in the following equation:
\begin{equation}
D: dilation_{SE}(f)(x) = max_{y \in SE}{f(x-y)}
\end{equation}
\begin{equation}
E: erosion_{SE}(f)(x) = min_{y \in SE}{f(x+y)}
\end{equation}
where $f$ is a gray-scale image and $SE$ is binary structure element. Then, they use the Mean Square Error (MSE) between the reference and synthesized images in all pyramids' sub-bands to quantify the distortion. As shown in Fig~\ref{fig:mppsnr}, during the decomposition, the dilation is used as expand{ing} operation and the erosion is used as {reducing} operation. The detail image of each scale is calculated as the difference between the original and processed (erosion and dilation) images. Finally, the overall quality is calculated by averaging the MSE of detail images in all the sub-bands and expressing it {in terms of} PSNR.

In \cite{sandic2016dibr}, Sandi\'c-Stankovi\'c \textsl{et al.} also proposed the reduced version of MP-PSNR, and MW-PSNR. Only detail images from higher decomposition scales are taken into account to measure the difference between the synthesized image and the reference image. The reduced version achieved significant improvement over the original FR metrics with lower computational complexity.

%
%\begin{equation}
%MSE_j = \frac{1}{N_j \times K_j} \sum_{k=1}^{K_j}\sum_{n=1}^{N_j}(x_j(k,n)-y_j(k,n))^2
%\end{equation}
%where $x_j$ and $y_j$ denote the reference and distorted image at scale $j$ with size $K_j\times N_j$. Multi-scale mean squared error MP-MSE is calculated as weighted product of MSE of all pyramid images with equal weights:
%\begin{equation}
%MP-MSE = \prod_{j=1}^{M} [MSE_j]^{\beta_j}
%\end{equation}
%where $\beta_j$ indicates a weight value parameter. Finally, the MP-PSNR score is calculated as:
%\begin{equation}
%MP-PSNR = 10 \times log_{10}(\frac{R^2}{MP-MSE})
%\end{equation}
%where $R$ is the maximum dynamic range of the image.

\subsubsection{{Dis-occlusion region based FR metrics}}
Since the DIBR view synthesis distortions mainly occur in the dis-occlusion regions, some of the FR metrics improve the performance of 2D FR metrics by using dis-occlusion maps\cite{zhou2016quality,tian2018dibrfr} instead of using weighting maps.

The SDRD metric proposed by Zhou in \cite{zhou2016quality} detects the dis-occlusion regions by simply comparing the absolute difference between the synthesized and reference images. Before that, a self-adaptive scale transform model is used to eliminate the effect of view distance, and a SIFT flow-based warping is adopted to compensate the global shift in the synthesized view image. The final quality score is obtained by weighting the dis-occlusion regions with their size since the distortions with bigger size are more annoying to human vision system.

Tian \textsl{et al.} proposed a full-reference quality assessment model (SCDM) for 3D synthesized views by considering global shift compensation and dis-occlusion regions \cite{tian2018dibrfr}. This model can be used on any pixel-based FR metrics. SCDM firstly compensates the shift by using a SURF + RANSAC approach instead of the SIFT flow used in SDRD. Then, the dis-occlusion regions are directly extracted from the depth map. It is more precise and uses more resources compared to SDRD.
%considering the fact that the DIBR view synthesis distortion mainly locate in the dis-occlusion region, a depth map is used to extract a dis-occlusion mask. 
The final quality score is obtained as a weighted PSNR or weighted SSIM. It is reported to improve the performance of PSNR and SSIM by 36.85\% and 13.33\% in terms of Pearson Linear Correlation Coefficients (PLCC).

Since the distortions in the DIBR-synthesized views are not restricted in the dis-occlusion regions only, they may occur around these regions as well. In \cite{wang2019qualityaccess}, Wang \textsl{et al.} proposed a critical region based metric by dilating the dis-occlusion region with a morphological operator. Similar to SDRD, the dis-occlusion region map is extracted by a SIFT-flow based approach. Then a Discrete Cosine Transform (DCT) decomposition method is used to partition and classify the critical regions into edge blocks, texture blocks and smooth blocks. Based on the perceptual properties of these three types of blocks, their distortions are measured differently. The  edge and texture blocks contain more complex edges or texture information, the blur distortions in these regions would be much more annoying than that in the smooth regions. On the other hand, the smooth regions are sensitive to color degradations. Thus, the texture similarity and color contrast similarity between the synthesized and reference images are calculated to measure the local distortions in the edge, texture and smooth blocks respectively. Finally, a global sharpness detection is combined with the local distortion measurement to obtain the overall quality score.

\subsubsection{{2D related FR metrics}}
The main reason of the ineffectiveness of 2D quality assessment metrics on DIBR-synthesized views can be {analyzed} as follows. Firstly, there {are} large object shift{s} in the synthesized views and this kind {of} shift{s} can be easily penalized by 2D metrics even though the HVS is not sensitive to the global shift in the image. The second reason is the distribution of distortions. The distortions in traditional 2D images often scatter over the whole image while the DIBR view synthesis distortions are mostly local, {mainly} in the dis-occluded regions. The 2D related metrics are based on the traditional 2D FR metrics, {e.g.} PSNR, SSIM{,} etc. They try to improve the performance of 2D metrics by considering HVS and the characteristics of DIBR view synthesis distortions. 

The VSQA metric proposed by Conze \textsl{et al.} in \cite{conze2012objective} tries to improve the performance of SSIM \cite{Wangssim} by taking advantage of known characteristics of the HVS. It aims to handle areas where disparity estimation may fail, such as thin objects, object borders, transparency{,} by applying three weighting maps on the SSIM distortion map. The main purpose of these three weighting maps is to characterize the image complexity in terms of texture\st{s}, diversity of gradient orientations and presence of high contrast since the HVS is more sensitive to the distortions in such areas. For example, the distortions in an untextured area are much more annoying than the ones located in a high texture complexity area. It is reported that this method approaches a gain of 17.8\% over SSIM in correlation with subjective measurements. 

Zhao \textsl{et al.} proposed the PSPTNR metric to measure the perceptual temporal noise of the synthesized sequence \cite{zhao2010perceptual}. The temporal noise is defined as the the difference between inter-frame change in the processed sequence and that in the reference sequence:
\begin{equation}
TN_{i,n} = ((P_{i,n} - P_{i, n-1}) - (R_{i,n} - R_{i,n-1})^2 ,
\end{equation}
where $TN$ indicates the temporal noise, $P$ and $R$ represent the distorted and reference sequence respectively. In order to better predict the perceptual quality of synthesized videos, temporal noise is filtered by a Just Noticeable Distortion (JND) model and a motion mask \cite{westerink1995perceived}, since the human can observe noise only beyond certain level and motion may decrease the texture sharpness in the video.

The shift compensation methods included in SDRD and SCDM only consider the global shift. {B}ut according to the recent research\cite{moorthy2009visual}, {the HVS} is more sensitive to local artefacts compared to the global object shift. In \cite{tian2018sc}, Tian \textsl{et al.} proposed a shift-compensation based image quality assessment metric (SC-IQA) for DIBR-synthesized views. The same as SCDM, a SURF + RANSAC approach is used to roughly compensate the global shift. In addition, a multi-resolution block matching method is proposed to precisely compensate the global shift and penalize the local shift at the same time. A saliency map \cite{jiang2013salient} is also considered to weight the distortion map of the synthesized view. Furthermore, only the blocks with the worst quality are used to calculate the final quality score since HVS tends to perceive poor regions in an image with more severity than the good ones \cite{moorthy2009visual, liu2015subjective}. SC-IQA achieves the performance of SCDM without access to the depth map.

The metrics introduced above consider only the view synthesis and compression artefacts which occur on applications that show the synthesized views on a 2D display, the binocular effect in the synthesized stereoscopic images is not taken into consideration. In \cite{jung2016critical}, Jung \textsl{et al.} proposed a SSIM-based FR metric to measure the critical binocular asymmetry (CBA) in the synthesized stereo images. Firstly, the disparity inconsistency between the two different views is generated to detect the critical areas in terms of Left-Right image mismatches. Then, only the SSIM value on the critical areas of each view are computed to measure the asymmetry in the corresponding view image. The final binocular asymmetry score is obtained by averaging the asymmetry score in the left and right views. 

\subsection{Side view based FR metrics}
The major limitation of the FR metrics is that they always need the reference view which may be unavailable in some circumstances {\textsl{e.g.}} FVV). In other words, there is no ground truth for a full comparison with the distorted synthesized view. In this part, four side view based FR metrics will be reviewed. This kind of metrics use the real image/video at the original viewpoint, from which the virtual view is synthesized, as {the} reference to evaluate the quality of DIBR-synthesized virtual views. These metrics are named {``}side view based FR metrics{''} in this paper. 

Solh \textsl{et al.} proposed a side view based FR metric 3VQM \cite{solh20113vqm} to evaluate synthesized view distortions by deriving an ``ideal'' depth map from the virtual synthesized view and the reference view at a different viewpoint. The ``ideal'' depth is the depth map that would generate the distortion-free image given the same reference image and DIBR parameters.
Three distortion measurements, spatial outliers, temporal outliers and temporal inconsistency are calculated from the difference between the ``ideal'' depth map and the distorted depth map:
\begin{equation}
SO = STD(\triangle Z)
\end{equation}
\begin{equation}
TO = STD(\triangle Z_{t+1} +\triangle Z_{t})
\end{equation}
\begin{equation}
TI = STD(Z_{t+1} + Z_{t})
\end{equation}
where $SO$, $TO$ and $TI$ denote the spatial outliers, temporal outliers and temporal inconsistencies respectively, $STD$ represents the standard deviation. $\triangle Z$ is the difference between the ``ideal'' and the distorted depth maps and $t$ is the frame number. These three measurements are then integrated into a final quality score.
Since the calculation of the ``ideal'' depth map is based on the assumption that the horizontal shift of the synthesized view and the original view is small, this metric would not work well when the baseline distance increases. 

Li \textsl{et al.} proposed a side view based FR metric for DIBR-synthesized views by measuring local geometric distortions in dis-occluded regions and global sharpness (LOGS) \cite{li2018quality}. This metric consists of three parts. Firstly, the dis-occlusion regions are detected by using SIFT-flow based warping. These dis-occluded regions are extracted from the absolute difference map between the synthesized view $I_{syn}$ and the warped reference view $I_{ref}^{w}$ followed by an additional threshold. Then, the distortion size and strength in the local dis-occlusion regions are combined to obtain the overall local geometric distortion. The distortion size is simply measured by the number of pixels in the dis-occluded regions and the distortion strength is defined as the mean value of the dis-occluded regions in the whole difference map $M$. The next part is to measure the global sharpness by using a reblurring-based method. The synthesized image is firstly blurred by a Gaussian smoothing filter. Both the synthesized image and its reblurred version are divided into blocks. The sharpness of each block is calculated by its textural complexity, which is represented by its variance $\sigma^2$. Then, the overall sharpness score is computed by averaging the textural distance of all blocks. Finally, the local geometric distortion and the global sharpness are pooled to generate the final quality score.

Farid \textsl{et al.} proposed a side view based FR metric (DSQM) for the DIBR-synthesized view in \cite{farid2017perceptual}. A block matching is firstly used to estimate the shift between the reference and synthesized image. Then the difference of Phase congruency (PC) in these two matched blocks is used to measure the quality of the block in the synthesized image, which is defined as follows:
\begin{equation}
PC(x) = \max_{\bar{\phi(x)} \in [0,2\pi]}{\frac{\sum_n{A_n cos(\phi_n(x)} - \bar{\phi(x)})}{\sum_n{A_n}}}
\end{equation}
where $A_n$ and $\phi_n(x)$ represent the amplitude and the local phase of the $n$-th Fourier component at position $x$ respectively. The implementation of phase congruency is based on {a} logarithmic Gabor wavelet method proposed in \cite{kovesi1999image}.
The quality score of each block is calculated as the absolute difference between the mean values of the phase congruency maps of the matched blocks in the synthesized and reference image:
\begin{equation}
Q_i = |\mu(PC_{si} - PC_{ri})|
\end{equation}
where $\mu()$ represents the mean value of the corresponding phase congruency map, the $PC_{si}$ and $PC_{ri}$ indicate the PC map of the matched blocks in the synthesized and reference image. The final image quality is obtained by averaging the quality score of all the blocks.

Farid \textsl{et al.} proposed a cyclopean eye theory \cite{julesz1972cyclopean} and divisive normalization (DN) transform \cite{teo1994perceptual} based Synthesized Image Quality Evaluator (SIQE) in \cite{farid2015objective}. The DIBR-synthesized view image associated with the left and right side views are firstly transformed by DN. Then, the statistical characteristics of the cyclopean image are estimated from the DN representations of the left and right side views while the statistical characteristics of the synthesized image are obtained directly from itself. The similarity (Bhattacharyya coefficient \cite{patra2015new}) between the distribution of the cyclopean and the synthesized image's DN representations is computed to measure the quality score of the synthesized image.

The SIQE metric only considers the texture information, in \cite{farid2018evaluating}, Farid \textsl{et al.} proposed an extended version of SIQM by considering both the texture and depth information. The depth distortion estimation is based on the fact that the edge regions in a depth image are more sensitive to noise than the flat homogeneous regions since the distorted edge in the depth map may cause very annoying structural distortions in the synthesized image. Firstly, the pixels in the depth map with a high gradient value are extracted as noise sensitive pixels (NSP). Then, for each NSP, a local histogram from the distorted depth map is constructed and analysed to estimate the distortion in the depth image. The overall depth distortions are calculated by averaging the  distortions in the left and right depth image. The final quality of the synthesized view is pooled from the texture and depth distortions.

\subsection{Depth image quality metrics}
{The quality of depth images is crucial for generating high-quality synthesized views. A few metrics have been proposed to predict the depth image quality in DIBR view synthesis.}

{Le \textsl{et al.} proposed a RR depth image quality metric (RR-DQM) \cite{le2017new} which requires a pair of color and depth images. The depth image quality is measured depending on the edge directions based on the fact that the local depth distortion and the local image characteristic are strongly correlated. A Gabor filter is applied to generate a weighting map which are then used to adaptively weight the local depth distortion.}

{Li \textsl{et al.} proposed a FR depth image quality metric based on weighted edge similarity \cite{li2019depth}. Based on their observation that the distortions in the DIBR-synthesized views are mainly concentrated in the edge regions of depth maps, the proposed metric is designed with emphasis on the distortions in depth edge regions. The similarity between the distorted and reference depth map is calculated in both intensity and gradient domains. Then, a weighting map is generated by combining a location prior and a depth distance measure. Finally, the edge indication is used as a guidance to pool the overall quality of depth map.
}

{Farid \textsl{et al.} proposed a blind depth quality metric (BDQM) \cite{farid2015blind} to evaluate the compression distortions in depth images. They noticed that the compression flattens the sharp transitions of the depth image. Therefore, the shape of the histogram around the depth boundaries are used to predict the depth quality.}

{In \cite{xiang2015no}, Xiang \textsl{et al.} proposed a NR depth image quality metric by calculating the misalignment errors between the edges of texture and depth images. The misalignments are evaluated from three similarities: the edge orientation similarity, the spatial similarity and the segmentation length similarity. Finally, the misalignments are used to calculate the final quality scores. 
}

{Li \textsl{et al.} proposed a NR depth image quality index based on the statistics of edge profiles (SEP) \cite{li2020no}. The first-order and second-order statistical features are firstly extracted based on edge profiles which are the neighbouring regions around the depth edges. Then, the random forest (RF) is applied to build a quality assessment model for depth maps.
}

{The depth image quality metrics can evaluate the quality of synthesized view before performing actual rendering and is thus more computational friendly. It can also be used in the rate distortion optimization of depth map compression.
The same as the texture IQA metrics, the NR depth image quality metrics are more practical than the FR ones since the depth maps are usually acquired by depth cameras or depth estimation approaches and are not always available.
}

\subsection{NR metrics}
In this part, we will review the NR metrics which do not need ground truth images/videos to evaluate the quality of DIBR-synthesized views.

\subsubsection{{Local image description based NR metrics}}
Due to the distorted depth map and imperfect rendering method, there exists a large number of structural and geometric distortions in the DIBR-synthesized views. As introduced in the RRLP metric \cite{jakhetiya2017prediction}, the structural distortions may result in local disorder in the image. Similarly, several local image description based NR metrics  have been proposed to evaluate the structural distortions by measuring the local inconsistency via different models. 

Gu \textsl{et al.} proposed an auto-regression (AR) based model (APT) to capture the geometric distortions in the DIBR-synthesized views. For each pixel, a local AR model (3$\times$3) is first used to construct a relationship between this pixel and its neighbouring pixels.
\begin{equation}
x_i = \Omega(x_i)s + d_i
\end{equation}
where $\Omega(x_i)$ denotes a vector which is composed of the neighbouring pixels of $x_i$ in the (3$\times$3) patch, $s$ is a vector of AR parameters and $d_i$ represents the error difference between the current pixel value and its corresponding AR prediction. The AR parameters are solved on the assumption that the 7$\times$7 local patch, which consists of the current pixel and its 48 adjacent pixels, shares the same AR model. The error difference map between the synthesized and the reconstructed images is obtained as the distortion map. Then, a Gaussian filter and a saliency map \cite{gu2015visual} associated with a maximum pooling are used to obtain the final image quality score. Due to its computational complexity, this method owns a high computing cost.

Different from the APT metric, the OUT (outliers) metric \cite{jakhetiya2018highly} proposed by Jakhetiya \textsl{et al.} uses a median filter to calculate the difference map. Then, two thresholds are used to extract the structural and geometric distortion regions. The quality score is finally obtained from the standard deviation of the structural and geometric distortion regions. 

These local image description based metrics can only detect thin distortions or local noise, they do not work well on the large size distortions.

%\subsubsection{Learning based metrics}
%Machine learning has shown its effectiveness in several topics, including image quality assessment. 
\subsubsection{{Morphological operation based NR metrics}}
The morphological operations show their effectiveness in the FR metric MP-PSNR \cite{sandic2016multi}. In \cite{shishun2017, tian2018niqsv+}, Tian \textsl{et al.} proposed two metrics NIQSV and NIQSV+ to detect the local thin structural distortions through morphological operations. These two metrics assume that the ``perfect'' image consists of flat areas and sharp edges, so such images are insensitive to the morphological operations while the local thin structural distortions can be easily detected by these morphological operations. The NIQSV metric firstly uses an opening operation to detect the thin distortions and followed by a closing operation with larger Structural Element (SE) to file the black holes. The NIQSV+ extend the NIQSV by proposing two additional measurements: black hole detection and stretching detection. The black hole distortion is estimated by counting the black hole pixels proportion in the image while the stretching distortion is evaluated by calculating the gradient decrease of the stretching region and its adjacent non-stretching region. 

Due to the limitation of the assumption and the SE size, these two metrics do not work well on the distortions in complex texture and the distortions with large size.
%These two metrics are 

\subsubsection{{Sharpness detection based NR metric}}
Sharpness detection has been widely used in 2D image quality assessment \cite{tang2016blind, zhang2017reduced, gu2015nosharpness} and also in the side view based FR metric LOGS \cite{li2018quality}. In this part, we will introduce its usage in NR metrics. Sharpness is one of the most important measurements in NR image quality assessment \cite{Gu2015noref, Min2018blind, Ferzli2009}. The DIBR view synthesis may introduce multiple distortions such as blur, geometric distortions around the object edges, which may significantly result in the degradation of sharpness. 
%The edge/gradient change between the synthesized and reference images can be used to represent the object deformation in the synthesized views. Besides, in the DIBR-synthesized videos, the gradient variation may result in temporal flickering which is one of the main distortions in DIBR-synthesized videos. Since some of the sharpness and edge detection metrics are achieved with similar methods, we classify them as one category in this part.

Nonlinear morphological wavelet decomposition can extract high-pass image content while preserving the unblurred geometric structures \cite{sandic2016multi, sandic2015dibrw}. In the transform domain, geometry distorted areas introduced by DIBR-synthesis are characterized by coefficients of higher value compared to the coefficients of smooth, edge and textural areas. In \cite{sandic2019fast}, Sandi\'c-Stankovi\'c \textsl{et al.} proposed a wavelet-based NR metric (NR\_MWT) for the DIBR-synthesized view videos. The sharpness is measured by quantifying the high frequency components in the image, which are represented by the high-high wavelet sub-band. The final quality is obtained from the sub-band coefficients whose value are higher than the threshold. Similar to MW-PSNR and MP-PSNR \cite{sandic2016multi, sandic2015dibrw}, the NR\_MWT also {has} a very low computational complexity.

Differently, in CLGM \cite{yue2018combining}, the sharpness is measured as the distance of standard deviations between the synthesized image and its down-sampled version.
%, which is quite similar to the self similarity used in the MNSS model \cite{gu2019multiscale}. 
Besides, two additional distortions, dis-occluded regions and stretching, are also taken into consideration in CLGM. The dis-occluded regions are detected through an analysis of local image similarity. Similar to NIQSV+ \cite{tian2018niqsv+}, the stretching distortion is estimated by computing the similarity between the stretching region and its adjacent non-stretching region.

%\subsubsection{\bf{Edge based metric}}
%Several FR use the edge/gradient change between the synthesized and reference images to evaluate the quality of synthesized images. At the same time, in the DIBR-synthesized videos, the gradient variation may result in temporal flickering which is one of the main distortions in DIBR-synthesized videos. 

In \cite{wang2019blind}, Wang \textsl{et al.} also proposed a NR metric (GDIC) to measure the geometric distortions and image complexity. Firstly, different from the wavelet transform based metrics introduced above, this GDIC metric uses the edge map of wavelet sub-bands to obtain the shape of geometric distortions. Then, the geometric distortion is measured {in terms of} edge similarity between the wavelet low-level and high-level sub-bands \cite{cohen1992biorthogonal}. Besides, the image complexity is also an important factor in human visual perception. In order to evaluate the image complexity of the DIBR-synthesized images, hybrid filter \cite{gu2017no, gu2016no}, which combines the Autoregressive (AR) and bilateral (BL), is used. The final image quality score is computed by normalizing the geometric distortion with image complexity. Furthermore, in \cite{wang2019blindtip}, this metric is extended to achieve higher performance by adding a log-energy based sharpness detection module. 

\subsubsection{{Flicker region based video NR metrics}}
In DIBR-synthesized videos, temporal flicker is one of the most annoying distortions. Extracting the flicker regions may help to evaluate the quality of DIBR-synthesized videos. 

In \cite{kim2016measurement}, Kim \textsl{et al.} also proposed a NR metric (CTI) to measure the temporal inconsistency and flicker regions in the DIBR-synthesized video. First, the flicker regions are detected from the difference between motion-compensated consecutive frames. Then, the structural similarity between consecutive frames are calculated on the flicker regions to measure the structural distortions in each frame. At the same time, the number of pixels in the flicker regions is used to weight the distortion of each frame. The final quality score is obtained as the weighted sum of the quality scores of all the frames in the DIBR-synthesized video.

In \cite{zhou2018no}, Zhou \textsl{et al.} proposed a NR metric FDI to measure the temporal flickering distortion in the DIBR-synthesized videos. Firstly, the gradient variations between each frame are used to extract the potential flickering regions. Followed by a refinement to precisely obtain the flickering regions through calculating the correlation between the candidate flickering regions and their neighbours. Then, the flickering distortion is estimated in SVD domain from the difference between the singular vectors of the flickering block and their associated block in the previous frame. The final video quality is computed as the average quality of all the frames.

\subsubsection{{Natural Scene Statistics based NR metrics}}
Natural Scene Statistics (NSS) based approaches, which assume that the natural images contain certain statistics and these statistics may be changed by different distortions, have achieved great success in the quality assessment of traditional 2D images \cite{saad2012blind, saad2011dct, moorthy2011blind, LIU2014494}. Due to the big difference between the DIBR view synthesis distortions and the traditional 2D ones, these NSS based metrics do not work well on the quality assessment of DIBR-synthesized views. Recently, several efforts have been made to fix this gap.

As introduced in the previous Edge/Contour based FR metrics part, the edge image is significantly degraded by structural and geometric distortions in DIBR-synthesized images, and the edge based FR metrics have shown their superiority. With this {consideration}, Zhou \textsl{et al.} proposed a NR metric (SET) for DIBR-synthesized images via edge statistics and texture naturalness based on Difference-of-Gaussian (DoG) in \cite{zhou2019no}. The orientation selective statistics (similar to the metric in \cite{moorthy2011blind}) are extracted from DoG images {at different scales} while the texture naturalness features are obtained based on the Gray level Gradient Co-occurrence Matrix (GGCM) \cite{Li2015blind} which represents the joint distribution relation of pixel gray level and edge gradient. A Random Forest (RF) regression model is finally trained based on these two groups of features to predict the quality of DIBR-synthesized images.

Gu \textsl{et al.} proposed a self-similarity and main structure consistency based Multiscale Natural Scene Statistics (MNSS) in \cite{gu2019multiscale}. The multiscale analysis on the DIBR-synthesized image and its associated reference image indicates that the distance (SSIM value \cite{Wangssim}) between the synthesized and the reference image decreases significantly when the scale reduces. It is assumed that the synthesized image at a higher scale holds a better quality, which means the {higher-scale} images can be approximately used as reference{s}. Thus, the similarity between the lower scale image (first scale is used in this metric) and the higher scale images (self similarity) are used to measure the quality of DIBR-synthesized image. Besides, in the main structure, the authors use 300 natural images from the Berkeley segmentation dataset \cite{Martin2001adatabase} to obtain the general statistical regularity of main structure in natural images. The similarity between the main structure map of the synthesized image and the obtained prior NSS vector is calculated to evaluate the structure degradation of the DIBR-synthesized image. Finally, the statistical regularity of main structure and the structure degradation are combined to get the overall quality score. 

Shao \textsl{et al.} propose a NR metric (HEVSQP) for DIBR-synthesized videos based on color-depth interactions in \cite{shao2017no}. Firstly, the video sequence is divided into Group of Frames (GoF). Through an analysis of color-depth interactions, more than 90 features from both texture and depth videos, including gradient magnitude, asymmetric generalized Gaussian distribution (AGGD) \cite{saad2011dct}, local binary pattern (LBP), are extracted. Then, a principal component analysis (PCA) is applied to reduce the feature dimension. Then, two dictionaries, color dictionary and depth dictionary, are learned to establish the relationship between the features and video quality. The final quality score is pooled from the color and depth quality.

In \cite{ling2018learn}, Ling \textsl{et al.} proposed a NR learning based metric for DIBR-synthesized views, which focuses on the non-uniform distortions. Firstly, a set of convolutional kernels are learned by using the improved fast convolutional sparse coding (CSC) algorithms. Then, the convolutional sparse coding (CSC) based features of the DIBR-synthesized images are extracted, from which the final quality score is obtained via support vector regression (SVR).

Although the NSS models have made great progress for the NR IQA, the hand-craft features may not be sufficient to represent complex image textures and artefacts, there {is} still a large gap between objective quality measurement and human perception \cite{yang2019survey}. 

\subsubsection{{Deep feature based NR metrics}}
The deep learning techniques, especially the Convolutional Neural Networks (CNN), have shown their great advantages in various computer vision tasks \cite{ye2020drm, lei2020face}. They make it possible to directly learn the representative features from image \cite{kim2017deep, zhuang2019recognition}. Unfortunately, {due} to the limitation of {the number of images in the} DIBR-synthesized view datasets, there is not enough data to train the deep model{s} straightforwardly. However, it is shown in the recent published literature that the deep neural network models trained on large-scale datasets, \textsl{e.g.} ImageNet \cite{ILSVRC15}, can be used to extract effective representative features of human perception. 
%Thus, the following two methods were proposed recently to evaluate the quality of DIBR-synthesized view images using pre-trained deep features. 

In \cite{wang2019deepicip}, Wang \textsl{et al.} proposed a NR metric SIQA-CFP which uses the ResNet-50 \cite{he2016resnet} model pre-trained on ImageNet to extract multi-level features of DIBR-synthesized images. Then, a contextual multi-level feature pooling strategy is designed to encode the high-level and low-level features, and finally to get the quality scores. 

%Differently, the metric GANs-NRM proposed by Ling \textsl{et al.} in \cite{ling2019gannrm} is based on a Generative Adversarial Networks (GANs) \cite{NIPS2014_5423}. 
As introduced in Section {1}, various distortions may be introduced during the dis-occlusion region filling stage. Meanwhile, in current literature, several Generative Adversarial Networks (GAN) \cite{goodfellow2014generative} based models have been proposed for image in-painting. As the generator is trained to in-paint the missing part, the discriminator is supposed to have the capability to capture the perceptual information which reflects the in-painted image quality. Based on this assumption, Ling \textsl{et al.} proposed a GAN based NR metric (GANs-NRM) \cite{ling2019gannrm} for DIBR-synthesized images. In GANs-NRM, a generative adversial network for image in-painting is firstly trained on two large-scale datasets (PASCAL \cite{everingham2015pascal} and Places \cite{zhou2018places}). Then, the features extracted from the pre-trained discriminator are used to learn a Bag-of-Distortion-Word (BDW) codebook. A Support Vector Regression (SVR) is trained on the encoded information of each image to predict the final quality of DIBR-synthesized images. Instead of simply using the general models trained for other tasks, \textsl{e.g.} object detection, this metric is more targeted, and it also proposes a new way to obtain the semantic features for image quality assessment.

\subsection{{Summary}}
In this section, 19 FR, 3 RR, 4 SV-FR and 15 NR DIBR quality metrics have been reviewed and categorized based on their used approaches and on the amount of reference information used. As shown in Table~\ref{tab:metrics}, most of the metrics consist of multiple parts. {It} is thus difficult to classify them into a single specific category thoroughly, {that is why} we just classify them into the most related one instead.
Besides, there are also some other ways to do the classification. For example, if we focus on the image structural representation used in these metrics, they can be classified into low-level \cite{liu2015subjective}), mid-level \cite{ling2017icme, ling2017image} and high-level \cite{ling2018learn, wang2019deepicip, ling2019gannrm} metrics. As introduced in \cite{manassi2013crowding}, the low-level representations indicate the pixel level edges or contours; the mid-level representations mean the shapes and texture information; the high-level representations refer to the complex features \textsl{e.g.} objects, unnatural structures.
Besides, there are also some hierarchical metrics which combine the above features, such as the LMS metric proposed in \cite{zhou2019quality} which uses both low-level and mid-level features \cite{ling2017icme} and the metric in \cite{ling2019perceptual} which integrates the features on each level. 
%In this section, we surveyed the existing state-of-the-art quality assessment metrics for DIBR-synthesized views. In the next section, we'll provide the experimental results and some discussions.

\begin{table*}%[!htbp]
\scriptsize
\centering
\caption{Performance of the DIBR dedicated metrics on DIBR-synthesized image dataset. }
\label{tab:plcc_ivc}
\begin{threeparttable}
%\begin{center}
\begin{tabular}{|c|c|c|c|c|c|c|c|c|c|c|c|c|c|}\toprule
\hline
\multicolumn{2}{|c|}{\multirow{2}{*}{Metric}} & \multicolumn{3}{c|}{IVC image dataset} & \multicolumn{3}{c|}{IETR image dataset}& \multicolumn{3}{c|}{MCL 3D image dataset}& \multicolumn{3}{c|}{IVY dataset}\\ \cline{3-14}
\multicolumn{2}{|c|}{}  & PLCC & RMSE & SROCC & PLCC & RMSE & SROCC & PLCC & RMSE & SROCC & PLCC & RMSE & SROCC \\ \hline
\multirow{2}{*}{FR 2D}&PSNR & 0.4557 & 0.5927 & 0.4417 & 0.6012 & 0.1985 & 0.5356& {0.7852} & {1.6112} & {0.7915} & \bf{0.6311} & \bf{19.1227} & \bf{0.6668} \\ \cline{2-14}
& SSIM \cite{Wangssim} & 0.4348 & 0.5996 & 0.4004 & 0.4016 & 0.2275 & 0.2395 & 0.7331 & 1.7693 & 0.7470 & 0.3786 & 22.8172 & 0.3742 \\ \hline
\multirow{2}{*}{NR 2D}&BIQI \cite{moorthy2009modular}& 0.5150 & 0.5708 & 0.3248 & 0.4427 & 0.2223 & 0.4321& 0.3347 & 2.4516 & 0.3696 & 0.5686 & 20.2791 & 5754 \\ \cline{2-14}
& BLIINDS2 \cite{saad2012blind} & 0.5709 & 0.5467 & 0.4702 & 0.2020 & 0.2428 & 0.1458 & 0.6338 & 2.0124 & 0.5893 & 0.3508 & 23.0855 & 0.2569 \\ \hline
\multirow{9}{*}{\makecell*[c]{FR \\ DIBR}}& Bosc \cite{bosc2012edge} & 0.5841 & 0.5408 & 0.4903 & --- &  --- & --- & 0.4536 & 2.2980 & 0.4330 & --- &  --- & --- \\ \cline{2-14}
&3DSwIM \cite{battisti2015objective}& 0.6864 & 0.4842 & 0.6125 & --- &  --- & --- & 0.6519 & 1.9729 & 0.5683 & --- &  --- & --- \\ \cline{2-14}
&VSQA \cite{conze2012objective}& 0.6122 & 0.5265 & 0.6032 & 0.5576 & 0.2062 & {0.4719} & 0.5078 & 2.9175 & 0.5120 &   ---& --- & --- \\ \cline{2-14}
%&CT-IQA  & 0.6809 & 0.4877 & --- & --- &  --- & ---&   &   &  &   &   &  \\ \cline{2-14}
&ST-SIAQ \cite{ling2017icme} & 0.6914 & 0.4812 & 0.6746 & 0.3345 & 0.2336 & 0.4232 &  0.7133 & 1.8233 & 0.7034  & --- & --- & --- \\ \cline{2-14}
&EM-IQA \cite{ling2017image} & 0.7430 & 0.4455 & 0.6282 & 0.5627 & 0.2020 & 0.5670 &--- &  --- & --- & --- &  --- & --- \\ \cline{2-14}
&MP-PSNR \cite{sandic2016multi} & 0.6729 & 0.4925 & 0.6272 & {0.5753} & {0.2032} & {0.5507}  & 0.7831 & 1.6179 & 0.7899 & {0.5947} & {19.8182} & {0.5707} \\ \cline{2-14}
&MW-PSNR \cite{sandic2015dibrw}& 0.6200 & 0.5224 & 0.5739 & 0.5301 & 0.2106 & {0.4845} & 0.7654 & 1.6743 & 0.7721 & 0.5373 & 20.7910 & 0.5051 \\ \cline{2-14}
&SCDM \cite{tian2018dibrfr} & 0.8242 & 0.3771 & 0.7889 & \bf{0.6685} & \bf{0.1844} & \bf{0.5903} & 0.7166 & 1.8141 & 0.7197 &   --- & --- & --- \\ \cline{2-14}
%&SSIM'(SCDM) & 0.5681 & 0.5479 & 0.5475 &   &   &  & 0.6000 & 2.0814 & 0.5451 &   &   &  \\ \cline{2-14}
&{{SC-IQA} \cite{tian2018sc}} & \bf{0.8496} & \bf{0.3511} & {0.7640} & \bf{0.6856} & \bf{0.1805} & \bf{0.6423} & \bf{0.8194} & \bf{1.4913} & \bf{0.8247} & 0.4326 & 22.2256 & 0.3135 \\ \cline{2-14}
&{Wang \cite{wang2019qualityaccess}} & \bf{0.8512} & \bf{0.3146} & \bf{0.8346} & {0.6118} & {0.1961} & {0.6136} & \bf{0.7910} & \bf{1.5917} & \bf{0.7929} & --- & --- & --- \\ \cline{2-14}
&CBA \cite{jung2016critical} & --- & --- & --- & --- & --- & --- & --- & --- & --- & \bf{0.826} & \bf{8.181} & \bf{0.829} \\
\hline
{RR}&MP-PSNRr \cite{sandic2016dibr} & 0.6954 & 0.4784 & 0.6606 & {0.6061} & {0.1976} & {0.5873} & 0.7740 & 1.6474 & 0.7802 & 0.5384 & 20.7733 & 0.5454 \\ \cline{2-14}
%&RRLP  & 0.7523 & 0.4386 & -- &   &   &  &   &   &  &   &   &  \\ \cline{2-14}
DIBR & MW-PSNRr \cite{sandic2016dibr} & 0.6625 & 0.4987 & 0.6232 & 0.5403 & 0.2090 & {0.4946} & 0.7579 & 1.7012 & 0.7665  & 0.5304 & 20.8993 & 0.5138 \\ \hline
\multirow{3}{*}{\makecell*[c]{SV-FR\\ 
                       DIBR}}&{SIQE \cite{farid2015objective}}& 0.7650 & 0.5382 & 0.4492 & 0.3144 & 0.2353 & 0.3418 & 0.6734 & 1.9233 & 0.6976 & --- & --- & --- \\ \cline{2-14}
&{LOGS \cite{li2018quality}}& \bf{0.8256} & \bf{0.3601} & {0.7812} & \bf{0.6687} & \bf{0.1845} & \bf{0.6683} & 0.7614 & 1.6873 & 0.7579 & \bf{0.6442} & \bf{18.8553} & \bf{0.6385} \\ \cline{2-14}
%&{SIQM}& --- & --- & --- &  --- & --- & --- & 0.7744 & 1.6461 & 0.7756 & --- & --- & ---\\ \cline{2-14}
& {DSQM \cite{farid2017perceptual}}& 0.7430 & 0.4455 & 0.7067 & 0.2977 & 0.2367 & 0.2369 & 0.6995 & 1.8593 & 0.6980 &   --- & --- & --- \\ \hline
\multirow{8}{*}{\makecell*[c]{NR \\ DIBR}}&{APT \cite{gu2017model}}& 0.7307 & 0.4546 & 0.7157 & 0.4225 & 0.2252 & 0.4187  & 0.6433 & 1.9870 & 0.6200 & 0.5156 & 21.1239 & 0.4754 \\ \cline{2-14}
&{OUT \cite{jakhetiya2018highly}}& 0.7243 & 0.4591 & 0.7010 & 0.2007 & 0.2429 & 0.1924 & 0.4208 & 2.3601 & 0.3171 & 0.2525 & 23.8530 & 0.2409 \\ \cline{2-14}
&{MNSS \cite{gu2019multiscale}}& 0.7700 & 0.4120 & 0.7850 & 0.3387 & 0.2333 & 0.2281 & 0.3766 & 2.4101 & 0.3531 &   0.3834 & 22.7681 & 0.2282 \\ \cline{2-14}
&{NR\_MWT \cite{sandic2019fast}}& 0.7343 & 0.4520 & 0.5169 & 0.4769 & 0.2179 & 0.4567 & 0.1373 &   2.5771 & 0.0110 & 0.4848 & 21.5614 & 0.4558 \\ \cline{2-14}
&{NIQSV \cite{shishun2017}}& 0.6346 & 0.5146 & 0.6167 & 0.1759 & 0.2446& 0.1473  & 0.6460 & 1.9820 & 0.5792 &   0.4113 & 22.4706 & 0.2717 \\ \cline{2-14}
&{NIQSV+ \cite{tian2018niqsv+}}& 0.7114 & 0.4679 & 0.6668 & 0.2095 & 0.2429 & 0.2190 & 0.6138 & 2.0375 & 0.6213 & 0.2823 & 23.6491 & 0.3823 \\ \cline{2-14}
&{SET \cite{zhou2019no}}& \bf{0.8586} & \bf{0.3015} & \bf{0.8109} & --- & --- & --- & \bf{0.9117} & \bf{1.0631} & \bf{0.9108} &  --- & --- & ---\\ \cline{2-14}
&{GANs-NRM \cite{ling2019gannrm}}& {0.826} & {0.386} & \bf{0.807} & 0.646 & 0.198 & 0.571 & --- & --- & --- &  --- & --- & --- \\ \hline
%\cline{2-14}
%&{CSC-NRM}& \bf{0.8302} & \bf{0.3233} & \bf{0.7827} & --- & --- & --- & --- & --- & --- &   --- & --- & --- \\ 
%\hline
\end{tabular}
%\end{center}
\begin{tablenotes}
\footnotesize
\item[``---''] : Due to the unavailability of source code or reference resources \textsl{e.g.} depth map and side view reference image, we just use the reported results in their corresponding publications instead, their associated results on other datasets are marked by the symbol ``---'' in the table.
\end{tablenotes}
\end{threeparttable}
\end{table*}

\section{Experimental results and discussions}
%In this section, the main methodologies to evaluate the performance of objective quality metrics will be introduced and 
In this section, the performance{s} of different objective quality assessment metrics are presented and analysed. Besides, some potential challenges and possible directions for future work will be discussed.

\subsection{Performance evaluation methodologies}
The subjective test results can be recognized as the ground truth visual quality since the human observer is the ultimate receiver of image/video content. The accuracy of an objective quality metric can be evaluated based on its consistencies with the subjective quality scores. In this part, we will introduce the Video Quality Expert Group (VQEG) \cite{vqeg} recommended correlation based methods and the recently proposed Krasula' model \cite{krasula2016accuracy} in detail.
\subsubsection{\bf{Correlation coefficients based methods}}
The reliability of objective metrics can be evaluated through their correlation with subjective test scores. Three widely used criteria, Pearson Linear Correlation Coefficients (PLCC) and Root-Mean-Square-Error (RMSE) and Spearman Rank-Order Correlation Coefficients (SROCC), are recommended by VQEG to evaluate the prediction accuracy, prediction monotonicity and prediction consistency of the objective metrics respectively, which are defined as follows:
%with respect to three aspects of their ability to estimate subjective assessment of video quality: prediction accuracy, prediction monotonicity and prediction consistency. 
%The prediction accuracy is the ability of the model to predict the observers' ratings (subjective scores) with a minimum error on average. 

\begin{equation}
PLCC(X,Y) = \frac{\sum_{i=1}^n(X_i - \bar{X})(Y_i - \bar{Y})}{\sqrt{\sum_{i=1}^n(X_i - \bar{Y})^2}\sqrt{\sum_{i=1}^n(Y_i - \bar{Y})^2}}
\end{equation}
\begin{equation}
RMSE(X,Y) = \sqrt{\frac{1}{m}\sum_{i=1}^m(X_i - Y_i)^2}
\end{equation}
\begin{equation}
SROCC(X,Y) = 1- \frac{6 \sum{d_i^2}}{n(n^2 - 1)}
\end{equation}
where $d_i$ indicates the difference of ranking of $X$ and $Y$. Higher PLCC and SROCC value{s} indicate higher accuracy and better monotonicity respectively. On the contrary, a higher RMSE value refers to a lower prediction accuracy. 

\begin{figure}%[!htbp]
  \centering
  \begin{minipage}[b]{0.48\linewidth}
  \centering
  \includegraphics[width=1\linewidth]{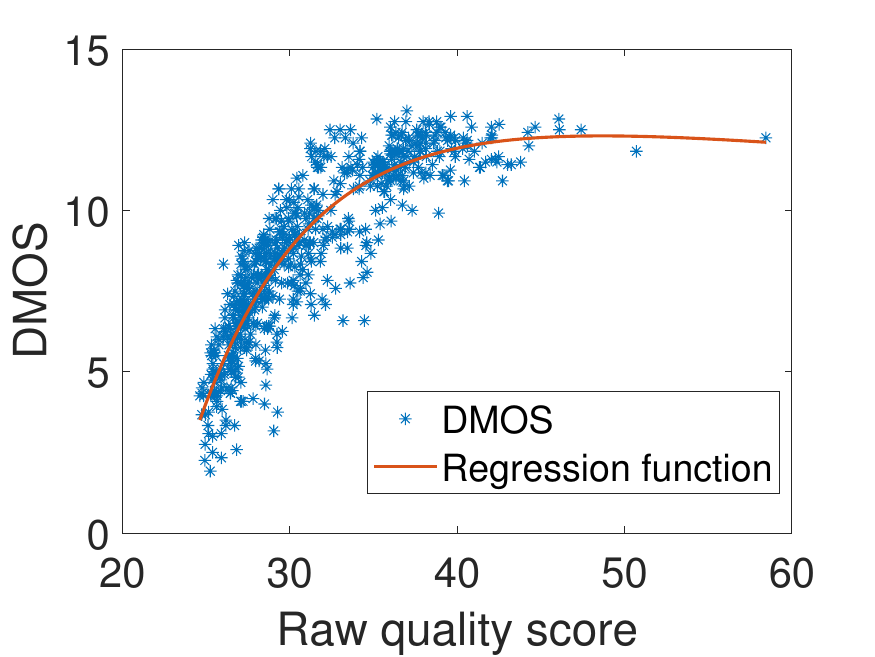}
  \centerline{(a) Before regression}\medskip
  \end{minipage}
  \begin{minipage}[b]{0.48\linewidth}
  \centering
  \includegraphics[width=1\linewidth]{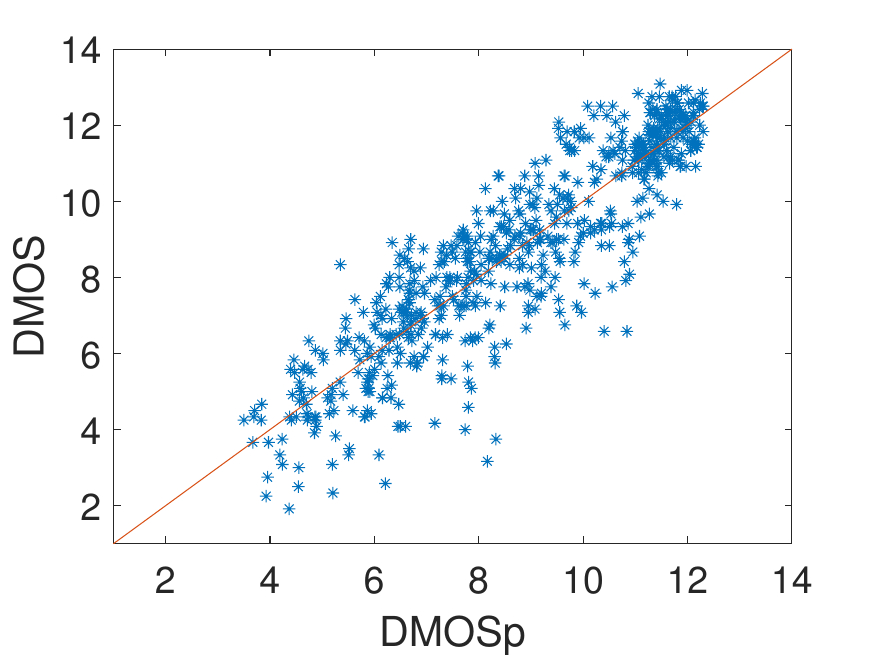}
  \centerline{(b) After regression}\medskip
  \end{minipage}
  \caption{Example relationship between DMOS and objective quality scores. This figure is from \cite{tian2019image}.}
  \label{Fig: fitting}
\end{figure}

Before computing these three criteria, the objective scores are recommended by VQEG to be mapped to the predicted subjective score $DMOS_p$ to remove the nonlinearties due to the subjective rating processing and to facilitate comparison of the metrics in a common analysis space \cite{vqeg}. The nonlinear function for regression mapping is shown as follows:
\begin{equation}
DMOS_p = \beta_1 (0.5-\frac{1}{(1+e^{(\beta_2 (s-\beta_3))})})+\beta_4 s+\beta_5
\label{eq:regression}
\end{equation}
where $s$ is the score obtained by the objective metric and $\beta_1, \beta_2, \beta_3, \beta_4, \beta_5$ are the parameters of these regression functions.
They are obtained through regression to minimize the difference between $DMOS_p$ and $DMOS$. As shown in Fig.~\ref{Fig: fitting}, the nonlinearity has been removed after the regression.

\begin{figure}%[!htbp]
  \centering
  \begin{minipage}[b]{\linewidth}
  \centering
  \centerline{\includegraphics[width=1\linewidth]{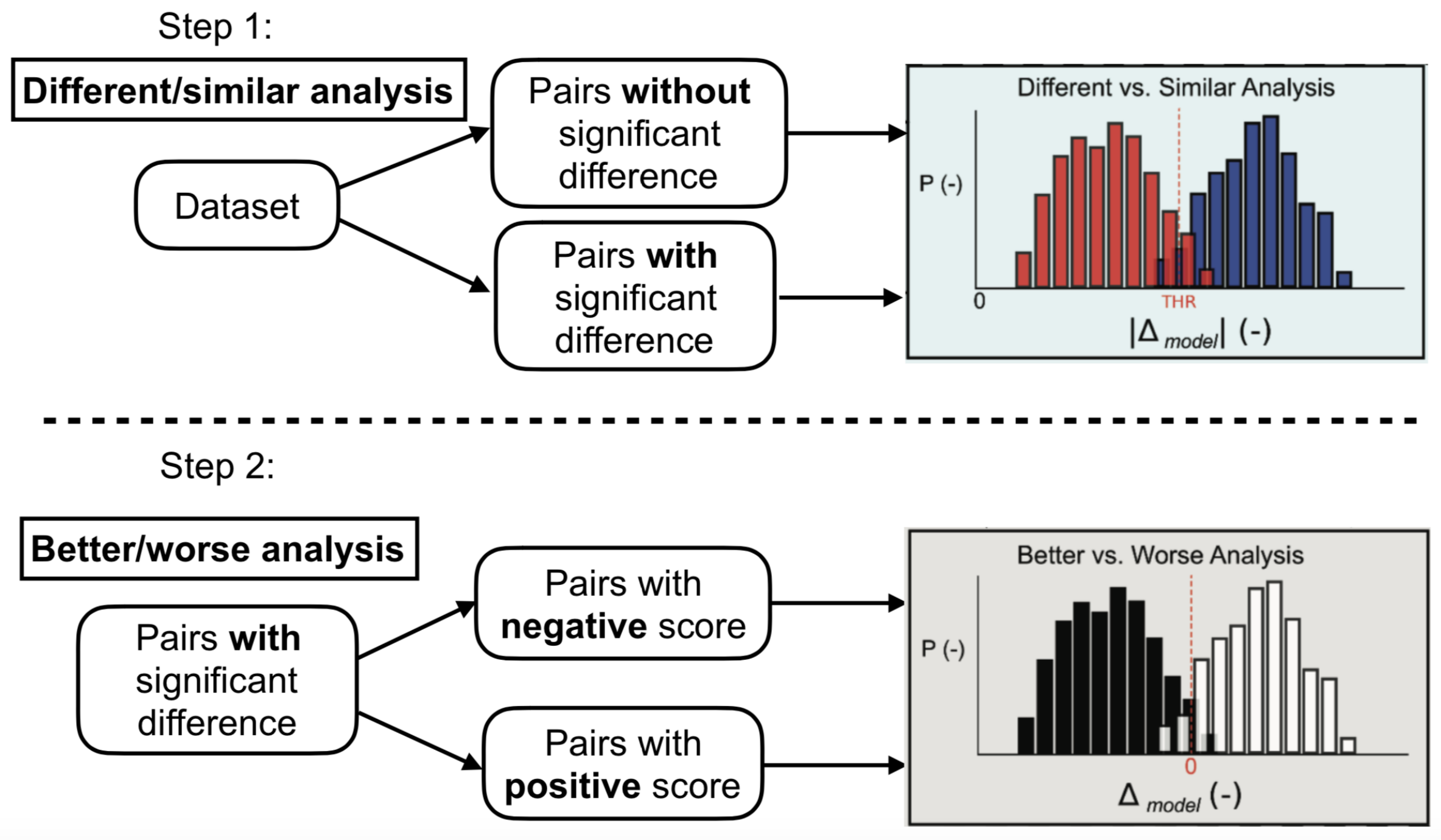}}
%  \vspace{1.5cm}
%  \centerline{(a) different / similar}\medskip
  \end{minipage}
  \caption{Krasula's model for performance evaluation of objective quality metrics \cite{krasula2016accuracy}.}
\label{fig:Krasula}
\end{figure}

\subsubsection{\bf{Analysis of Krasula's model}}
The above methods compare the performance of each metric by calculating their correlations with the subjective results. However they only consider the mean value of subjective scores, the uncertainty of the subjective scores are ignored. In addition, the quality scores need to be regressed by a regression function \textsl{cf.} Eq.~\ref{eq:regression}, that is not the way they are exactly used in real scenarios. Thus, we further conduct a statistical test proposed by Krasula \textsl{et al.} in \cite{krasula2016accuracy} which does not suffer from the drawbacks of the above methods. The performances of objective metrics are evaluated by their classification abilities.

As shown in Fig. \ref{fig:Krasula}, firstly, the tested image pairs in the dataset are divided into two groups: different and similar according to their subjective scores. The cumulative distribution function ({CDF}) of the normal distribution is used to calculate the probability of image pairs. Then, we consider the pairs with {a probabilty} higher than the selected significance level 0.95 to be significantly different. The others will be recognized as similar.

There are two performance analyses. The first performance analysis is conducted by \st{by} evaluating how well the objective metric succeeds to distinguish significantly different image pairs from {non-significantly} different video pairs, in a consistent way with subjective evaluation of significant difference.
%how well the objective metric can distinguish the video pair are significant different or not. 
The second analysis determines whether the objective metric can correctly identify the image of higher quality in the pair. 

Compared to simply calculating the correlation coefficients, this model considers not only the mean value of subjective scores, but also their uncertainties. Besides, since no regression is used, this model less depends on the quality ranges of different datasets. Another advantage of Krasula's model is that it can easily combine the data from multiple datasets and evaluate a comprehensive performance on multiple datasets instead of simply averaging the results on different datasets. 

\subsection{Performance on DIBR image datasets}
\subsubsection{\bf{Results of PLCC, RMSE and SROCC}}
The obtained PLCC, RMSE and SROCC values of the objective image quality assessment metrics on the DIBR-synthesized image datasets are given in Table~\ref{tab:plcc_ivc}, in which four 2D metrics \cite{moorthy2009modular, saad2012blind, Wangssim} and 24 DIBR metrics are tested. The best three performances among the blind IQA methods are shown in bold. 
% we can easily observe that none of the tested metrics can achieve a satisfactory performance on these two datasets since the highest PLCC SROCC values can only achieve 0.85 and 0.81 respectively. Among these metrics, 
We can easily observe that the DIBR-synthesized view dedicated metrics significantly outperform the traditional 2D metrics on the IVC and IETR image datasets which focus on the DIBR view synthesis distortions. In other words, the metrics initially designed for traditional 2D image distortions can not well evaluate the DIBR view synthesis distortions.

The shift compensation based FR and SV-FR metrics obtain great improvement compared to the original 2D FR metrics, {\textsl{e.g.}} the SC-IQA compared to PSNR. One main reason is that the global object shift existing in the DIBR-synthesized images may not be perceived by human observers but can be easily detected by the original 2D pixel-based FR metrics. {Thus} this shift distortions are often overestimated by the 2D pixel-based FR metrics. 

If we focus on the wavelet transform-based metrics (NR\_MWT and MW-PSNR), the NR metric (NR\_MWT) perform{s} better than the FR metric (MW-PSNR) on the IVC dataset. It is surprising that the FR metric performs even worse than the NR metric since these metrics use similar features and {the} FR metric has access to the ground truth. While on the IETR dataset, the NR metric perform{s} worse than the FR metric\st{s}. The main reason probably {lies in} the global shift distortion in the IVC image dataset.

To further explore the object shift effect, we have made an additional experiment on the IVC dataset while excluding the A1 view synthesis algorithm \cite{telea2004image} which causes great object shift in the synthesized views. The A1 algorithm fills the black holes in the dis-occlusion regions by simply stretching the adjacent texture which may cause great global object shift in the synthesized views. The results are shown in Table~\ref{tab:IVC_exclude_A1}. We can observe that the performance of FR and RR metrics increase significantly when large global shift artefacts are excluded.

\begin{table}%[!htbp]
\caption{Performance on the IVC DIBR image dataset excluding A1 algorithm.}
\label{tab:IVC_exclude_A1}
\begin{center}
\begin{tabular}{|c|c|c|c|c|}
\hline
\multicolumn{2}{|c|}{Metric}  & PLCC & RMSE & SROCC \\ \hline
\multirow{2}{*}{FR 2D }&PSNR & 0.7519 & 0.4525 & 0.6766 \\ \cline{2-5}
&SSIM & 0.5956 & 0.5513 & 0.4424 \\ \hline
FR DIBR & MW-PSNR & 0.8545 & 0.3565 & 0.7750 \\ \hline
RR DIBR & MW-PSNRr & 0.8855 & 0.3188 & 0.8298 \\ \hline
%\multirow{2}{*}{FR DIBR image metrics}
%&MP-PSNR & 0.8549 & 0.3561 & 0.7759 \\ \cline{2-5}
%&MW-PSNR & 0.8545 & 0.3565 & 0.7750 \\ \hline
%\multirow{2}{*}{RR DIBR image metrics}&MP-PSNRr & 0.9039 & 0.2936 & 0.8634 \\ \cline{2-5}
%&MW-PSNRr & 0.8855 & 0.3188 & 0.8298 \\ \hline

\end{tabular}
\end{center}
\end{table}

The edge/contour based metrics also perform much better than the 2D pixel-based FR metrics since the edge/contour features can better represent the geometric degradations in the DIBR-synthesized images compared to simple pixel information.

The NR metrics do not need any reference information to evaluate the image quality, {thus} the global shift does not have effect on the NR metrics. Besides, since the real reference images at virtual viewpoints are not always available in real applications, the NR metrics are more practical and useful. From Table~\ref{tab:plcc_ivc}, we can easily find that the performance{s} of the DIBR-synthesized view dedicated metrics decrease greatly {on the} IETR dataset compared to {theirs on the} IVC dataset. Among these metrics, the NR ones decrease the most, especially the learning based NR metrics. This is because of the fact that these NR metrics {are designed for} the distortions in the IVC dataset{. However,} in the IETR dataset, many ``old fashioned'' distortions are excluded. 

\begin{figure*}%[!htbp]
  \centering
  \begin{minipage}[b]{0.48\linewidth}
  \centering
  \includegraphics[width=1\linewidth]{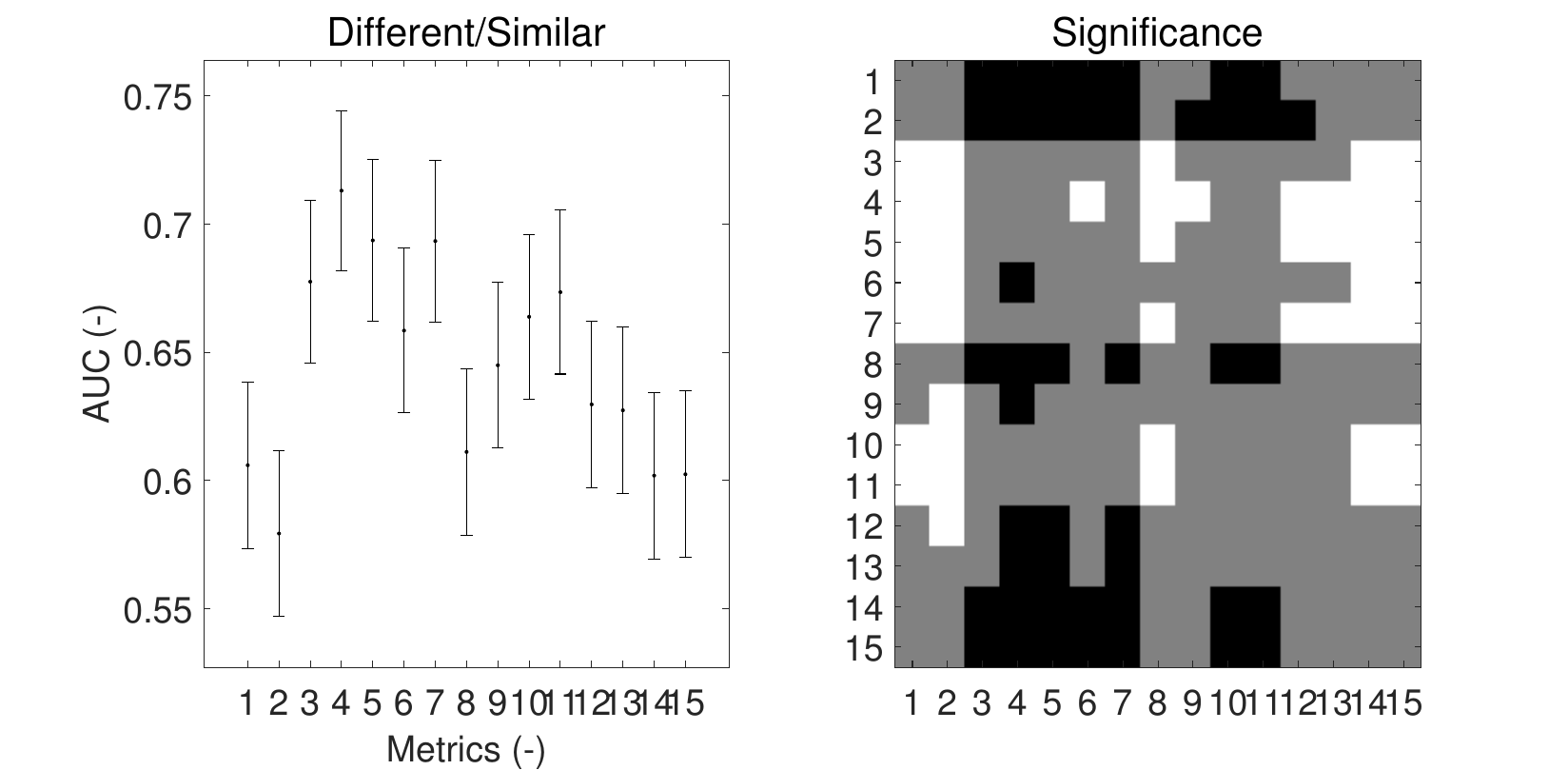}
  \centerline{(a) Different / Similar analysis on IVC image dataset}\medskip
  \end{minipage}
  \begin{minipage}[b]{0.48\linewidth}
  \centering
  \includegraphics[width=1\linewidth]{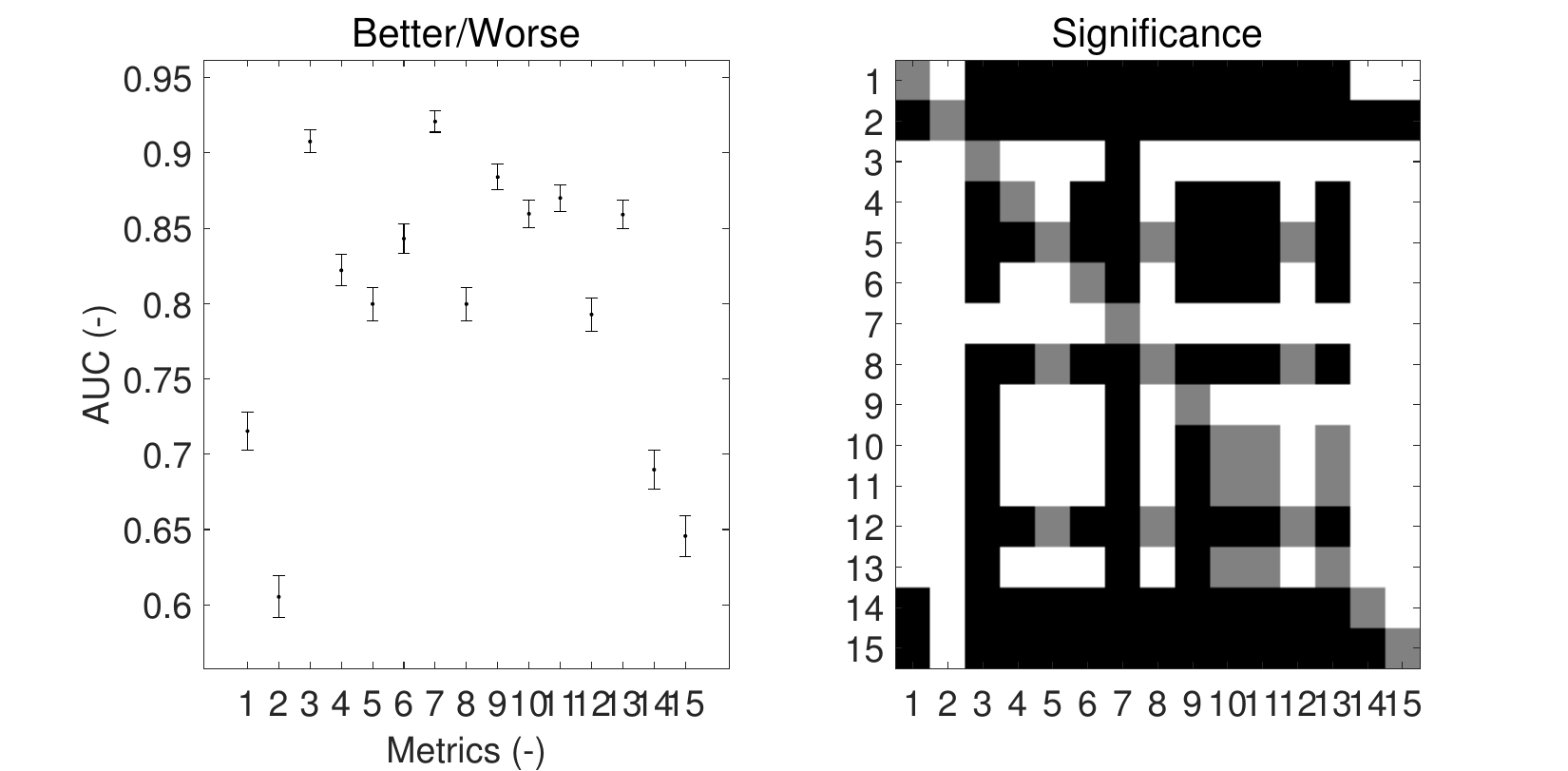}
  \centerline{(b) Better / Worse analysis on IVC image dataset}\medskip
  \end{minipage}
  
  \centering
  \begin{minipage}[b]{0.48\linewidth}
  \centering
  \includegraphics[width=1\linewidth]{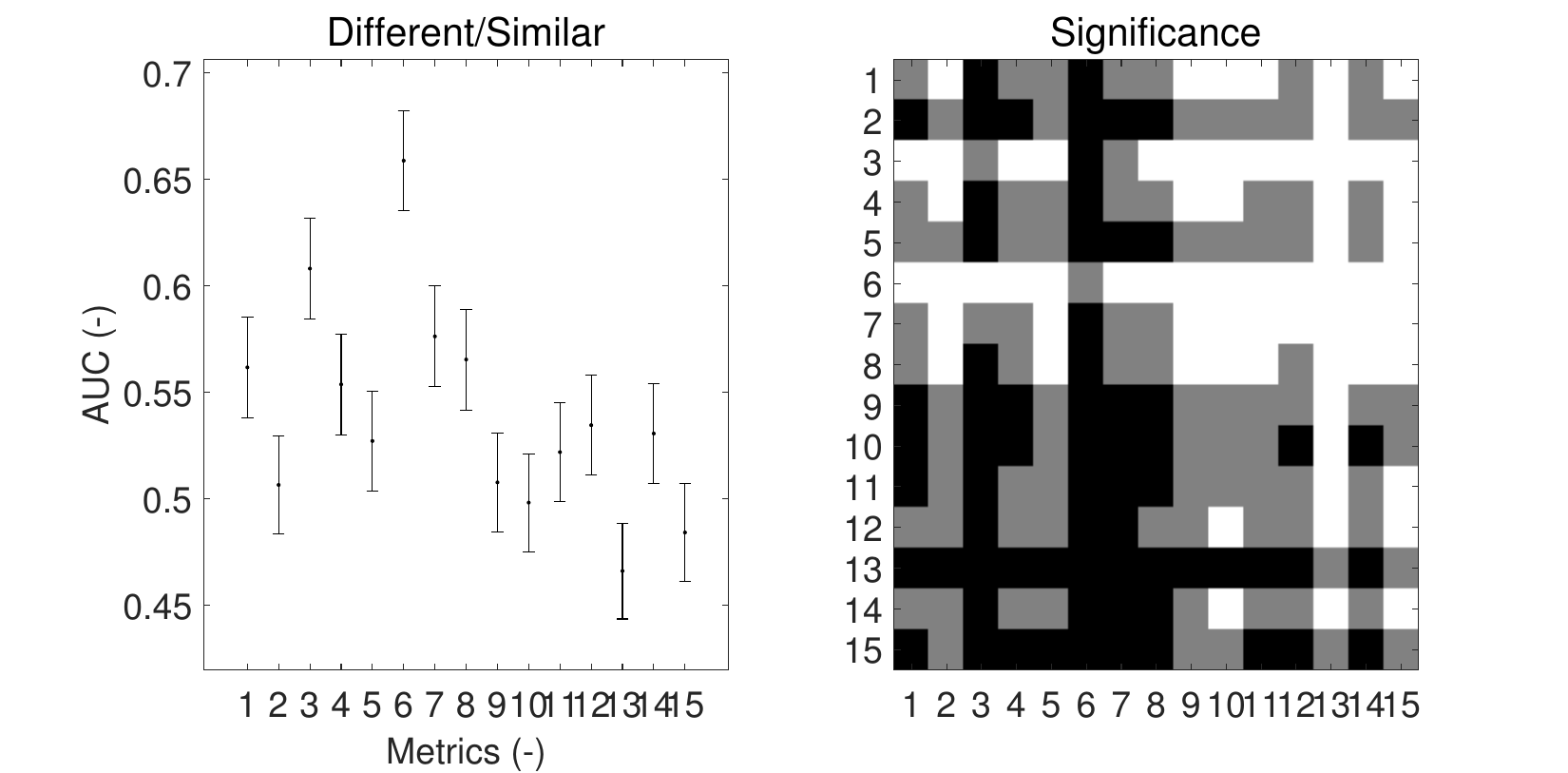}
  \centerline{(c) Different / Similar analysis on IETR image dataset}\medskip
  \end{minipage}
  \begin{minipage}[b]{0.48\linewidth}
  \centering
  \includegraphics[width=1\linewidth]{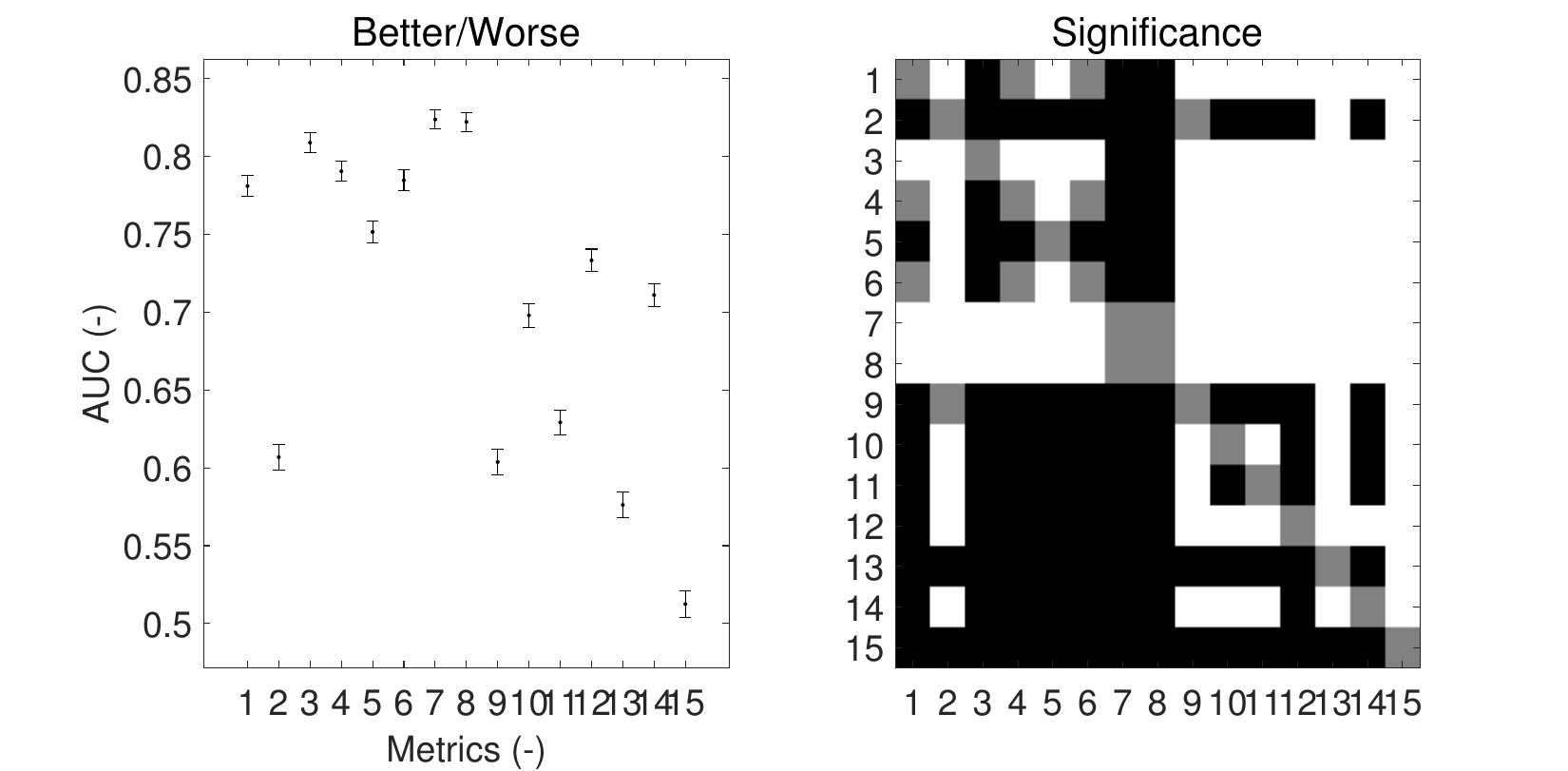}
  \centerline{(d) Better / Worse analysis on IETR image datasets}\medskip
  \end{minipage}
  
  \centering
  \begin{minipage}[b]{0.48\linewidth}
  \centering
  \includegraphics[width=1\linewidth]{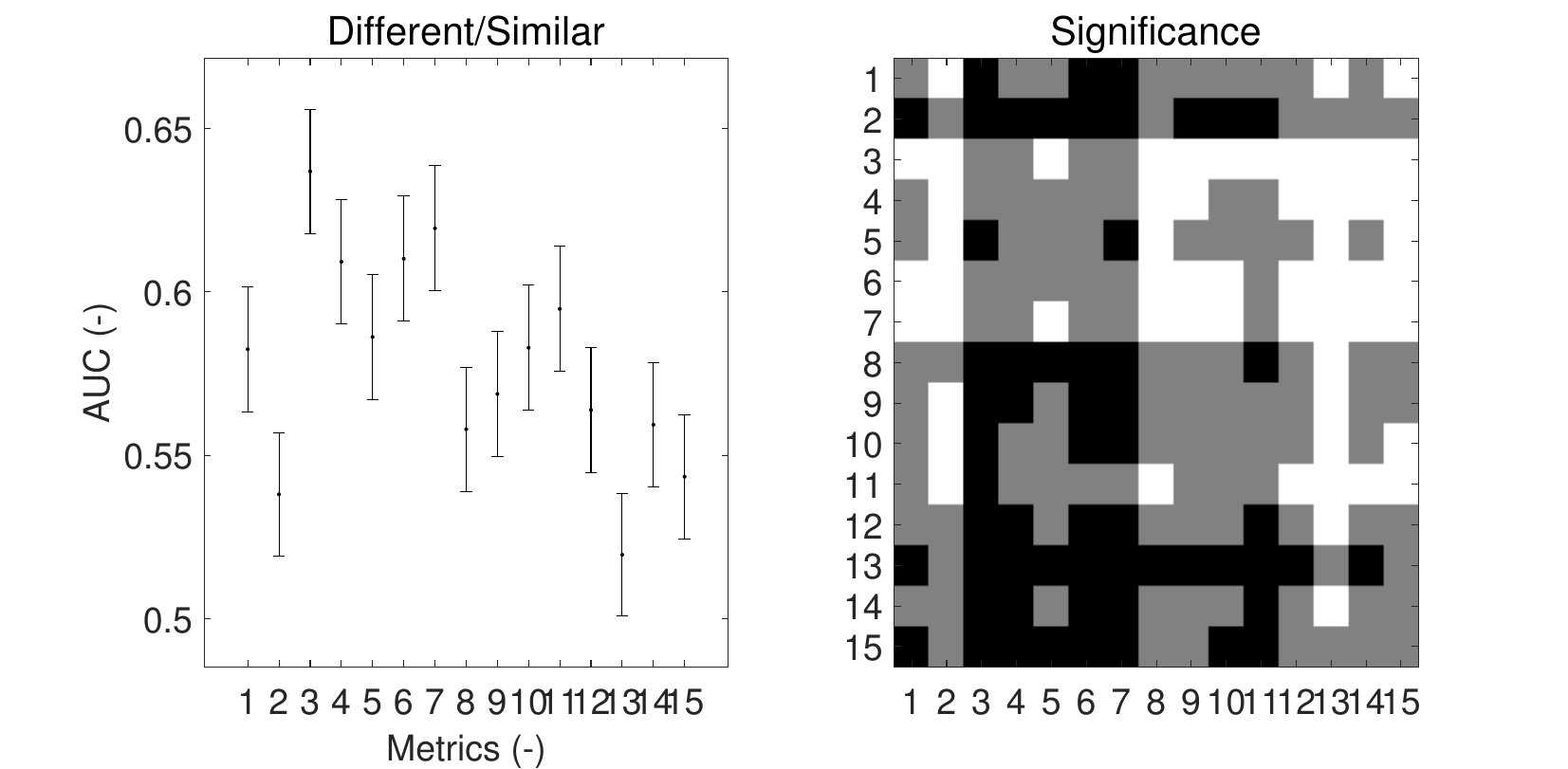}
  \centerline{(e) Different / Similar analysis combining two datasets}\medskip
  \end{minipage}
  \begin{minipage}[b]{0.48\linewidth}
  \centering
  \includegraphics[width=1\linewidth]{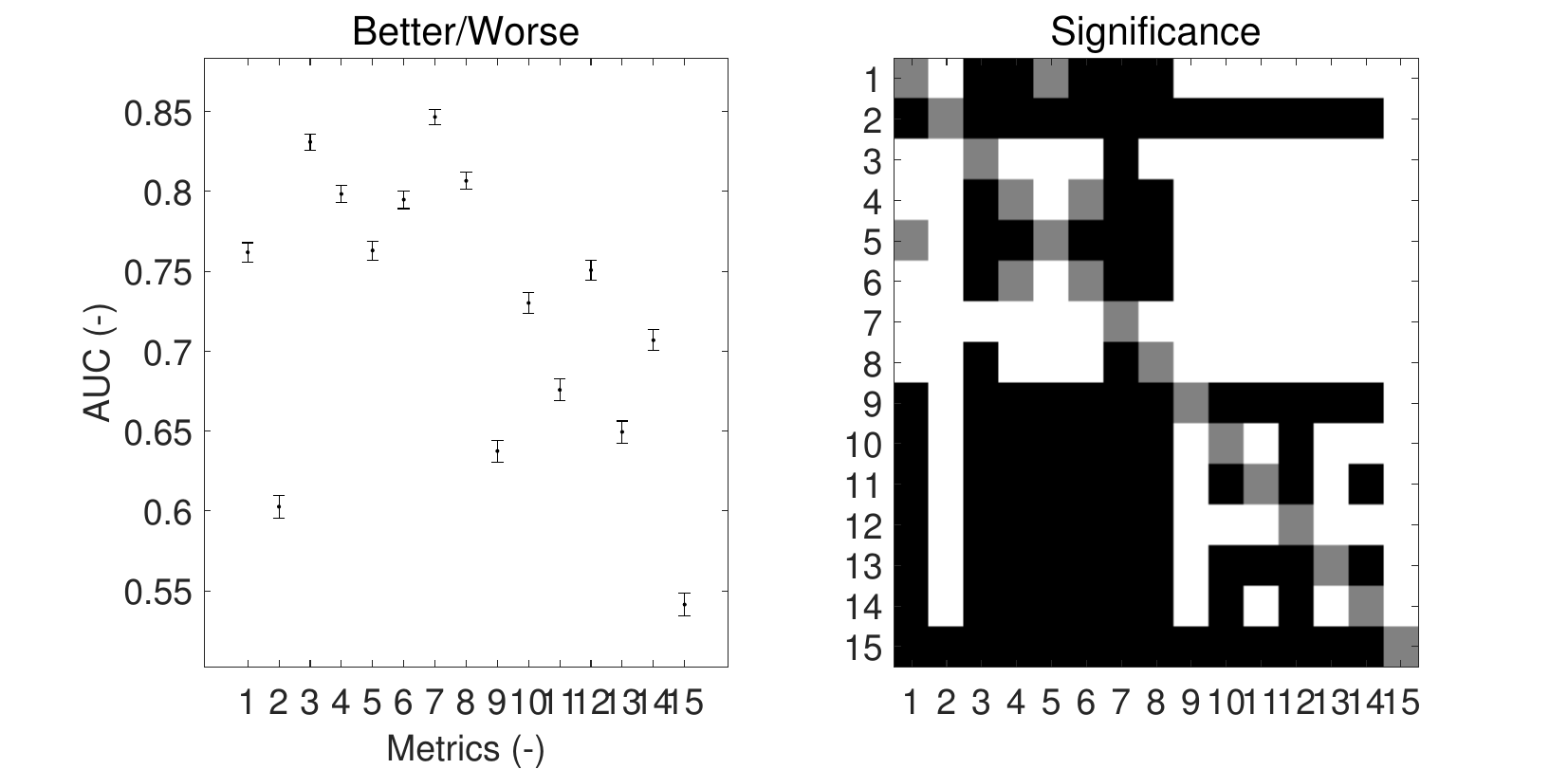}
  \centerline{(f) Better / Worse analysis combining two datasets}\medskip
  \end{minipage}
  \caption{Performance on IVC and IETR image datasets using Krasula's model. The metrics 1-15 indicate PSNR, SSIM, SCDM. MP-PSNRr, MW-PSNRr, EM-IQA, SC-IQA, LOGS, NIQSV+, APT, MNSS, NR\_MWT, OUT, BIQI, BLiindS2 respectively. In the significant test results, the white block indicates that the metric in the row performs significantly better that the metric in the column and vice versa for the black block. The gray block means these two metrics are statistically equivalent.}
  \label{Fig: IVC_IETR_lucas}
\end{figure*}

As introduced in Section II, the MCL-3D dataset does not focus on the DIBR view synthesis distortions, but on the traditional distortion effects on the synthesized views. Thus, the performance{s} of the tested objective metrics are quite different. Some of the metrics (Bosc, VSQA and NR\_MWT) that only consider the DIBR view synthesis distortions perform not as good as the traditional 2D metrics. Some 2D related FR metrics perform even worse than their {backbones}. For instance, VSQA and 3DSwIM can not achieve the performance of SSIM; SCDM, MP-PSNR and MW-PSNR perform worse than PSNR. Among these metrics, the feature-based FR metrics perform better than the simple edge/contour based metrics. It can be inferred that the frequency domain features can represent not only the edge/contour information, but also some other texture characteristics. The SET metric contains not only the DoG features for the DIBR view synthesis distortions, but also the GGCM based features for the texture naturalness. That may explain its good performance on both IVC and MCL-3D datasets.

The IVY dataset considers not only the view synthesis distortion, but also de binocular asymmetry in synthesized stereoscopic images. The baseline distance between the virtual viewpoint and the original viewpoint is much bigger than that in \st{the} other datasets. Thus, the metrics which do not consider the binocular asymmetry perform not well on this dataset.

\subsubsection{\bf{Results of Krasula's model}}
Only the IVC and IETR datasets are tested in this part since the MCL-3D and IVY datasets do not provide the standard deviation which represents the subject uncertainty. The obtained Area Under the Curves (AUC) and significant test results on IVC and IETR are shown in Fig.~\ref{Fig: IVC_IETR_lucas} (a) (b) (c) (d). 
The Fig.~\ref{Fig: IVC_IETR_lucas} (e) and (f) demonstrate the results on the combination of IVC and IETR datasets. A higher AUC value indicates a higher performance. In the significant test results, the white block indicates that the metric in the row performs significantly better that the metric in the column and vice versa for the black block. The gray block means these two metrics are statistically equivalent.

In the first different / similar analysis on the IVC dataset \textsl{cf.} Fig.~\ref{Fig: IVC_IETR_lucas} (a), none of these metrics perform well since most AUC values are below 0.7 and there even exist some metrics whose AUC values are under 0.5. Generally, the DIBR FR metrics perform better than the other metrics. 

In the second different / similar analysis on the IVC dataset \textsl{cf.} Fig.~\ref{Fig: IVC_IETR_lucas} (b), the DIBR-synthesized view dedicated metrics perform significantly better than the 2D metrics (first and last 2 metrics) since the DIBR metrics can achieve higher AUC values. Among these metrics, the SCDM and SC-IQA metrics perform the best, they achieve AUC values higher than 0.9. 

The results on the IETR dataset \textsl{cf.} Fig.~\ref{Fig: IVC_IETR_lucas} (c) (d) and the combination of the two datasets \textsl{cf.} Fig.~\ref{Fig: IVC_IETR_lucas} (e) (f) show that most of the FR metrics outperform the NR metrics except the SSIM metric. The 2D NR metrics achieve similar results compared to their performance on IVC dataset, while the performance of the DIBR NR metrics decrease greatly compared to their performance on IVC dataset. The results of Krasula's model are consistent with the correlation coefficients results in the previous part.

\begin{table*}%[!htbp]
\caption{Performance on the IVC and SIAT DIBR video dataset.}
\label{tab:VIDEO_plcc}
\begin{center}
\begin{tabular}{|c|c|c|c|c|c|c|c|}\toprule
\hline
\multicolumn{2}{|c|}{\multirow{2}{*}{Metric}} & \multicolumn{3}{c|}{IVC video dataset} & \multicolumn{3}{c|}{SIAT video dataset} \\\cline{3-8}
\multicolumn{2}{|c|}{}  & PLCC & RMSE & SROCC & PLCC & RMSE & SROCC\\ \hline
\multirow{2}{*}{FR 2D image metrics}& PSNR & 0.5104 & 0.5690 & 0.4647 &  0.6525 & 0.0972 & 0.6366 \\ \cline{2-8}
&SSIM \cite{Wangssim} & 0.4081 & 0.6041 & 0.3751 &  0.4528 & 0.1144 & 0.4550 \\ \hline
\multirow{2}{*}{FR 2D video metrics}& MOVIE \cite{seshadrinathan2009motion} & 0.4971 & 0.4903 & 0.3877 &  0.646 & 0.097 & 0.693 \\ \cline{2-8}
&ST-RRED \cite{soundararajan2012video} & 0.2025 & 0.6480 & 0.5777 &  \bf{0.7164} & \bf{0.0895} & \bf{0.6971} \\ \hline
%&VQM & 0.--- & 0.--- & 0.--- &  0.669 & 0.095 & 0.655 \\ \hline
\multirow{2}{*}{NR 2D video metrics}& SpEED \cite{bampis2017speed} & 0.3771 & 0.6128 & 0.5952 &  \bf{0.7236} & \bf{0.0885} & \bf{0.6987} \\ \cline{2-8}
&VIIDEO \cite{mittal2015completely} & 0.5971 & 0.5308 & 0.5877 & 0.2586 & 0.1239 & 0.2535 \\ \hline
\multirow{4}{*}{FR DIBR image metrics}
&Bosc \cite{bosc2012edge} & 0.5856 & 0.4602 & 0.2654 & 0.453 & 0.114 & 0.431 \\ \cline{2-8}
&MP-PSNR \cite{sandic2016multi} & 0.5026 & 0.5720 & 0.5478 &  0.5681 & 0.1056 & 0.5044 \\ \cline{2-8}
&MW-PSNR \cite{sandic2016multi} & 0.4911 & 0.4638 & 0.4558 &  0.5745 & 0.1050 & 0.5024 \\ \cline{2-8}
&3DSwIM \cite{battisti2015objective}& 0.4822 & 0.4974 & 0.3320 & 0.5677 & 0.1057 & 0.2762 \\ \hline
%&SIQE & 0.4084 & 0.5138 & 0.0991 &  0.3627 & 0.1195 & 0.2586
%\\ \hline
\multirow{2}{*}{RR DIBR image metrics}&MP-PSNRr \cite{sandic2016dibr}& 0.4617 & 0.5869 & 0.5307 &  0.5640 & 0.1059 & 0.5040 \\ \cline{2-8}
&MW-PSNRr \cite{sandic2016dibr}& 0.4802 & 0.5804 & 0.5038 & 0.5757 & 0.1049 & 0.5853 \\ \hline
\multirow{2}{*}{SV-FR DIBR image metrics}& SIQE \cite{farid2015objective}& 0.4084 & 0.5138 & 0.0991 &  0.3627 & 0.1195 & 0.2586 \\ \cline{2-8}
&DSQM \cite{farid2017perceptual}& 0.5241 & 0.4857 & 0.3157 & 0.4001 & 0.1071 & 0.3994 \\ \hline
\multirow{4}{*}{NR DIBR image metrics}& OUT \cite{jakhetiya2018highly}& 0.6762 & 0.4874 & 0.6151 &  0.0945 & 0.1277 & 0.0926 \\ \cline{2-8}
& NR\_MWT \cite{sandic2019fast}& \bf{0.7530} & \bf{0.4354} & \bf{0.7145} &  0.5051 & 0.1107 & 0.3092 \\ \cline{2-8}
& NIQSV \cite{shishun2017}& 0.6505 & 0.5025 & 0.5963 &  0.5144 & 0.1100 & 0.4562 \\ \cline{2-8}
&MNSS \cite{gu2019multiscale}& 0.5180 & 0.5660 & 0.5371 & 0.1591 & 0.1266 & 0.2463 \\ \hline
\multirow{3}{*}{FR DIBR video metrics}
&CQM \cite{sun2012efficient}& 0.4102 & 0.5101 & 0.3265 & 0.4021 & 0.1070 & 0.4064 \\ \cline{2-8}
&PSPTNR \cite{zhao2010perceptual} & 0.4321 & 0.5002 & 0.4152 &  0.4461 & 0.1069 & 0.4305 \\ \cline{2-8}
&VQA-SIAT \cite{liu2015subjective}& 0.5943 & 0.5321 & 0.5879 & \bf{0.8527} & \bf{0.0670} & \bf{0.8583} \\ \hline
\multirow{2}{*}{NR DIBR video metrics}&CTI \cite{kim2016measurement} & \bf{0.6821} & \bf{0.4372} & \bf{0.6896} &  0.5736 & 0.1053 & 0.5425 \\ \cline{2-8}
&FDI \cite{zhou2018no}& \bf{0.7576} & \bf{0.4319} & \bf{0.7162} & 0.5952 & 0.1033 & 0.5425 \\ \hline
\end{tabular}
\end{center}
\end{table*}

\begin{figure*}
\centering
  \begin{minipage}[b]{0.48\linewidth}
  \centering
  \includegraphics[width=1\linewidth]{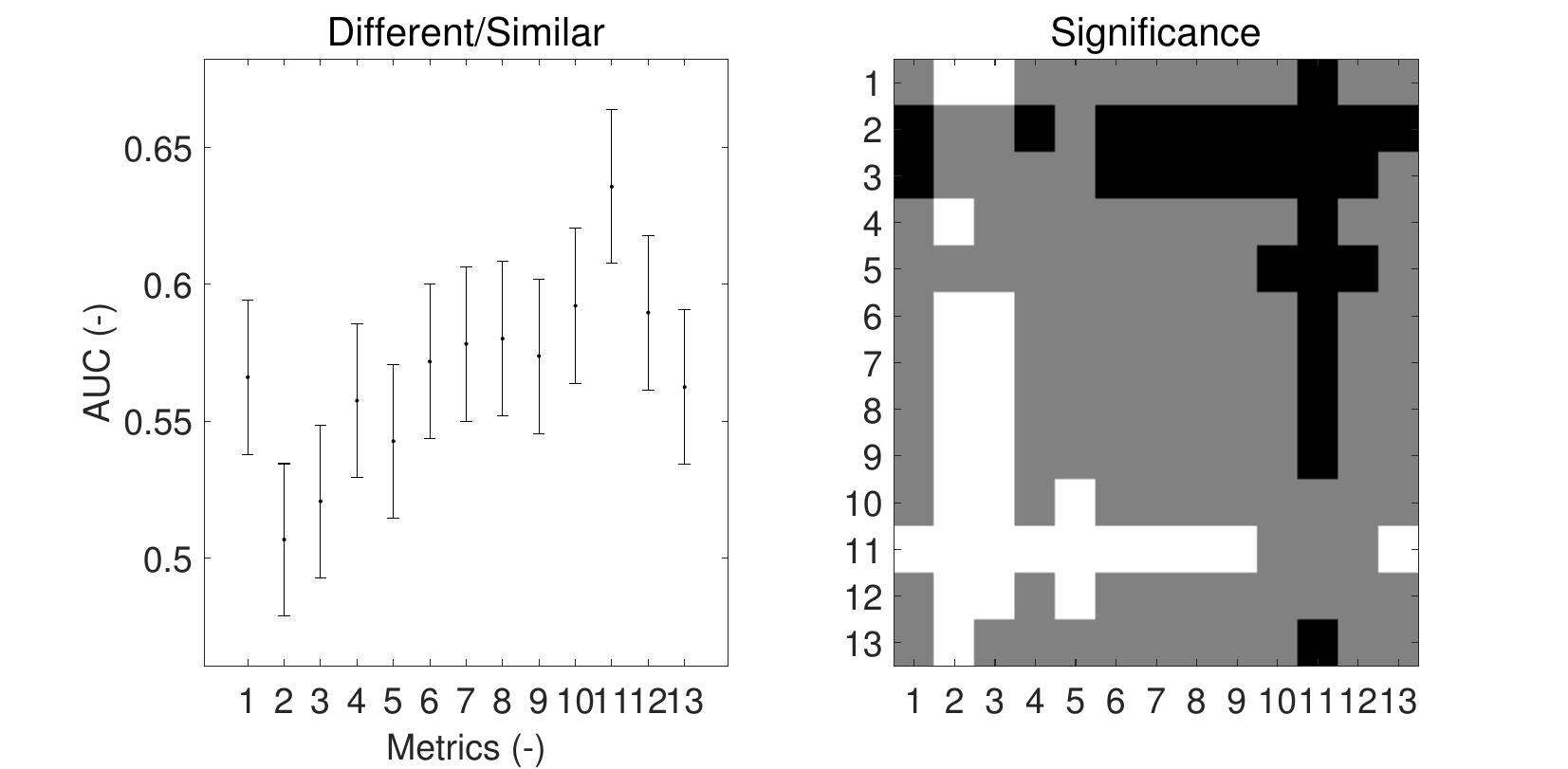}
  \centerline{(a) Different / Similar analysis on IVC Video dataset}\medskip
  \end{minipage}
  \begin{minipage}[b]{0.48\linewidth}
  \centering
  \includegraphics[width=1\linewidth]{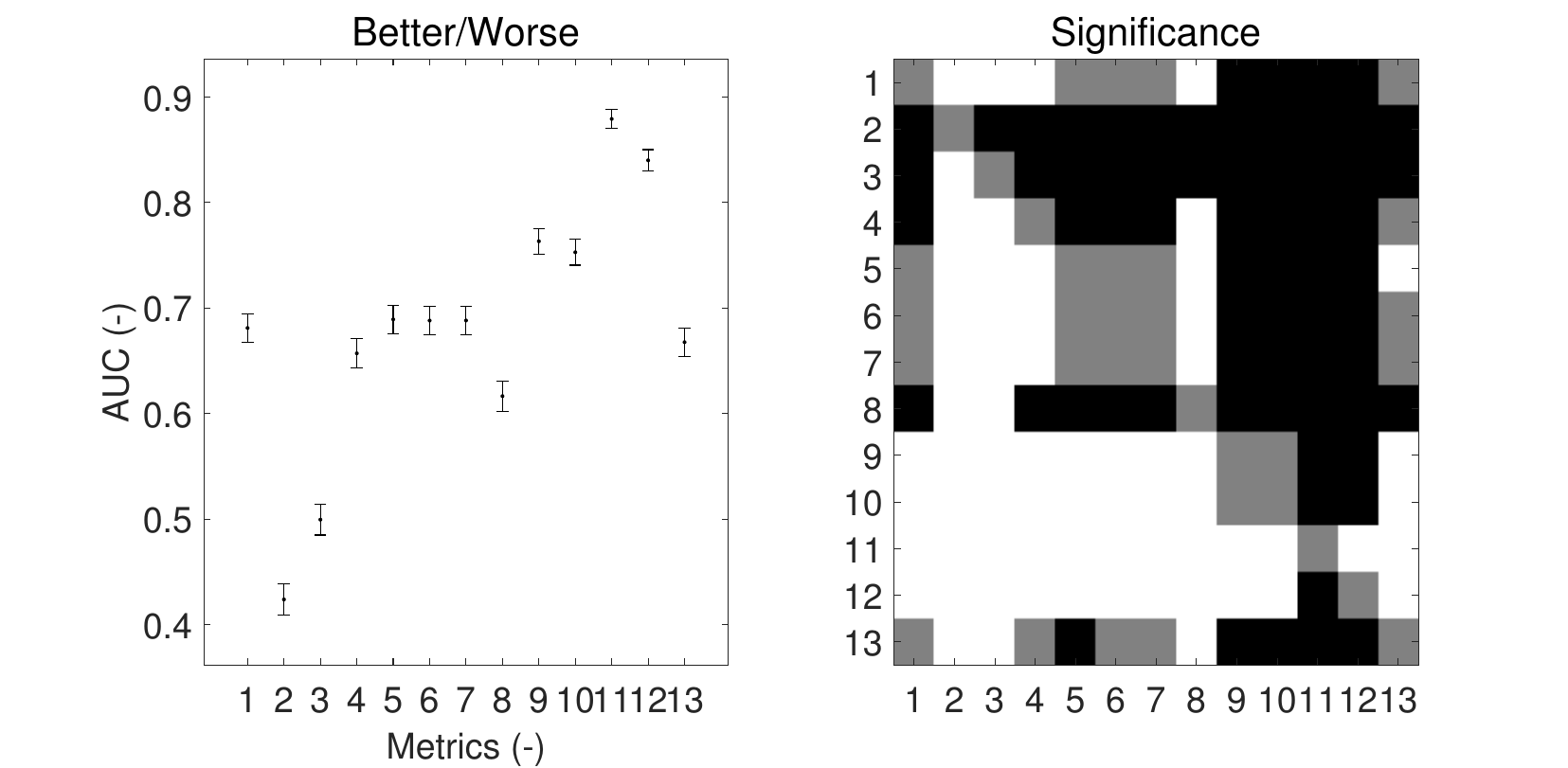}
  \centerline{(b) Better / Worse analysis on IVC Video dataset}\medskip
  \end{minipage}
  \caption{Performance on IVC video dataset using Krasula's model. The metrics 1-13 represent: PSNR, SSIM, SpEED, ST-RRED, VIIDEO, MP-PSNRr, MW-PSNRr, NIQSV, OUT, MNSS, NR\_MWT, FDI, SIAT-VQA respectively. In the significant test results, the white block indicates that the metric in the row performs significantly better that the metric in the column and vice versa for the black block. The gray block means these two metrics are statistically equivalent.}
  \label{Fig: IVC_Video_lucas}
\end{figure*}

\subsection{Performance on DIBR video datasets}
The DIBR-synthesized videos contain some temporal distortions, such as flickering, in addition to the spatial distortions in images. In this experiment, 12 state-of-the-art DIBR image metrics in addition to 5 DIBR video metrics are tested. To compare the performance of DIBR metrics and traditional 2D metrics, 5 widely used 2D video metrics and 2 2D image metrics are tested. The quality scores of image metrics are obtained by averaging the quality of all the frames. The three metrics which performance the best among the BIQA methods are marked in bold. 

The obtained PLCC, RMSE and SROCC values on IVC video and SIAT video datasets are given in Table~\ref{tab:VIDEO_plcc}. Only the results of Krasula's model on IVC video dataset are shown in Fig.~\ref{Fig: IVC_Video_lucas} since the SIAT video dataset does not provide the uncertainty of subject ratings.

The IVC video dataset focuses on the DIBR view synthesis distortions while the SIAT dataset focuses on the compression effect{s} on the synthesized views. {We can easily observe that the best three metrics on IVC dataset are all DIBR metrics while the best three metrics on SIAT dataset are VQA-SIAT and two 2D metrics.} The VQA-SIAT metric mainly focuses on the compression effect which may lead obvious flicker in the DIBR-synthesized views. The spatial view synthesis distortions considered in this metric are very limited. That may explain why it significantly outperforms the other metrics on SIAT dataset while it can not obtain a very good performance on the IVC dataset. When we focus on the IVC video dataset, none of FR metrics achieve{s a} high correlation with the subjective results. Moreover, there is no significant difference between the performance{s} of DIBR FR and 2D FR metrics. However, the DIBR NR metrics perform the best compared to other metrics{, also due to} the global shift effect. 

\subsection{Discussions}
The experimental results show that although great progress has been made towards the quality assessment of synthesized views, there is still {a large} room for improvement.

\subsubsection{\bf{Synthesized video quality assessment}} 
The DIBR-synthesized videos contain not only the compression distortions but also the distortions induced by DIBR. The VQA-SIAT metric works well on capturing the temporal flicker caused by video compression, but it fails to assess the DIBR view synthesis distortions in the synthesized video frames. In addition, the imperfect view synthesis algorithms may also result in great miss-match between the adjacent frames in the synthesized video, which causes very annoying temporal distortions that the 8 by 8 block matching (in VQA-SIAT) may fail to detect. Therefore, we could try to further analyse the specific spatial-temporal distortions in the synthesized videos and design a complete metric for the DIBR-synthesized videos.

\subsubsection{\bf{Quality assessment of synthesized views in real applications}}
As introduced previously, DIBR can be used in various applications, but the quality assessment for these applications are rarely researched. For example, the free viewpoint videos (FVV) and multi-view videos (MVV) provide the images from multiple viewpoints at the same time instant. The temporal distortions in FVV or MVV are mainly introduced by the changing of viewpoints instead of timeline \cite{ling2019prediction, ling2019perceptual}. This type of distortions are different from that in normal DIBR-synthesized views videos. Besides, in order to provide immersive perception for the observer, the AR or VR applications need to generate multiple synthesized images and change the viewpoint with the motion of the observer. The synthesized video contains both the inter-frame and inter-viewpoint temporal distortions, as well as the binocular asymmetric distortions which may happen in stereoscopic applications \cite{jung2016critical}. It could be interesting to try to design the metrics for these applications since they are currently rarely explored.

\subsubsection{\bf{Deep learning approaches}}
The main limitation of the usage of deep learning on the quality assessment of DIBR-synthesized views is the limited size of available dataset{s}. Unlike the homogeneous distortions in the traditional 2D images, the distortions in the DIBR-synthesized views mostly occur in the dis-occlusion regions. In other words, the major part of the DIBR-synthesized view holds a perfect quality. {Thus we cannot split the synthesized image into several patches and then directly use the quality of the whole image as the quality of the patches.} Creating a very large-scale dataset may significantly help to train a deep model. But unlike the datasets for other tasks {, \textsl{e.g.}} object recognition, creating an image quality dataset necessarily requires subjective tests which are quite expensive and time-consuming. Thus, exploring how to train a comprehensive model on limited data could be more practical, {maybe via} one-shot learning or few-shot learning \cite{fei2006one, snell2017prototypical}. 
%\textcolor{blue}{Besides, since the ranking of predicted image quality scores (monotonicity) is also an important index to evaluate the performance of an IQA metric, in addition to their individual scores (precision), 
{Besides, in addition to the individual predicted image quality scores (precision), the ranking of the predicted scores (monotonicity) is also an important index to evaluate the performance of an IQA metric. Therefore, 
learning from rankings \cite{liu2017rankiqa, jiang2019no} may help to solve the problem of IQA dataset size limit. Firstly, the ranked image sets can be automatically generated without subjective tests \cite{liu2017rankiqa}. We can pre-train our model on the generated ranked image sets and then fine-tune it on the target IQA datasets. Secondly, a reliable ranking loss can enhance the ability of the model to rank images in terms of quality and thus help to generate more precise quality scores \cite{jiang2019no}.} The fact that quality score of the whole synthesized image can not directly be distributed to all the image patches does not mean that the image can not be processed patch by patch. The main challenge is to find a proper pooling method to get the overall quality score.
Although the pre-trained deep features have been successfully used in metrics \cite{wang2019deepicip, ling2019gannrm}, more effort{s} could be made to create a more general and effective end-to-end deep model.

\section{Conclusion}
In this paper, we present an up-to-date overview for the quality assessment methods of DIBR-synthesized views. We firstly described the existing DIBR-synthesized view datasets. Secondly, we analysed and discussed the recently proposed state-of-the-art objective quality metrics for DIBR-synthesized views, and classified them into different categories based on their \st{used} approaches. Then, we conducted a reliable experiment to compare the performance of each metric, and analysed their advantages and disadvantages \st{at the same time}. Furthermore, we discussed the potential challenges and directions for future research. We hope this overview can help to better understand the state-of-the-art of this research topic and provide insights to design better metrics and experiments for effective DIBR-synthesized images/videos quality evaluation.

\section*{Acknowledgment}

The authors would like to thank Dr. Suiyi Ling and Dr. Yu Zhou for sharing their code. We would also like to thank Prof. Patrick Le Callet and Dr. Lucas Krasula for their kind advices on the experiment.
This work was supported in part by the NSFC Project under Grants 61771321 and 61872429, in part by the Guangdong Key Research Platform of Universities under Grants 2018WCXTD015, in part by the Natural Science Foundation of Guangdong Province, China, under Grants 2020A1515010959, and in part by the Interdisciplinary Innovation Team of Shenzhen University. 

\bibliographystyle{elsarticle-num}
\bibliography{elsarticle}

\end{document}